\documentclass[onecolumn]{els-mrw} 
\usepackage[utf8]{inputenc}
\usepackage{amsmath,amssymb,amsfonts,amsthm,makeidx,graphicx}
\usepackage{txfonts}
\usepackage{helvet}
\usepackage{color}
\usepackage{hyperref}
\renewcommand{\vec}[1]{\mathbf{#1}}  % -> for vectors, roman and bold

\newcommand{\init}[1]{{#1}^{(\textrm{i})}}
\newcommand{\fin}[1]{{#1}^{(\textrm{f})}}

\newcommand{\vecfin}[1]{\fin{\vec{#1}}}
\newcommand{\vecinit}[1]{\init{\vec{#1}}}

\newcommand{\tE}{{t_{\mathrm{E}}}}    % -> Ehrenfest time
\newcommand{\tH}{{t_{\mathrm{H}}}}    % -> Heisenberg time
\newcommand{\heff}{{\hbar_{\mathrm{eff}}}} % -> effective hbar
\newcommand{\lsp}{{\lambda_{\mathrm{sp}}}} % -> mean field Lyapunov exponent
\newcommand{\rhosp}{{\rho_{\mathrm{sp}}}} % -> rho-single particle

\newcommand{\ud}{\mathrm{d}}

\begin{document}

\chapter{Quantum chaos in many-body systems of indistinguishable particles}\label{chapUK}

\author[1]{Juan-Diego Urbina}
\author[1]{Klaus Richter}
\address[1]{\orgname{Institut f\"ur Theoretische Physik, Universit\"at Regensburg, 93040 Regensburg, Germany}}

\maketitle

\begin{glossary}[Keywords]
Quantum chaos, many-body systems, discrete bosonic fields, nonlinear field equations, thermodynamic limit, equilibration.
\end{glossary}

\begin{abstract}[Abstract]
In quantum systems with a classical limit, advanced semiclassical methods provide the crucial link between phase-space structures, reflecting the distinction between chaotic, mixed or integrable classical dynamics, and the corresponding  quantum properties. Well established techniques dealing with ergodic wave interference in the usual semiclassical limit $\hbar \rightarrow 0$, where the classical limit is given by Hamiltonian mechanics of particles, constitute a now standard part of the toolkit of theoretical physics. During the last years, these ideas have been extended into the field theoretical domain of systems composed of $N$ indistinguishable particles, aka quantum fields, displaying a different type of semiclassical limit $\heff = 1/N \rightarrow 0$ and accounting for genuine many-body quantum interference. The foundational concept behind this idea of many-body interference, the many-body version of the van Vleck-Gutzwiller's  semiclassical propagator, is explained in detail. Based on this the corresponding semiclassical many-body theory is reviewed. It provides a unified framework for understanding a variety of quantum chaotic phenomena addressed, including random-matrix spectral correlations in many-body systems, the universal morphology of many-body eigenstates, interference effects kin to mesoscopic weak localization, and the key to the scrambling of many-body correlations characterized by out-of-time-order correlators.
\end{abstract}

\section{Introduction: the different facets of emergent classicality}
\label{sec:Introduction}

\subsection{Dynamical complexity in systems of identical particles}
\label{sec:}

Even before the successes that the field of quantum chaos has gained during the last several decades, marked by the insight that dynamical complexity does not require a large number of degrees of freedom, Bohr’s compound nucleus model~\cite{Bohr36} may be viewed as the first quantum chaotic system, although at that time there was no concrete association with classically chaotic dynamics. Later, Wigner's foundational work on random matrix ensembles~\cite{Wigner55, Wigner58} evolved into a definitional property of quantum chaos in a broader sense~\cite{Haake06, Bohigas88, Guhr98, RevModPhys.69.731, StockmannBook, Mehta04} in terms of random matrix universality, based on Random Matrix Theory (RMT). Remarkably, the description of systems of indistinguishable particles, fermions in the case of nuclei, lies at the very heart of what eventually became the field of quantum chaos.

The missing link, namely that the essential dynamical mechanism responsible for the emergence of RMT behavior happens already at the level of individual systems (as opposed to ensembles) is provided by the notion of classically chaotic dynamics (absent in RMT), imprinting itself into the quantum properties of the system, i.e. a direct link between quantum and chaos. The way this happens was revealed in a seminal series of papers~\cite{Gutzwiller70,Gutzwiller71,Gutzwiller71a} where, starting from Feynman's path integral, Gutzwiller derived a semiclassical trace formula expressing a single-particle (SP)  quantum spectrum as a sum over contributions from unstable classical periodic orbits. More than fifty years ago, he thereby set the cornerstones of the bridge connecting the classical and quantum mechanics of non-integrable systems~\cite{Haake06,StockmannBook,Gutzwiller90,BraBha03Book}. This constitutes a spectacular example of the use of semiclassical methods --the study of the regime of large actions compared with $\hbar$ by means of asymptotic analysis of path integrals -- in systems that allow for an unambiguous identification of their classical limit, and whose Hilbert space contains a sector of large actions, the semiclassical regime.

The impressive success of Gutzwiller's theory and related ideas in SP systems opened the door for its use to address~\cite{Ezra91} foundational questions that were left unanswered since the failure of the "old quantum theory", that can be understood as an incomplete precursor of the semiclassical program. During the decades after Gutzwiller's seminal papers the new form of the semiclassical approximation was essential to study quantum properties near the quantum-classical transition, from atoms~\cite{Wintgen87} to mesoscopic electronic systems~\cite{Klaus1}, while providing the first glimpses of the dynamical origin of universal spectral fluctuations, among many other features.

As in any foundational enterprise, focus was mostly, but by no means uniquely, put on conceptually simple systems where the signatures of quantum chaos could be cleanly isolated and studied by means of exact numerical simulations or anayltical methods. In particular, quantum chaos in the realm of many-body (MB) systems, with their field theoretical description naturally based on discrete or continuous quantum fields, was largely unexplored and has remained as a notable exception left untouched by the powerful methods of modern semiclassics. This happened despite the obvious interest that semiclassical quantization a l'a Gutzwiller had created both, in the high energy community or in view of an extension of the old quantization rules applicable only to integrable systems.

After the methods of semiclassical quantization reached full maturity in the 2000's, and simultaneously with the swift development of MB quantum chaos ideas in the framework of systems without a semiclassical regime, a strong interest in the semiclassical quantization of discrete quantum fields emerged. The development of these methods and the way they lift the notion of quantum chaos for systems where the classical limit describes particle degrees of freedom into the Fock space of discrete quantum fields is the main subject of this chapter.

\subsection{Semiclassical regimes of quantum many-body systems of identical particles}
\label{subsec:limits}

Referring to chaos, an inherently classical concept, requires properly defined notions of classical and semiclassical limits in MB physics. The term ''semiclassical'' is here being used in the very sense, just as in quantum chaos, as the crossover regime between the classical and quantum worlds.  Semiclassical theory is then formally based on asymptotic (effective) $\hbar$ expansions of quantum mechanical (MB) Feynman propagators. The resulting semiclassical expressions, although based on classical quantities for input, nevertheless fully account for quantum (or wave) interference as an integral part of the theory. In this sense, semiclassical methods take a special place in the toolbox of approximate schemes, as their use of classical structures does not clash with the kinematical structure of quantum mechanics embodied in fundamental concepts like unitarity of quantum evolution and linearity of the space state. This is because the semiclassical approximation respects and incorporates the coherent nature of quantum mechanical amplitudes.

There are several types of MB quantum chaotic systems with a sector of their state space that reside at such a semiclassical interface between non-integrable MB quantum and classical dynamics. This occurs in a two-fold way:  First, far out-of-equilibrium quantum dynamics is associated with high-energy excitations and thereby with the usual short-wavelength limit, {\em i.e.}\ small $\hbar$.  Alternatively, the  limit of large particle numbers $N$ can also be regarded as semiclassical, governed by an effective Planck constant $\heff = 1/N$. Whereas this analogy can be made rigorous in the case of non-dilute systems where $N/L \gg 1$, with $L$ the number of possible SP microstates, in which case it simply corresponds with the notion of a thermodynamic limit with large enough densities, there is some evidence of its validity even for dilute regimes characteristic of fermionic systems.  We therefore consider MB chaotic quantum systems in the limits where either $\hbar$ or $\heff$ is small but nonzero.  Both types of quantum-classical transitions are singular implying disruptive changes and complexity. Typically, these systems require exceedingly difficult numerical quantum simulations due to vastly growing Hilbert space dimensions. Thus, there has been a quest for MB methods specifically devised for these two complementary crossover regimes. In the following, the underlying concepts and challenges of a corresponding semiclassical MB theory are outlined.

%%%%%%%%%%%%%%%%%%%%%%%%%%%%%%%%%%%%%%%%%%%%%%%%%%%%%%%%%%%%%%%%%%

\begin{figure}
    \centering
  \includegraphics[width=0.8\linewidth]{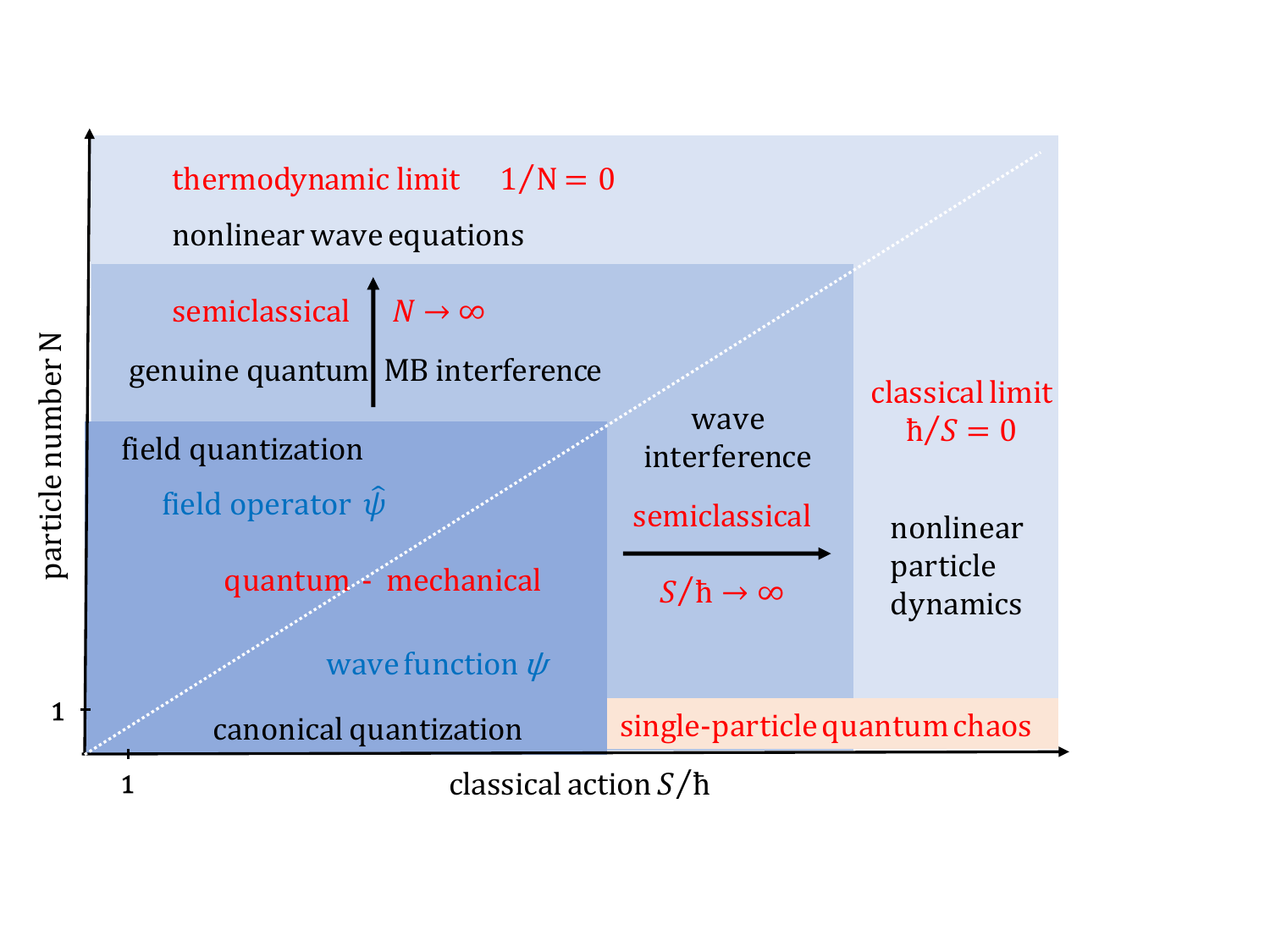}
    \caption{\label{fig:sc-limits}
   {\bf
   Emergent classical limits and semiclassical regimes} | When the number of particles $N$ is kept fixed, a semiclassical regime of large actions $S/\hbar \rightarrow \infty$, often loosely referred to as $\hbar \rightarrow 0$, is related with a classical limit that emerges from the transition from quantum mechanical waves to classical particles, nearly always involving nonlinear Hamiltonian dynamics. Semiclassical theory within the field of quantum chaos has traditionally addressed single-particle systems, i.e.~the lower right zone of the graph.  In quantum field theory another classical limit and corresponding semiclassical regime emerges, this time along 
   the vertical direction of increasing particle number $N$, consistent with what is usually considered as thermodynamic limit (for non-dilute systems). The quantum-classical transition corresponds now to a different, but complementary, semiclassical regime with effective $\heff= 1/N \rightarrow 0$ where nonlinear wave equations, also with a Hamiltonian structure, play the role of classical limit.
   }
\end{figure}
%%%%%%%%%%%%%%%%%%%%%%%%%%%%%%%%%%%%%%%%%%%%%%%%%%%%%%%%%%%%%%%%%%%

Consider the familiar case of a SP quantum system with an existing classical limit that is approached in the limit $\hbar \rightarrow 0$.  More precisely, the semiclassical limit is one in which the dimensionless ratio $\hbar/S \ll 1$ with $S=\int {\bf p} d{\bf q}$, a typical classical action of the particle with momentum $p$. This is the standard limit of short wave lengths $\lambda$ in view of the de Broglie relation $\lambda = h/p$.
In the schematic Fig.~\ref{fig:sc-limits} with horizontal scale $S/\hbar$ and vertical scale denoting the particle number $N$, this semiclassical limit corresponds to the horizontal crossover  from the deep quantum regime at $S/\hbar \sim 1$ into the semiclassical range where (for $N=1$) SP wave mechanics approaches classical mechanics, {\em i.e.} classical particles usually possessing nonlinear, possibly chaotic dynamics.

Since Gutzwiller's and Berry's early works~\cite{Gutzwiller71, Berry76}, semiclassical approaches in quantum chaos have nearly exclusively been focused on the case $N=1$, {\em i.e.}~in the lower right region of the $S$-$N$-landscape of Fig.~\ref{fig:sc-limits}.  This region is the subject of several textbooks~\cite{Haake06, StockmannBook, Gutzwiller90,BraBha03Book, Reichl21}.  However, the limit $\hbar/S \rightarrow 0$  also formally applies to systems with more than one particle in $D$ dimensions by considering semiclassical methods applied in a respective $2D\!\cdot\! N$-dimensional phase space. In Fig.~\ref{fig:sc-limits} this corresponds to moving vertically upwards in the right semiclassical regime and considering the limit $\hbar \rightarrow 0$ for given $N$. Extending such semiclassical approaches from one to $N$ particles is accompanied by a variety of notable challenges, from the difficulty in visualizing high-dimensional phase spaces, to the correct incorporation of genuinely MB effects like particle symmetry (bosonic vs fermionic) and interactions. Consequently, corresponding attempts have been rare, even for (non-integrable) few-particle systems.  

An early example is the successful semiclassical quantization of the correlated Coulomb dynamics of the electrons in the helium atom, a longstanding problem dating prior to Schrödinger and his equation, i.e.~to the `old quantum theory'~\cite{Bohr13a},
% Bohr13b, Bohr13c, Langmuir21}
 see \cite{KlausH1,Kragh12} for reviews of the history.  By applying a  cycle expansion to chaotic dynamics in 6-dimensional phase space ground and excited states of helium could be semiclassically computed with high precision~\cite{Ezra91}. 
 In Ref.~\cite{Primack98} the accuracy of the semiclassical trace formula for systems with more than two degrees of freedom was considered and numerically studied for the three-dimensional Sinai-billiard.
 More recently, the dynamics of quantum maps with up to 8-dimensional phase spaces has been visualized and thoroughly investigated~\cite{Richter14}. Furthermore, interesting collective MB dynamics were recently semiclassically identified in kicked spin-chains up to particle numbers of order $N\sim 20$~\cite{Akila17,Waltner17} making use of a remarkable particle number-time duality in the Ising spin chain~\cite{Akila16}. 

Summarizing previous attempts for for generalizations of the van Vleck-Gutzwiller propagator~\cite{Gutzwiller71} to genuinely MB systems, in~\cite{Weidenmueller1993} Gutzwiller’s trace formula for the density of states $\rho(E,N)$ for systems of non-interacting identical particles, in particular fermions that, however, being based on classical SP phase space, remains purely formal without a direct interpretation in terms of MB phase space. Starting from a semiclassical theory for many interacting fermions (for a review see~\cite{Ullmo_2008}), in~\cite{Ullmo98} it was shown that the orbital magnetic response can be greatly enhanced by the combined effects of interactions and finite size.  In the context of MB scattering~\cite{Urbina16} contains a semiclassical calculation of the transmission probabilities through mesoscopic cavities for systems of many non-interacting particles, in particular photons, thereby generalizing advanced semiclassical SP techniques~\cite{Richter02, Mueller09, Berkolaiko12} for scattering in chaotic conductors.  There, high complexity, arising from the interplay between interference at the SP level and quantum indistinguishability, leads to specific universal correlations in MB scattering properties with relevance for boson sampling~\cite{Aaronson10} and the Hong-Ou-Mandel effect~\cite{Hong87}. In the context of large-spin chains, remarkable progress has been achieved by Gutkin and co-authors providing certain classical foundations of many particle chaos based on models for coupled cat maps~\cite{Gutkin16,Gutkin21}. Very recently, a chaotic scalar lattice field theory (in one dimension) has been proposed~\cite{Laksh11,Gutkin21}, complementary in spirit to Gutzwiller's periodic-orbit approach for low-dimensional chaotic dynamics~\cite{Gutb}.

%%%%%%%%%%%%%%%%%%%%%%%%%%%%%%%%%%%%%%%%%%%%%%%%%%%%%%%%%%%%%%%%%%%

Besides the usual notion of $S/\hbar\rightarrow \infty$ discussed so far, in quantum field theory there is the complementary limit of large particle number $N$, but not necessarily small $\hbar$. In the limit $N=\infty$, and as long as the typical occupations of SP states are large enough (the non-dilute case), a new asymptotic regime emerges where nonlinear wave equations characteristic of the thermodynamic limit appear as a kind of `classical' fluid dynamics, such as governed by the Gross-Pitaevskii equation. This regime of large but not infinite $N$, corresponding to the upper (left) region in Fig.~\ref{fig:sc-limits}, can be associated with an effective Planck constant $\heff =1/N \ll 1$ and hence also be considered semiclassical.

Wave interference is built into semiclassical propagators as coherent sums over classical paths in configuration space with interfering amplitudes, leading for instance to the van Vleck propagator~\cite{Vanvleck28} or the Gutzwiller trace formula~\cite{Gutzwiller70} 
for the oscillatory part of the level density in terms of unstable periodic orbits of classical particles.  In the complementary limit, $\heff = 1/N \ll 1$, MB propagators are in turn formally described also by means of semiclassical sums over paths defined by classical field solutions, with a completely different interpretation and meaning.  The summations are taken over collective modes of MB densities in a phase space where a continuum version of the MB Fock space plays the role of configuration space~\cite{Engl14}, instead of particle trajectories in real space, as is outlined in Sec.~\ref{sec:SC-MB}. These Fock-space paths represent the various, in principle infinitely many, time-dependent solutions of the nonlinear wave equations in the classical limit $1/N = 0$ (upper region of Fig.~\ref{fig:sc-limits}).  Quantum MB interactions turn into nonlinearities in these wave equations and may result in unstable, possibly chaotic MB mode dynamics.  In this way chaos at the level of these classical-limit nonlinear waves implies genuine MB interference and {\em many-body quantum chaos} at the level of quantum fields\footnote{There are other conceivable routes to MB quantum chaos not considered further in this contribution. This issue is revisited for in the outlook Sec.~\ref{sec:persp}.}.  
This is entirely analogous to signatures of chaotic classical particle dynamics in wave functions at the Schrödinger equation level, i.e. quantum chaos in the limit $\hbar \rightarrow 0$.  In a sense, such an approach promotes Gutzwiller's semiclassical theory of the time evolution operator from the level of ``first quantization'' to that of ``second quantization''.  Note, however, that the classical quantities entering semiclassical path integrals have different meanings in the two complementary limits: for instance different Lyapunov exponents quantify the instability of particle trajectories and collective modes, respectively.
Remarkably, the semiclassical theory in the limit $\heff = 1/N \ll 1$ also applies to ground or low-lying excited MB states.

Although by no means obvious,  the classical paths in MB space, i.e.~the time-dependent solutions of the nonlinear wave equations, turn out to just represent mean-field solutions of the full MB problem.  This opens a totally new perspective on the connections between chaotic mean-field dynamics, quantum correlations due to MB interactions, scrambling, and the generation of entanglement.  MB interaction effects beyond mean-field are commonly considered as correlation effects~\cite{Fulde95}. Hence, as will be explained in Sec.~\ref{sec:SC-MB}, the interpretation of quantum MB amplitudes in the semiclassical regime as a coherent sum over different collective mean-field modes implies that massive MB interference between these chaotic modes describes or explains quantum correlations in the MB propagator. Hence MB quantum chaos and quantum correlation phenomena are intimately intertwined.  To highlight the difference between (SP) wave and MB quantum interference we coin the term for the latter case {\em genuine MB quantum interference}.

%%%%%%%%%%%%%%%%%%%%%%%%%%%%%%%%%%%%%%%%%%%%%%%%%%%%%%%%%%

\section{Universality in many-body quantum chaos from a semiclassical perspective}
\label{subsec:outline}

The semiclassical methods reviewed below in this chapter address these leading-order (in $\heff=1/N$) MB quantum mechanical contributions to the thermodynamic limit for discrete, bosonic quantum fields. They can be applied to broad dynamical regimes from fully chaotic, partially chaotic, or even to integrable mean-field dynamics. Thereby, on the one hand these semiclassical techniques can be applied to systems not behaving ergodically and in a universal manner and hence allow for addressing system specific individual, possibly quite atypical properties.  On the other hand, the MB semiclassical approach provides dynamical foundations for universal aspects of quantum chaotic MB systems. We focus here on the way the new machinery of MB semiclassics lifts the well established quantum signatures of chaos in SP systems into the MB domain. We will review essential aspects such as the emergence of universal, RMT-like spectral fluctuations, universal Fock-space correlations of MB chaotic eigenstates and the saturation of the scrambling as witnessed by the Out-Of-Time-Ordered correlators, primarily based on the recent accomplishments in Refs.~\cite{Engl14,Engl15,TF_Remy,Rammensee2018,Schoeppl2025}. 
Starting from fully chaotic MB mean-field dynamics, this approach invokes ergodic properties and corresponding sum rules for the exponentially numerous classical paths, i.e.~collective modes, entering into semiclassical expressions for the various MB correlation functions. Such assumptions often enable an analytical treatment of the arising multiple sums over chaotic MB modes. 

%%%%%%%%%%%%%%%%%%%%%%%%%%%%%%%%%%%%%%%%%%%%%%%%%%%%%%%%%%%%%%%%%
\subsection{Semiclassical theory of single-particle quantum chaos}
\label{sec:SC-SP}
\begin{figure}
  \centering
  \includegraphics[width=0.8\linewidth]{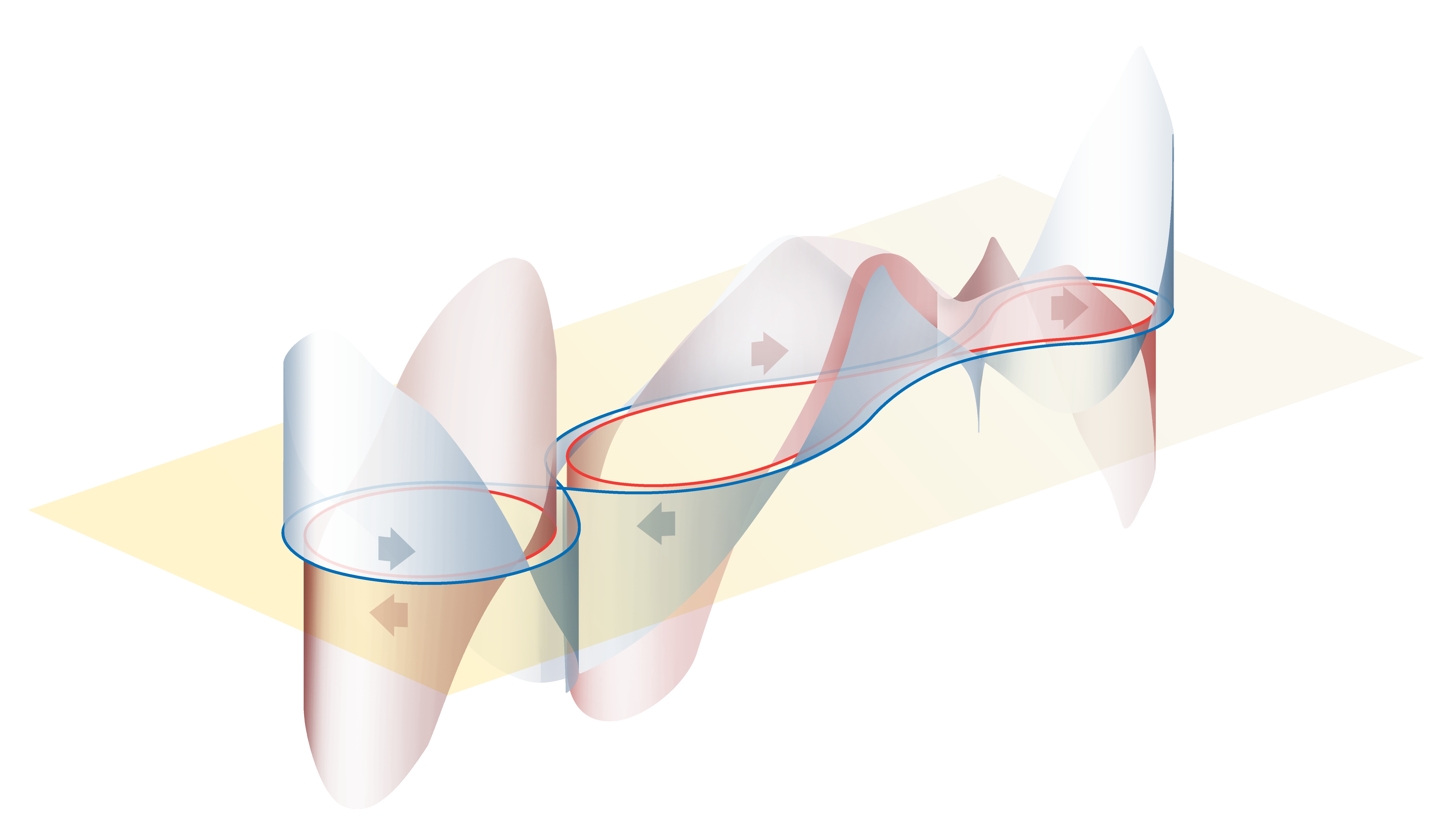}
    \caption{
   {\bf
   Correlated periodic orbits} | Phase-space sketch of classical periodic single-particle orbits that are nearly identical up to encounter regions, located at self-crossings in a two-dimensional configuration space (yellow $xy$-plane). Vertical components indicate the respective momenta in $y$-direction  
  (taken from Ref.~\cite{Haake11} with permission). \label{fig:po-pair-HR}}
\end{figure}

Starting from Feynman's path integral and invoking corresponding stationary phase approximations in the semiclassical regime $\hbar \to 0$ naturally leads to expressing unitary quantum time evolution in terms of sums over phase-carrying classical solutions. Quantum interference, as a consequence of the principle of superposition, is captured by the existence of multiple classical solutions and their coherent summation. Dependening on the structure of the quantum observables to be considered, for instance spatial or spectral $n$-point correlation functions, this leads to a corresponding number of summations over classical paths. Modern semiclassical theory is built from  the challenge of how such multiple summations can be carried out based on the (ergodic) properties of the classical dynamics while preserving the inherent underlying quantum interference mechanisms. In this respect the Ehrenfest time $\tE$~\cite{Ehrenfest27} plays a key role. As discussed below, it has turned out to be of tremendous importance as it separates quantum propagation in phase space around {\em one} dominant (unstable) classical trajectory at short time scales from subsequent times governed by strong wave interference, {\em i.e.}, involving propagation of amplitudes along {\em many} trajectories.  Due to the exponential sensitivity of chaotic dynamics to initial conditions, $\tE$ is a logarithmically short time scale as a function of $\hbar$. Therefore, quasi-classical approaches that do not account for such interference contributions are strongly limited in the corresponding (time) range of validity.

%%%%%%%%%%%%%%%%%%%%%%%%%%%%%%%%%%%%%%%%%%%%%%%%%%%%%%%%%%%%%%%%%%%%%%%

\subsubsection{Single sums over paths: van Vleck propagator and Gutzwiller trace formula}
\label{sec:SP-Gutzwiller}

The van Vleck propagator $K_{\rm sp}(t)$, and its refinement by Gutzwiller, represent  semiclassical approximations to the quantum time evolution operator $U(t) = \exp{(-(i/\hbar) H t)}$ in configuration space, assuming for simplicity a time-independent Hamiltonian, generalization is straightforward.  

As the derivation of $K_{\rm sp}(t)$ and Gutzwiller's periodic orbit theory can be found in various excellent text books~\cite{Haake91,StockmannBook,Gutzwiller90,BraBha03Book,Reichl21}, here only the relevant expressions are directly introduced. Evaluating the Feynman propagator $\langle {\bf r}_{f}| U(t)| {\bf r}_{\rm i}\rangle $ in a stationary phase approximation yields the 
van Vleck-Gutzwiller propagator for the time evolution of a quantum particle  between initial and final coordinates ${\bf r}_{\rm i}$ and ${\bf r}_{\rm f}$ in $d$ dimensions:
\begin{equation}
\label{eq:vVG}
    K_{\rm sp} ({\bf r}_{\rm f},{\bf r}_{\rm i},t)
     = \sum_{\gamma} 
     \left(\frac{1}{(2\pi i \hbar)^d} \left| \frac{ \partial^2
     R_{\gamma}({\bf r}_{\rm f},{\bf r}_{\rm i},t)}{\partial {\bf r}_{\rm f} \partial {\bf r}_{\rm i}}
     \right|\right)^\frac{1}{2} 
     \
     {\rm e}^{i(R_{\gamma}({\bf r}_{\rm f},{\bf r}_{\rm i},t)/\hbar - \nu_\gamma \pi/2)} \, 
\end{equation}
with classical action
$ R_{\gamma}({\bf r}_{\rm f},{\bf r}_{\rm i},t) = \int_0^t L(\dot{\bf r}, {\bf r}, t) d t$ along the trajectory $\gamma$ connecting ${\bf r}_{\rm i}$ to ${\bf r}_{\rm f}$ in time $t$, and index $\nu_\gamma$, which appropriately tracks the correct phase of the determinant's square root; see, e.g.,~\cite{Keller58}.  For simplicity, all the various topological indices like $\nu$ are referred to as Maslov indices.
The semiclassical approximation, Eq~(\ref{eq:vVG}) for the exact propagator holds generally for either chaotic, mixed or integrable classical dynamics. Computing the energy-dependent Green function via a Laplace transform of $K_{\rm sp}(t)$ and upon calculating the spatial trace integral by means of further stationary phase approximations, Gutzwiller derived the famous trace formula for the density of states $\rhosp(E)$ of a {\em classically chaotic} quantum SP system,
thereby laying the foundations of periodic-orbit approaches to quantum chaos~\cite{Gutzwiller71}:
\begin{equation}
    \rhosp(E) \simeq \bar{\rho}_{\rm sp}(E) \ + \ \rho_{\rm sp}^{\rm (osc)}(E) =
    \bar{\rho}_{\rm sp}(E) \ + \frac{1}{\pi\hbar} {\rm Re}\left\{
    \sum_{\rm po} A_{\rm po}{\rm e}^{(i/\hbar) S_{\rm po}(E)} \right\} \, .
    \label{eq:SP-Gutzwiller}
\end{equation}
The Weyl term $\bar{\rho}_{\rm sp}(E)$ is a smooth function of energy $E$. It is obtained, to leading order, by calculating the volume of the classical phase space energy shell,
\begin{equation}
    \bar{\rho}_{\rm sp} (E)=\left(\frac{1}{2\pi\hbar}\right)^{d}\int \ d{\bf r} \ d{\bf p}\ \delta(E-H_{\rm sp} ({\bf r},{\bf p})) \, ,
    \label{eq:SP-Weyl}
\end{equation}
where $H_{\rm sp}$ is the classical Hamiltonian. Further higher $\hbar$-corrections are also smooth and can be systematically included \cite{Balian1970}. The oscillatory part $\rho_{\rm sp}^{\rm (osc)}(E)$ of the density of states is given through a coherent sum over 
all {\it periodic orbits} (po) of the corresponding classical system at energy $E$. A given periodic orbit contributes to $\rho_{\rm sp}^{\rm (osc)}(E)$ through an oscillatory amplitude with a phase
\begin{equation}
    S_{\rm po}(E) = \int_{\rm po}  \ {\bf p} \cdot d {\bf q} -\hbar \mu_{\rm po}  \ \pi /2 \, .
    \label{eq:SP-action}
\end{equation}
containing its classical action and Maslov index  $\mu_{\rm po}$.
Note that the sum over all periodic orbits includes also all higher repetitions with periods $T_{\rm po}=nT_{\rm ppo}$ (for $n\ge2$) of a given primitive periodic orbit (ppo) with period $T_{\rm ppo}$.
The amplitudes in Eq.~(\ref{eq:SP-Gutzwiller}) read
\begin{equation}
    A_{\rm po}(E) = \frac{T_{\rm ppo}(E)}{|{\rm d
    et}({\bf M}_{\rm po}(E) - {\bf I} |^{1/2}}  \, 
    \label{eq:SP-stab}
\end{equation}
where the monodromy (or stability) matrix ${\bf M}_{\rm po}(E)$ characterizes the linearized phase space structure in the vicinity of the periodic orbit in terms of stability exponents (similar to Lyapunov exponents) for chaotic dynamical systems. 
 
 The Gutzwiller trace formula (\ref{eq:SP-Gutzwiller}) expresses the quantum spectrum in a Fourier-type way. In this way, whereas short periodic orbits contribute with long-ranged cosine-type spectral modulations, accounting for contributions from longer and longer periodic orbits, in principle, increases the spectral resolution~\footnote{There is an extensive literature about convergence properties of the trace formula and the challenges associated with semiclassically computing individual energy levels; see~\cite{Cvitanovic89}}. 
 To resolve the quantum density of states at the scale of the mean level spacing $1/\bar{\rho}(E)$ requires, in turn, to control semiclassical wave interference in the time domain on scales in the range of or longer than the so-called Heisenberg time, $\tH=\hbar \bar{\rho}(E)$, the longest time scale involved. The challenge of coping with this {\em late-time behavior} leads to partially solved issues~\cite{Mueller2009}.
%%%%%%%%%%%%%%%%%%%%%%%%%%%%%%%%%%%%%%%%%%%

\subsubsection{Multiple sums over paths: spectral two-point correlation function}
\label{sec:SPcorrelations}

Within the statistical analysis of the spectrum, which proves adequate in complex systems where energy-by-energy enumeration is essentially impossible, the quantities of interest are usually not the bare densities of states $\rhosp(E)$, but rather spectral $n$-point correlation functions. Using the trace formula (\ref{eq:SP-Gutzwiller}) a semiclassical approach to $n$-point correlators naturally leads to $n$-fold coherent summations over amplitudes evolving along, in principle, infinitely many orbits. Further evaluations of such multiple sums seems, at first glance, hopeless. However, an intrinsic strength of semiclassical theory lies in the fact that systems with ergodic classical dynamics often do not require the computation of specific trajectories (nor is it always desirable). Instead, invoking ergodicity and uniformity of chaotic phase space implies powerful classical sum rules that permit a treatment of the orbits in a statistical manner. As no system-specific information is required, such approaches naturally lead to universal features of quantum-chaotic dynamics and may provide physical laws applicable to whole classes of quantum systems, exclusively characterized by means of their respective symmetry class.

The semiclassical evaluation of multiple sums over paths is illustrated for the prominent case of the two-point correlator (a corresponding treatment of four-point objects is given in Sec.~\ref{sec:OTOC} for out-of-time-order correlators)
\begin{equation}
 C(\epsilon) = \frac{1}{\bar{\rho}_{\rm sp}(E)^2} 
 \left\langle 
 \rho_{\rm sp}^{\rm (osc)}
 \left(E+\frac{\epsilon}{2\bar{\rho}_{\rm sp}(E) }\right)\
  \rho_{\rm sp}^{\rm (osc)}
  \left(E-\frac{\epsilon}{2\bar{\rho}_{\rm sp}(E)} \right)
  \right\rangle_E 
    \label{eq:SP-2-point}
\end{equation}
with the angular brackets denoting a running local average over energy $E$. $C(\epsilon)$ is a simple but fundamental measure of spectral correlations. The dimensionless variable $\epsilon$ stands for a spectral energy distance in units of the mean level spacing $1/\bar{\rho}_{\rm sp}(E)$. Replacing $\rhosp$ in Eq.~(\ref{eq:SP-2-point}) by its semiclassical approximation, Eq.~(\ref{eq:SP-Gutzwiller}), one obtains
\begin{equation}
 C(\epsilon) \propto \left\langle 
    \sum_{\gamma} \sum_{\gamma'} A_\gamma A_{\gamma'}^\ast \ {\rm e}^{(i/\hbar) 
    [S_\gamma(E) - S_{\gamma'}(E)) + (T_\gamma(E)+T_{\gamma'}(E))\epsilon/ (2\pi \bar{\rho}_{\rm sp})]} 
  \right\rangle_E \ .
    \label{eq:SP-2-point-sc}
\end{equation}
Here, $S(E+E') \simeq S(E) + T(E) E'$ with the orbit's period $T(E) =    \partial S/  \partial E$. The contributions of periodic orbit pairs $\gamma, \gamma'$ that exhibit an action difference $\Delta S(E)$ in the phase factor are handled separately from those that do not.  Of course, $\Delta S(E)$ vanishes for the joint contributions of the specific orbit pairs $\gamma = \gamma'$, i.e. {\it the diagonal contributions}. In addition, if a system is invariant with respect to some symmetry, for in stance time-reversal symmetry, then that symmetry is reflected in multiplicities of symmetry related orbits with classical action degeneracies as well. The resulting constructive interference encodes the symmetry's influence on the quantum system.  In effect, this can be considered as part of the diagonal contributions.

In the semiclassical limit, the phases $S_\gamma(E)/\hbar$ oscillate rapidly upon varying $E$, and only terms with sufficiently small action differences 
\begin{equation}
\Delta S(E) = S_\gamma(E)-S_{\gamma'}(E)
\label{eq:delta_S}
\end{equation}
can survive the energy averaging $\langle \ldots \rangle_E$. Using the classical sum rule of
Hannay and Ozorio de Almeida~\cite{Hannay84},
\begin{equation}
    \sum_{\gamma} 
    \frac{f_\gamma(T_\gamma) }{{\rm det}({\bf M}_\gamma-{\bf I})}
    \simeq \int_{T_0} dT \frac{f(T)}{T} \, , 
    \label{eq:SP-HOdA}
\end{equation}
that follows from the assumption of uniform phase space exploration by unstable periodic orbits, Berry computed the  diagonal contribution
\begin{equation}
 C_d(\epsilon) \simeq  \frac{1}{\bar{\rho}_{\rm sp}(E)^2}  \left(\frac{1}{\pi\hbar}\right)^2
 \int_{T_0} dT \ T \  e^{iT\epsilon/ (2\pi \hbar \bar{\rho}_{\rm sp})} 
    \label{eq:SP-2-point-diag}
\end{equation}
to the two-point correlator~\cite{Berry85}. He thus derived the spectral rigidity found in RMT semiclassically. For the spectral form factor $K(\tau)$ (with $\tau \!=\! T/\tH$), the Fourier transform of $C(\epsilon)$ in diagonal approximation leads to the linear "ramp": $K(\tau) = \eta\tau$, 
% where $\tau = T/\tH$ 
with $\eta =$ 2 and 1 for systems with and without time reversal symmetry, respectively. Although Berry's analysis provided the first intimate theoretical link between RMT and the semiclassical theory of chaos, the importance of the diagonal approximation goes well beyond spectral statistics, imprinting itself in corresponding contributions to the evolution of observables, the pre-Ehrenfest time behavior of OTOCs, among many more. 

\begin{figure}
    \centering
  \includegraphics[width=0.55\linewidth]{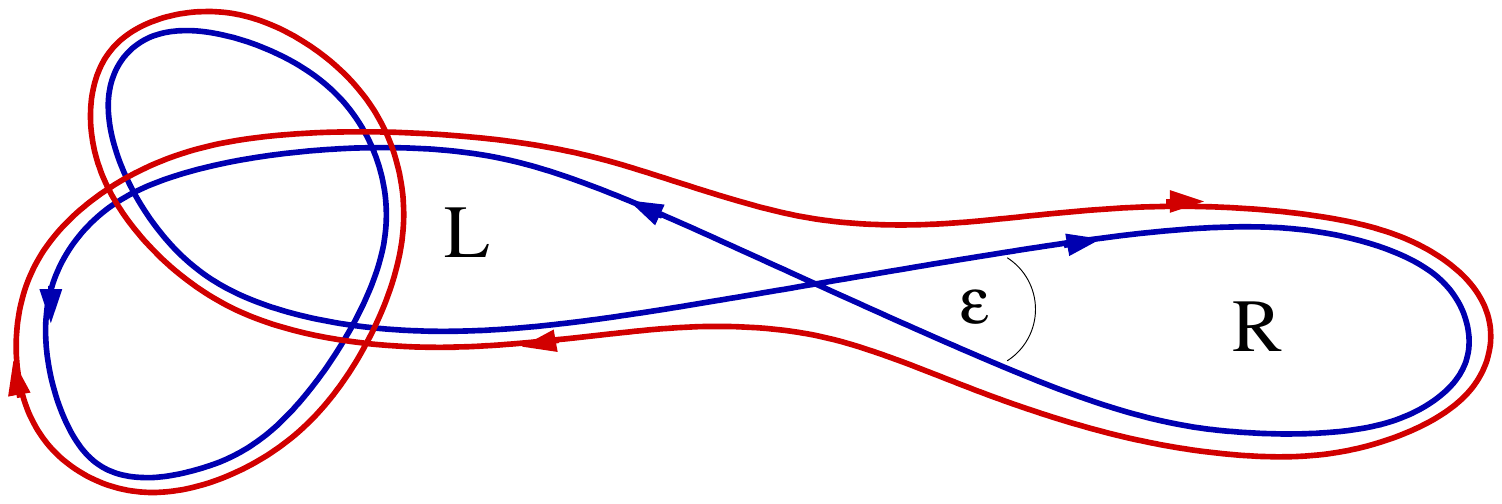}
       \caption{\label{fig:orbit-pair}
  Scheme of a fundamental pair of classically correlated long periodic orbits linked to each other through a common self-encounter region~\cite{Sieber01}.
   }
\end{figure}

Apart from the diagonal terms there is an enormous number of off-diagonal orbit pairs in a chaotic system due to the exponential proliferation of the number of periodic orbits with increasing period, $T_\gamma(E)$. Most of the orbit pairs consist of periodic orbits with actions that are uncorrelated.  Summing over them and performing the energy average, they collectively have a vanishing average, including the effects of ``accidentally'' quasi-degenerate  actions $S(E)$.  However, from RMT it had been known that for time-reversal invariant systems, there must exist further universal spectral correlations beyond those related to the diagonal term~\cite{Bohigas91}:
\begin{equation}
K^{\rm GOE} (\tau) = 
\left\{
\begin{array}{ll}
2\tau -\tau \log(1+2\tau) & {\rm if} \quad \tau < 1 \, , \\
2 - \tau \log \frac{2\tau +1 }{2\tau -1} & {\rm if} \quad \tau >1 \, . \\
\end{array}
\right. 
    \label{eq:SP-FormFac}
\end{equation}
Hence to describe such universal RMT features, one had to find orbit pairs with non-random action differences~\cite{Argaman93}. Although it was expected that such non-vanishing contributions come from a relatively small number of pairs of correlated orbits, for a long time it was unclear how these orbit correlations could emerge from an ergodic phase space structure.
%%%%%%%%%%%%%%%%%%%%%%%%%%%
Research since 2000 has brought to light that chaotic dynamics is subject to further principles of order: periodic orbits do not appear as independent individual entities but in pairs, as first discovered by~\cite{Sieber01,Richter02,Sieber02}, and more generally in densely packed bundles~\cite{Mueller09}. This hidden classical property of periodic orbits in chaotic systems turned out to play a central role for understanding universal spectral properties.
The fundamental trajectory doublet is schematically shown in  Fig.~\ref{fig:orbit-pair}. One of the paths has a crossing in the configuration space under a small angle$\epsilon$. From this the corresponding partner trajectory can be uniquely constructed by matching trajectory segments associated with the local stable and unstable manifold of the reference orbit. Then for each such long orbit, a partner orbit starting and ending (exponentially) close to the first one exists. 
%%%%%%%%%%%%%%%%%%%%%%%%%%%
%%%%%%%%%%%%%%%%%%%%%%%%%%%%

Due to the close similarity of the two members of the orbit pair, the members exhibit near-degenerate actions and are too highly correlated to ignore when energy averaging.  This was discovered and first worked out in Ref.~\cite{Sieber01} for the case of correlated orbits pairs forming a two-encounter (Fig.~\ref{fig:orbit-pair}).  This analysis provided the leading quadratic contribution to the GOE spectral form factor
Eq.~\ref{eq:SP-FormFac}. Based on these insights, the Essen group worked out the encounter calculus that allows one to classify and compute general encounter structures, and used it for systematically calculating the semiclassical theory for the two-point correlator and  spectral form factor, respectively~\cite{Mueller04,Heusler06,Mueller07}. All details can be found in Haake's textbook~\cite{Haake10}, and hence we do not summarize them here again.

%%%%%%%%%%%%%%%%%%%%%%%%%%%%
\subsubsection{Eigenstate properties: the random wave model}

Moving into the semiclassical description of quantum chaos signatures in the morphology of stationary states and following the classic paper of Berry \cite{Berr4}, we start with a correlation function between the projections of the eigenstates of a two-dimensional system onto two states $|\mathbf{r}_1 \rangle,|\mathbf{
r}_2\rangle$, whereby $\mathbf{r}_1,\mathbf{r}_2$ lie inside the classical allowed domain. Although there are different possible ensembles that can be used to define such a correlation 
function, a natural choice in the spirit of running energy averages in RMT is to define the amplitude-amplitude cross-correlator as a spectral average over a number of eigenstates inside a specific energy range
\begin{align}
    R(\mathbf{r_1},\mathbf{r_2},E,\eta) := \frac{1}{M} \sum_{E_n \in [E-\frac{\eta}{2},E+\frac{\eta}{2}]} \langle \mathbf{r_1}|\Psi_n\rangle\langle\Psi_n|\mathbf{r_2}\rangle. \label{DefCrossCorrelator}
\end{align}
over all $M$ eigenstates $|\Psi_n\rangle$, which eigenenergies $E_n$ lie inside the energy interval $[E-\frac{\eta}{2},E+\frac{\eta}{2}]$.

This definition of the correlation function allows us to express the cross-correlator in terms of the Wigner function, a fundamental object allowing for a phase-space representation of quantum mechanics (see \cite{Polkovnikov10} for a review in the MB context) 
\begin{align}
    R(\mathbf{r_1},\mathbf{r_2},E,\eta)=\frac{1}{M} \sum_{E_n \in [E-\frac{\eta}{2},E+\frac{\eta}{2}]} \int d^Dp \, W_n\biggl(\frac{\mathbf{r_1}+\mathbf{r_2}}{2},\mathbf{p}\biggr) e^{(\frac{i}{\hbar}(\mathbf{r_1}-\mathbf{r_2})\cdot \mathbf{p})}. \label{Cross-Correlator with Wigner function}
\end{align}

Eq.~(\ref{Cross-Correlator with Wigner function}) resembles a Fourier transformation of the spectral average of Wigner functions centered in between the two points $\mathbf{r}_1$ and $\mathbf{r}_2$. Since we are nterested in the transition regime between classical and semiclassical physics we now chose the center of the energy interval $E$ to be high enough, such that a semiclassical approximation for the {\it averaged}Wigner function \cite{Berr4} 
\begin{equation}
 \frac{1}{M} \sum_{E_n \in [E-\frac{\eta}{2},E+\frac{\eta}{2}]} W_n(\mathbf{r},\mathbf{p})\simeq \frac{\delta(E-H_{\text{cl}}(\mathbf{r},\mathbf{p}))}{\int d^Dr \int d^Dp \, \delta(E-H_{\text{cl}}(\mathbf{r},\mathbf{p}))} 
\end{equation}
can be utilized and correspondingly obtain he semiclassical version of the cross-Ccorrelator
\begin{align}
    R^{\rm scl}(\mathbf{r_1},\mathbf{r_2},E) = \frac{\int d^Dp \, \delta(E-H_{\text{cl}}(\mathbf{r},\mathbf{p})) \exp{(\frac{i}{\hbar}[\mathbf{r_1}-\mathbf{r_2}]\cdot \mathbf{p})}}{\int d^Dr \int d^Dp \, \delta(E-H_{\text{cl}}(\mathbf{r},\mathbf{p}))}, \label{SCL Cross-Correlator Billiard}
\end{align}
which is solely governed by classical quantities $\mathbf{r},\mathbf{p}$ and $H_{\text{cl}}(\mathbf{r},\mathbf{p})$.

Considering the simplest case 
\begin{align*}
    H_{\text{cl}}(\mathbf{r},\mathbf{p}) = \frac{\mathbf{p}^2}{2m}+V(\mathbf{r}),
\end{align*}
and assuming that the considered two points $\mathbf{r}_1,\mathbf{r_2}$ are far enough away from the boundaries defining the classically allowed region, the semiclassical cross-correlator becomes 
\begin{equation}
\label{eq:RWMBerry}
 R^{\rm scl}(\mathbf{r_1},\mathbf{r_2},E) =  \frac{1}{V(E)}\frac{J_{D/2-1}(k(E,\bar{{\bf r}})|{\bf r_1}-{\bf r_2}|)}{(k(E,\bar{{\bf r}})|{\bf r_1}-{\bf r_2}|)^{D/2-1}}
\end{equation}
that displays characteristic coherent oscillations as a function of the spatial distance $|{\bf r_1}-{\bf r_2}|$ fixed through the local wavenumber $k^{2}(E,\bar{\bf r})=2m(E-V(\bar{\bf r}))/\hbar^{2}$ at the energy $E$ and mean position $\bar{{\bf r}}=({\bf r_1}+{\bf r_2})/2$. The cross-correlator is then normalized by the effective volume $V(E)$ of the classically allowed region at energy $E$.

The notion of {\it universality of eigenstate spatial fluctuations} comes from the functional form of the correlator, Eq.~\ref{eq:RWMBerry}, being largely independent on the particular form of the external potential. At the same time, the purely RMT limit of uncorrelated random eigenvectors corresponds only to the strict limit $E \to \infty$.

%%%%%%%%%%%%%%%%%%%%
\subsection{Semiclassical theory of bosonic many-body systems}
\label{sec:SC-MB}

%%%%%%%%%%%%%%%%%%%%%%%%%%%%%%%%%%%%%%%%%%%%%%%%%%%%%%%

\subsubsection{van Vleck propagator for bosonic systems}

%%%%%%%%%%%%%%%%%%%%%%%%%%%%%%%%%%%%%%%%%%%%%%%%%%%%%%

The semiclassical approximation, based on the unaltered kinematic structure of quantum mechanics, especially the unitarity and linearity of time evolution and the superposition principle, and supplemented by the asymptotic analysis of the propagator, seem to require only minimal modifications to be ushered into the realm of interacting MB systems. Due to the product form of the total Hilbert space in such systems, the corresponding modification of the MB propagator would be just adapting the classical limit into the high dimensional phase space of MB classical systems.  However, this picture leaves out a kinematic aspect of purely quantum origin, namely quantum indistinguishability. Contrary to classical distinguishability due to the fundamental existence of classical point-like states in phase space, quantum indistinguishability imposes severe restrictions on the physically meaningful MB states by demanding that they have specific transformation rules under particle label permutations~\cite{Sakurai94}, a restriction on the allowed states is completely alien to the world of classical mechanics. Even at the non-interacting level this quantum property has macroscopic effects, such as the stability of fermionic matter~\cite{Dyson67a, Dyson67b, Lieb76}, the phenomena of Bose-Einstein condensation~\cite{Bose24, Einstein24}, the Hong-Ou-Mandel effect~\cite{Hong87}, and related coherence effects of much recent interest, such as boson sampling~\cite{Aaronson10}. When suitably constructed,  direct symmetrization of the MB quantum state combined with the semiclassical formalism has proven very successful, especially in the framework of  non-interacting MB systems (or weakly interacting ones using perturbation theory), from the mesoscopic theory of quantum transport~\cite{richterSemiclassicalTheoryMesoscopic2000,Rodolfo} and quantum dots~\cite{Ullmo98} to the description of the scattering of bosonic states through chaotic cavities~\cite{Urbina16}. The problem is substantially more involved if interactions are taken into account due to the very different nature of the sum over classical paths inherent in the semiclassical propagator and the sum over permutations accounting for indistinguishability. This lack of compatibility is clearly seen in the limit $N \gg 1$ that becomes an essentially impossible task due to the (factorial) proliferation of terms in the (anti-)symmetrization process, even if one could efficiently account for the classical distinguishable limit.

As it is well known, the study of interacting quantum systems of identical particles naturally profits from the change of perspective given by the description of such systems in terms of {\it quantum fields}~\cite{Negele98}. Understanding particle states as excitations of a quantum field means that the individual identity of the distinguishable degrees of freedom, now an extra conceptual baggage without any physical relevance, is immediately absent from the description: instead of building MB states out of (anti-)symmetrized states of distinguishable particles, one specifies states by saying how many particles occupy any given set of SP orbitals irrespective of their individual identities. 

The space of quantum states labeled by such configurations is the familiar Fock space. The goal of this section is to review the conceptual and technical steps that enable the adaptation of the semiclassical program into this new framework, as well as evaluating its new regime of validity, advantages and limitations of this semiclassical approach. 

Following standard references~\cite{Negele98}, one begins with selecting an arbitrary set of SP states, denoted from now on as ``sites'' or ``orbitals'':
\begin{equation}
  \phi_{i} {\rm \ \ with \ \ \ }i=1,\ldots,d  
\end{equation}
where $d$ is the (possibly infinite) dimension of the SP Hilbert space. A physically allowed state $|\Psi\rangle$ is then an element of the Fock space $|\Psi\rangle \in {\cal F}$ 
\begin{equation}
    |\Psi\rangle=\sum_{\bf n}\Psi_{{\bf n}}|{\bf n}\rangle
\end{equation}
where the so-called Fock states, 
\begin{equation}
 |{\bf n}\rangle=|n_{1},\ldots,n_{d}\rangle {\rm \ \ \ \ with \ \ } n_{i}=0,1,2,\ldots 
\end{equation}
are labeled by occupation numbers $n_{i}$, themselves eigenvalues of the observables $\hat{n}_{i}$ with
\begin{equation}
    \hat{n}_{i}|{\bf n}\rangle=n_{i}|{\bf n}\rangle \, .
\end{equation}
It counts how many particles occupy the SP states $\phi_{i}$. Observables in ${\cal F}$ are written in terms of the corresponding creation/annihilation operators $\hat{b}^{\dagger},\hat{b}$ defined through,
\begin{eqnarray}
    \hat{b}_{i}|n_{1},\ldots,n_{i},\ldots,n_{d}\rangle&=&\sqrt{n_{i}-1}|n_{1},\ldots,n_{i}-1,\ldots,n_{d}\rangle \nonumber \\
    \hat{b}_{i}^{\dagger}|n_{1},\ldots,n_{i},\ldots,n_{d}\rangle&=&\sqrt{n_{i}}|n_{1},\ldots,n_{i}+1,\ldots,n_{d}\rangle \, ,
\end{eqnarray}
satisfying the Fock-space properties through the canonical commutation relations
\begin{equation}
   \left[\hat{b}_{i},\hat{b}_{j}\right]=0 {\rm \ \ , \ \ }\left[\hat{b}_{i},\hat{b}_{j}^{\dagger}\right]=\hat{1}\delta_{i,j} {\rm \ \ and \ \ } \hat{n}_{i}=\hat{b}_{i}^{\dagger}\hat{b}_{i}.
\end{equation}

The general form of the Hamiltonian describing a system of bosons evolving under the influence of external potentials (appearing as quadratic combinations of $\hat{b}^{\dagger},\hat{b}$) and pairwise interactions (of quartic order) is 
\begin{equation}
\label{eq:Ham}
    \hat{H}=\sum_{i,j}h_{i,j}\hat{b}^{\dagger}_{i}\hat{b}_{j}+\sum_{i,j,i',j'}v_{i,j,i',j'}\hat{b}^{\dagger}_{i}\hat{b}_{j}\hat{b}^{\dagger}_{i'}\hat{b}_{j'}\, .
\end{equation}
Correspondingly, the Fock-space propagator (assuming for simplicity time-independent external and interaction potentials) is defined as usual by
\begin{equation}
    K({\bf n}^{(f)},{\bf n}^{(i)},t)=\langle {\bf n}^{(f)}|{\rm e}^{-\frac{i}{\hbar}\hat{H}t}|{\bf n}^{(i)}\rangle \ .
\end{equation}
Our goal is to first identify the Fock-space version of the semiclassical regime $\hbar_{\rm eff}=1/N\to 0$, where $N$ is the eigenvalue of the operator representing the total number of particles 
\begin{equation}
 \hat{N}=\sum_{i}\hat{n}_{i}   \, .
\end{equation}
Then, starting from a suitable path integral representation of $K$, one performs the steps to derive the Fock space version of the van Vleck-Gutzwiller semiclassical sum over classical paths.

Historically, the obvious problem with regard to constructing a path integral from a discrete basis was addressed by introducing the coherent state basis on ${\cal F}$~\cite{Negele98} defined by
\begin{equation}
    \hat{b}_{i}|{\bf z}\rangle=z_{i}|{\bf z}\rangle \ 
\end{equation}
and labeled by a set of continuous complex numbers. This basis admits a realization of the unit operator in ${\cal F}$
\begin{equation}
  \frac{1}{(2\pi i)^{d}}\int \prod_{i}dz_{i}dz^{*}_{i} |{\bf z}\rangle \langle {\bf z}| = \hat{1} 
\end{equation}
in a form suitable for the usual time-slicing path integral construction. 

The resulting coherent state path integral for bosonic quantum fields has been extensively used as a basis for semiclassical approximations~\cite{Negele98, Klauder78, Baranger01}, so its use to derive a van Vleck-Gutzwiller type of approach is quite appealing. However, a fundamental drawback is that the resulting saddle point equations do not generally admit real solutions, thus requiring the complexification of the classical limit of the theory~\cite{Baranger01}. If one is interested in keeping the dynamical variables real, the widely used coherent state path integral turns out not to be the ideal starting point for the van Vleck-Gutzwiller derivation.

A path integral in Fock space that provides a semiclassical approximation with real paths can be based on a Fock space basis set given by eigenstates of  operators that have three key properties of the familiar configuration (or momentum) operators in SP systems, namely they must have a real, continuous, and unbounded spectrum and a classical limit where they play the role of {\it half} of a canonical pair. A very natural choice is then the Hermitian combination
\begin{equation}
\label{eq:quad1}
    \hat{q}_{i}=\frac{\hat{b}^{\dagger}_{i}+\hat{b}_{i}}{\sqrt{2}} {\rm \ \ , \ \ } \hat{p}_{i}=\frac{\hat{b}^{\dagger}_{i}-\hat{b}_{i}}{\sqrt{2}i} \, .
\end{equation}
In close analogy with the known algebra of the harmonic oscillator, these pairs can be easily shown to fulfill the relations
\begin{equation}
    \left[\hat{q}_{i},\hat{q}_{j}\right]=\left[\hat{p}_{i},\hat{p}_{j}\right]=0 {\rm \ \ and \ \ } \left[\hat{q}_{i},\hat{p}_{j}\right]=i\delta_{i,j}
\end{equation}
of canonically conjugate operators.  Together with their eigenbases that satisfy all the required conditions as can be seen by direct computation, they allow for the construction of the semiclassical approximation for bosonic matter fields. Following the terminology of quantum optics where these canonical pairs are of common use as Hermitian versions of the standard field operators, we refer to them as quadratures~\cite{Scully97}. 

Armed with the quadratures and their eigenbases
\begin{equation}
\label{eq:quad2}
    \hat{q}_{i}|{\bf q}\rangle=q_{i}|{\bf q}\rangle {\rm \ \ and \ \ } \hat{p}_{i}|{\bf p}\rangle=p_{i}|{\bf p}\rangle
\end{equation}
nicely satisfying \cite{Engl2016}
\begin{equation}
    \langle {\bf q}|{\bf p}\rangle=\frac{{\rm e}^{i{\bf q}\cdot {\bf p}}}{(2\pi)^{d/2}}
\end{equation}
it is possible to express the exact path integral representation of the propagator using the usual method of time-slicing.  Inserting the quadrature definitions (\ref{eq:quad1}) into the generic form of the bosonic Hamiltonian, Eq.~(\ref{eq:Ham}), leads to classical Hamiltonians of a very different character than the ``kinetic plus potential energy'' (mechanical type) often found in non-relativistic SP systems. This is a reminder that, despite their identical formal properties, quadratures do not represent anything like position and momentum, and they can be considered as just a technical tool used to develop a path integral with the desired properties. For systems of massive bosons, they are not observable in the strict sense (a property they share with coherent states~\cite{Bartlett07}). Therefore, the construction of the propagator between physical (Fock) states must also eventually be addressed, as we will show below.

After these caveats, the construction of the path integral form of the propagator between configuration quadratures formally follows the standard methodology~\cite{Schulman81} with a few key modifications. The evolution is first cut into a large number of factors of the form ${\rm e}^{-i\delta t \hat{H}/\hbar}$ with $\delta t \to 0$, and the representation of the unit operator in Fock space is inserted in terms of the $q$-quadratures. The form of the Hamiltonian (\ref{eq:Ham}) demands a careful treatment of the resulting matrix elements, achieved by 
\begin{equation*}
 \langle {\bf q}|{\rm e}^{-\frac{i}{\hbar}\delta t \hat{H}}|{\bf q}'\rangle = \int d{\bf p}\langle {\bf q}|{\rm e}^{-\frac{i}{\hbar}\delta t \hat{H}}|{\bf p}\rangle \langle{\bf p}|{\bf q}'\rangle = \int d{\bf p}\langle {\bf q}|{\bf p}\rangle{\rm e}^{-\frac{i}{\hbar}\delta t \frac{\langle {\bf q}|\hat{H}|{\bf p}\rangle}{\langle {\bf q}|{\bf p}\rangle}}\langle{\bf p}|{\bf q}'\rangle+{\cal O}(\delta t) \ .
\end{equation*}
Such explicit appearance of extra integrations over momentum quadratures naturally leads to the so-called Hamiltonian (or phase space) form of the propagator \cite{Negele98}
\begin{equation}
    K({\bf q}^{(f)},{\bf q}^{(i)},t):=\langle{\bf q}^{(f)}|{\rm e}^{-\frac{i}{\hbar}\hat{H}t}|{\bf q}^{(i)}\rangle=\int {\cal D}[{\bf q}(s),{\bf p}(s)]{\rm e}^{iR[{\bf q}(s),{\bf p}(s)]}
\end{equation}
where the integral is now defined over the space of paths $({\bf q}(s),{\bf p}(s))$. Note that, in sharp contrast to the coherent state case, only the paths in configuration quadrature endpoints are constrained, 
\begin{equation}
{\bf q}(s=0)={\bf q}^{(i)}, {\bf q}(s=t)={\bf q}^{(f)} \, ,
\end{equation}
whereas the momentum quadratures are completely unconstrained. Finally, the action functional is given in its Hamiltonian form by
\begin{equation}
  R[{\bf q}(s),{\bf p}(s)]=\int ds \left[{\bf p}(s) \cdot \dot{{\bf q}}(s)-H_{\rm cl}({\bf q}(s),{\bf p}(s))/\hbar\right]  
\end{equation}
and the classical symbol is obtained from the Hamiltonian operator expressed by moving all $\hat{p}$ operators to the right of the $\hat{q}$ ones as 
\begin{equation}
H_{\rm cl}({\bf q},{\bf p})=\frac{\langle {\bf q}|\hat{H}|{\bf p}\rangle}{\langle {\bf q}|{\bf p}\rangle}.
\end{equation}
Concerning the role of $\hbar$, one observes that the action functional in Fock space is dimensionless, an aspect that once again reflects that quadrature operators are not related to any physical coordinate/momentum in any sense beyond the formal analogies with their SP counterparts. In contrast to the first quantized approach, inspection of the action functional reveals that in Fock space $\hbar$ is simply a constant that transforms energies like $H$ into frequencies, but does not play the fundamental role of defining the small parameter upon which the asymptotic analysis is built. Preparing the way for the semiclassical approximation where ordering effects lead to subdominant contributions, and denoting 
\begin{equation}
\label{eq:Tom1}
\hat{H}=H(\hat{{\bf b}}^{\dagger},\hat{{\bf b}}),
\end{equation}
the phase space function appearing in the path integral, the Hamiltonian is given by  
\begin{equation}
\label{eq:Tom2}
H_{\rm cl}({\bf q},{\bf p})=H\left(\frac{{\bf q}+i{\bf p}}{\sqrt{2}},\frac{{\bf q}-i{\bf p}}{\sqrt{2}}\right)\, .
\end{equation}

In order to identify a suitable asymptotic parameter, we focus on the action of the quadrature propagator on the physical Fock states.  Consider first 
\begin{equation}
\label{eq:transfo}
    K({\bf n}^{(f)},{\bf n}^{(i)},t)=\int d{\bf q}^{(f)}d{\bf q}^{(i)}\langle {\bf n}^{(f)}|{\bf q}^{(f)}\rangle K({\bf q}^{(f)},{\bf q}^{(i)},t)\langle {\bf q}^{(i)}|{\bf n}^{(i)}\rangle
\end{equation}
where the transformation overlaps $\langle q|n\rangle$ are formally derived from the algebraic properties of quadrature states as shown in~\cite{Engl2016} to get
\begin{equation}
\label{eq:CT}
   \langle q|n\rangle=\frac{{\rm e}^{-q^{2}/2}}{\pi^{1/4}\sqrt{2^{n}n!}} H_{n}(q)\, .
\end{equation}

Armed with Eq.~(\ref{eq:CT}), one can obtain the propagator between Fock states based on the path integral in quadrature representation.  The semiclassical analysis, and the identification of a proper asymptotic regime, begins with the analysis of this kernel in the limit of large occupations $n \gg 1$. In this case, a well known asymptotic form \cite{Gradshteyn00}
\begin{equation}
   \langle q|n\rangle \simeq A(q,n)\cos{(F(q,n)+\pi/4)} 
\end{equation}
holds, where $A(q,n)$ is a smooth prefactor. Following \cite{Engl2016}
\begin{equation}
   F(q,n)=\int dq\sqrt{2n+1 - q^{2}}
\end{equation}
is identified as the generating function of the classical canonical transformation between the canonical pairs $(q,p)$ and $(n,\theta)$ with
\begin{equation}
    q+ip=\sqrt{2n}{\rm e}^{i\theta}.
\end{equation}
Using this generating function, the phase space variables $(q,p)$ labelling  quadrature eigenstates on each orbital are maximal for $q^{2}+p^{2}=2n$ if the quadrature propagator is applied to Fock states with large occupation numbers. For large occupations $n_{i} \gg 1$, the overlap between quadrature and Fock states naturally suggests the scaling $q_{i}\propto \sqrt{n_{i}},p_{i} \propto \sqrt{n_{i}}$. 

For most closed systems of interest, the Hamiltonian has an additional conservation property already easily seen in the general form of Eq.~(\ref{eq:Ham}), namely the operator $\hat{N}$, a constraint that is fundamental in the case of massive bosons~\cite{Bartlett07}.  A consequence of this symmetry is that the Fock state propagator is different from zero if and only if
\begin{equation}
    N^{(f)}=N^{(i)}=N \, . 
\end{equation}
Therefore physically allowed dynamical processes and amplitudes are restricted to the subspace of ${\cal F}$ fixed by $N$ that acts as a real numerical constant parameterizing the propagator. Also, simple combinatorial arguments imply that for asymptotically large $N$ and fixed number of orbitals $d$, the vast majority  of Fock states satisfy $n_{i} \simeq N/d$. The set of  Fock states with occupations bounded away from zero defines a subspace of the Fock space,  with the quadrature variables scaling as
\begin{equation}
    (q_{i},p_{i})=\sqrt{N}(Q_{i},P_{i}).
\end{equation}

This particular scaling with the total particle number in the quadrature propagator automatically carries over to the path integral, except for considerations about the functional measure that can be accounted for by a convenient regularization. In particular, it transforms the action functional as
\begin{equation}
  R[\sqrt{N}{\bf Q}(s),\sqrt{N}{\bf P}(s)]=\int ds \left[N{\bf P}(s) \cdot \dot{{\bf Q}}(s)-H_{\rm cl}(\sqrt{N}{\bf Q}(s),\sqrt{N}{\bf P}(s))\hbar\right] \ ,  
\end{equation}
given the specific homogeneity properties of the second-quantized Hamiltonian of Eq.~(\ref{eq:Ham}), 
\begin{equation}
  H_{\rm cl}(\sqrt{N}{\bf Q}(s),\sqrt{N}{\bf P}(s))=NH_{\rm cl}({\bf Q}(s),{\bf P}(s))  
\end{equation}
as long as the interaction matrix elements are rescaled by 
\begin{equation}
   v=\tilde{v}/N\ .
\end{equation}
This scaling of the coupling constant is actually a natural requirement arising from a very intuitive observation: For large occupations, $N \gg 1$, the interaction term in Eq.~(\ref{eq:Ham}) trivially dominates the dynamics given its natural scaling with $N^{2}$. 

All together a meaningful scaling limit is only achieved by 
\begin{equation}
    N \to \infty,\quad v \to 0,\quad vN=\tilde{v}=const\ ,
\end{equation}
when the quadrature propagator acts in the subspace of large occupations, and therefore 
\begin{equation}
\label{eq:action}
  R[\sqrt{N}{\bf Q}(s),\sqrt{N}{\bf P}(s)]=N\int ds \left[{\bf P}(s) \cdot \dot{{\bf Q}}(s)-{\cal H}({\bf Q}(s),{\bf P}(s))/\hbar\right] \, .  
\end{equation}
Here, ${\cal H}$ denotes the classical Hamiltonian with rescaled interaction strength making the dynamics {\it fully independent of $N$}. The form of Eq.~(\ref{eq:action}) implies the important formal identification $\hbar_{\rm eff}=1/N$ and $N \to \infty$ as the semiclassical regime of systems with a large number of interacting bosons.   

There is one remaining ingredient in Eq.~(\ref{eq:transfo}) to be written in terms of scaled variables, the part corresponding to the transformation kernels. One readily sees that
\begin{equation}
    F(\sqrt{N}Q,n)=NF(Q,\rho),\quad \rho=n/N\ ,
\end{equation}
thus bringing the quadrature propagator, when projected on initial and final Fock states with large total number of particles, into a linear combination of integrals of the form
\begin{eqnarray}
\label{eq:full}
   K({\bf n}^{(f)},{\bf n}^{(i)},t)&=&\int d{\bf Q}^{(f)}d{\bf Q}^{(i)} \prod_{i}A(Q_{i}^{(f)},\rho_{i}^{(f)}){\rm e}^{iNF(Q_{i}^{(f)},\rho_{i}^{(f)})}A(Q_{i}^{(i)},\rho_{i}^{(i)})  \nonumber \\
   &\times& {\rm e}^{iNF(Q_{i}^{(i)},\rho_{i}^{(i)})}\int_{{\bf Q}(0)={\bf Q}^{(i)}}^ {{\bf Q}(t)={\bf Q}^{(f)}}{\cal D}[{\bf Q}(s),{\bf P}(s)]{\rm e}^{iN{\cal R}[{\bf Q}(s),{\bf P}(s)]}\ .
\end{eqnarray}
The asymptotic limit for $N\to \infty$ naturally emerges since the corresponding action functional ${\cal R}[{\bf Q}(s),{\bf P}(s)]$ and phase functions $F(Q,\rho)$ are {\it independent of $N$}.  Therefore, both the stationary phase condition $\delta {\cal R}=0$ that defines a consistent classical limit when supplemented with the boundary conditions ${\bf Q}(0)={\bf Q}^{(i)},{\bf Q}(t)={\bf Q}^{(f)}$, and the canonical transformation performing the change of phase space coordinates $(Q,P) \to (\rho,\theta)$ can be performed using stationary phase analysis.

Performing explicitly the variations over the $Q,P$ paths we get the corresponding Hamilton equations
\begin{eqnarray}
    \frac{\delta {\cal R}}{\delta {\bf Q}}=0 &\rightarrow& \hbar\frac{d}{ds}{\bf P}=-\frac{\partial {\cal H}}{\partial {\bf Q}} \nonumber \\ 
    \frac{\delta {\cal R}}{\delta {\bf P}}=0 &\rightarrow& \hbar \frac{d}{ds}{\bf Q}=\frac{\partial {\cal H}}{\partial {\bf P}} \, .
\end{eqnarray}
Using Eqs.~(\ref{eq:Tom1}) and (\ref{eq:Tom2}), they are  recognized as the real and imaginary parts of the mean field equations 
\begin{equation}
\label{eq:MFE}
    i \hbar\frac{d}{ds}\psi_{i}(s)=\frac{\partial}{\partial \psi^{*}_{i}}{\cal H}_{\rm MF}({\bf \psi},{\bf \psi}^{*})
\end{equation}
that now emerge neatly as a classical limit of the quantum field theory, where the mean-field Hamiltonian is defined in terms of the classical symbol $H_{\rm cl}$ as 
\begin{equation}
 {\cal H}_{\rm MF}({\bf \psi},{\bf \psi}^{*})= H_{\rm cl}({\bf Q},{\bf P})\, .
\end{equation}
Finally, the mean-field Hamiltonian is a function of the complex classical fields
\begin{equation}
    \psi_{i}=\frac{Q_{i}+iP_{i}}{\sqrt{2}}
\end{equation}
that together with $\psi^{*}$ parameterize the manifold in the phase space of the classical limit with the constraint
\begin{equation}
    \sum_{i}\rho_{i}=1.
\end{equation}
We see then that in our construction, the classical limit is identical to the celebrated mean-field equations well known from the theory of interacting bosonic systems in their discrete (lattice) version. Armed with this key physical notion of the mean-field equations as the underlying theory playing the role of the classical limit of MB quantum theory in the asymptotic regime $N \gg 1$, the relationship between quantum, semiclassical, and mean-field descriptions becomes transparent. Very much as Hamilton's classical equations of motion for particles not being able to describe quantum interference simply because the (phase) space of classical states does not allow for superpositions, the mean-field solutions cannot by themselves contain genuine MB quantum interference. Here again, to understand the connection between multiple mean-field solutions and their interference requires the full machinery of the semiclassical approximation.

To appreciate the fundamentally different role played by the mean-field equations in a purely mean-field framework as opposed to the semiclassical program, we note that, invariably, mean-field methods are based on the propagation of a {\it single} solution of the mean-field equations, fully and uniquely determined by its initial condition $\psi(s=0)$. Quite to the contrary, within a van Vleck-Gutzwiller semiclassical approach the classical limit of the interacting MB quantum theory is {\it not} an initial value problem, but actually a two-point boundary value problem consistently determined by the mean-field equations supplemented with the boundary conditions
\begin{equation}
   {\bf Q}(s=0)={\bf Q}^{(i)},{\bf Q}(s=t)={\bf Q}^{(f)} 
\end{equation}
and thus generally admitting multiple solutions, except for the non-interacting case.  Transformations can be made to an initial value representation method, but the multiplicity of solutions remains a key feature. Following the conceptual framework of the semiclassical approximation, the role of these combined multiple mean-field solutions is clear: they describe genuine MB quantum interference. 

Following again the derivation of the semiclassical propagator for SP systems, MB quantum interference is made explicit by the application of the stationary phase approximation to the MB propagator, fully justified now by the emergence of the large parameter $N$. This renders all integrals involved  to be highly oscillatory and allows for following formally Gutzwiller's classic calculation \cite{Gutzwiller07}. First, the (scaled) quadrature path integral is calculated to obtain
\begin{equation}
\label{eq:semquad}
   \int_{{\bf Q}(0)={\bf Q}^{(i)}}^ {{\bf Q}(t)={\bf Q}^{(f)}}{\cal D}[{\bf Q}(s),{\bf P}(s)]{\rm e}^{iNR^{(r)}[{\bf Q}(s),{\bf P}(s)]}\simeq \sum_{\gamma({\bf Q}^{(i)},{\bf Q}^{(f)},t)}D_{\gamma}({\bf Q}^{(i)},{\bf Q}^{(f)},t){\rm e}^{iN{\cal R}_{\gamma}({\bf Q}^{(i)},{\bf Q}^{(f)},t)} 
\end{equation}
as a sum over the {\it mean field} solutions $\gamma$ with given boundary conditions. These solutions involving semiclassical amplitudes $D_{\gamma}$ that account for the stability of the solutions and other topological features of the Hamiltonian mean-field flow around them and, most importantly, the actions ${\cal R}_{\gamma}$ along each of them.

The semiclassical propagator in quadrature representation, Eq.~(\ref{eq:semquad}), is strictly speaking just an intermediate step in the construction of its Fock state version. Nevertheless, it is a very useful object with very desirable features. The first is the perfect analogy between the quadrature and coordinate representations in the semiclassics of  MB and SP  systems respectively, and the suggestive possibility of directly importing into the MB context several key ideas and results. Specifically, the derivation of the MB trace formula follows identical steps as in Gutzwiller's derivation for SP systems as discussed ahead. The second concerns the friendly way coherent states are treated in quadrature representation, allowing for a very tractable way to connect these two useful and natural sets of MB states. Last, but not least, it is the very natural way in which systems with negligible interactions are described by a mean field that defines a simple {\it linear} problem, a very appealing feature that is lost if physical Fock states are used instead~\cite{Engl14b}.

For this last change of representation the integrals are performed over quadratures, Eq.~(\ref{eq:full}), in the stationary phase approximation. Leaving aside lengthy technical details that can be found in~\cite{Engl2016,Engl2015}, the final result is~\cite{Engl14b}
\begin{equation}
   K({\bf n}^{(f)},{\bf n}^{(i)},t)\simeq \sum_{\gamma({\bf n}^{(i)},{\bf n}^{(f)},t)}A_{\gamma}({\bf n}^{(i)},{\bf n}^{(f)},t)){\rm e}^{iN{\cal R}_{\gamma}({\bf n}^{(i)},{\bf n}^{(f)},t)} 
   \label{eq:MB-propagator}
\end{equation}
where the sum extends over the solutions $({\bf n}_{\gamma}(s),{\boldsymbol \theta}_{\gamma}(s))$ of the mean-field equations with boundary conditions
\begin{equation}
    |\psi_{i}(s=0)|^{2}=n_{i}^{(i)} \quad , \quad |\psi_{i}(s=t)|^{2}=n_{i}^{(f)}
\end{equation}
and actions ${\cal R}_{\gamma}$. The semiclassical amplitudes are explicitly given by
\begin{equation}
   A_{\gamma}({\bf n}^{(i)},{\bf n}^{(f)},t)=\left[{\rm det}\left| \frac{N}{2\pi}\frac{\partial^{2} {\cal R}_{\gamma}({\bf n}^{(i)},{\bf n}^{(f)};t)}{\partial {\bf n}^{(i)} \partial {\bf n}^{(f)}}\right|\right]^{1/2}{\rm e}^{-i\frac{\pi}{2}\mu_{\gamma}}
\end{equation}
where the index $\mu_\gamma$ is the Maslov index of the trajectory $\gamma$~\cite{OzorioBook}.

The semiclassical approximation of the Fock state propagator is a fundamental result providing the starting point for the semiclassical analysis of both dynamical and stationary properties of MB quantum systems of interacting bosons.  The propagator~(\ref{eq:MB-propagator}) is not restricted to chaotic dynamics, but also allows, in principle, for investigating the imprint of more complex, {\it e.g.} mixed regular-chaotic, phase space dynamics and the consideration of system-dependent properties unique to individual bosonic MB systems.

Having at hand both the semiclassical propagator and a well defined classical (mean field) limit,  a fundamental conceptual aspect can be addressed, namely, the meaning of MB quantum chaos.  Since the asymptotic analysis automatically provides as a limit a theory with a well-defined Hamiltonian structure~\cite{OzorioBook} given by Eq.~(\ref{eq:Tom2}), the quantum ramifications of chaotic mean-field dynamics can be rather precisely investigated and interpreted. Therefore, for systems of interacting bosons with large occupations the MB quantum manifestations of MB mean-field chaos can be placed on a firm theoretical foundation.

The following passage summarizes the directions that, starting with the semiclassical propagator in Fock space or its variants, have been pursued by several groups during the last years in an attempt to understand the quantum signatures of mean-field integrability and chaos.

%%%%%%%%%%%%%%%%%%%%%%%%%%%%%%%%%%%%%%%%%%%%%%

\section{Applications}
\label{sec:appl}
\subsection{Dynamical aspects: the truncated Wigner approximation and beyond} 
\label{sec:twa}

The way the classical limit plays a role in the quantum mechanical description of a many-body system happens in three stages. At the most primitive level expectation values of time-dependent (Heisenberg) observables $\hat{A}(t)=A(\hat{{\bf b}}^{\dagger}(t),\hat{\bf b}(t))$, defined with respect to an initial coherent state $|{\bf z}\rangle=|{\boldsymbol \psi}\rangle$, are directly translated from the classical limit simply by transporting the classical symbol along the solution ${\boldsymbol \psi}(t), {\boldsymbol \psi}(0)={\boldsymbol \psi}$ of the mean-field equations (\ref{eq:MFE}),
\begin{equation}
    \langle {\boldsymbol \psi}|\hat{A}(t)|{\boldsymbol \psi}\rangle \simeq A({\boldsymbol \psi}(t),{\boldsymbol \psi}^{*}(t)) \ ,
\end{equation}
which defines the strict, textbook mean-field approximation.

In the second stage, the classical solutions are still used directly to guide the quantum evolution, but initial quantum fluctuations parametrized by the phase-space representation of the initial state are incorporated while still evolving through the mean-field flow,
\begin{eqnarray}
  \langle {\boldsymbol \psi}|\hat{A}(t)|{\boldsymbol \psi}\rangle \simeq \int d{\boldsymbol \Psi}d{\boldsymbol \Psi}^{*} {\rm e}^{-|{\boldsymbol \Psi}-{\boldsymbol \psi}|^{2}} A({\boldsymbol \Psi}(t),{\boldsymbol \Psi}^{*}(t)),
\end{eqnarray}
giving the celebrated truncated Wigner approximation (TWA) \cite{Polkovnikov10}. The pure mean-field approximation is then obtained as a particular case when the classical symbol $A_{\rm cl}$ is smooth and the integral is well approximated by taking ${\boldsymbol \Psi}\simeq {\boldsymbol \psi}$.  Both the mean field and TWA fail to account for coherent effects due to path interference. The former because it is based on a single unique classical solution, and the latter because it is based on adding probabilities instead of amplitudes. In essence, both approximations are {\it classical}. 

The third stage is to fully incorporate the semiclassical approximation.  It accounts for interference effects explicitly and completely in terms of the sum over amplitudes.  In the exact expression
\begin{equation}
    \langle {\boldsymbol \psi}|\hat{A}(t)|{\boldsymbol \psi}\rangle=\sum_{{\bf n},{\bf n}',{\bf m},{\bf m}'}{\boldsymbol \psi}_{{\bf n}}^{*}{\boldsymbol \psi}_{{\bf n}'}A_{{\bf m},{\bf m}'}K({\bf n},{\bf m},t)K^{*}({\bf n}',{\bf m}',t)
\end{equation}
where 
\begin{equation}
   {\boldsymbol \psi}_{{\bf n}}= \langle {\boldsymbol \psi}|{{\bf n}}\rangle {\rm \ \ , and \ \ }A_{{\bf m},{\bf m}'}=\langle {\bf m}|\hat{A}|{\bf m}'\rangle,
\end{equation}
the substitution of $K$ by its semiclassical approximation, given in Eq.~(\ref{eq:MB-propagator}), does the trick. The key object to analyze is the product
\begin{equation}
\label{eq:doublesum}
   K({\bf n},{\bf m},t)K^{*}({\bf n}',{\bf m}',t)\simeq \sum_{\gamma,\gamma'}A_{\gamma}A_{\gamma'}^{*}{\rm e}^{iN({\cal R}_{\gamma}-{\cal R}_{\gamma'})} 
\end{equation}
where $\gamma$ labels mean-field paths joining ${\bf n}$ with ${\bf m}$ in time $t$, and similarly for $\gamma'$. The TWA is readily obtained (in its polar form where $\psi=\sqrt{n}{\rm e}^{i\theta}$) from the diagonal approximation where the action  is linearized for the $\gamma'=\gamma$ terms~\cite{Engl14}.  Genuine MB interference arises from off-diagonal contributions $\gamma' \ne \gamma$. For a given case it is then a question of how much off-diagonal information, which demands a great effort to evaluate, is necessary to describe the physical phenomena of interest. This may range from the explicit and precise description of every quantum fluctuation around the classical background as done in \cite{Tomsovic18} well beyond the Ehrenfest time, to the selective use of restricted off-diagonal contributions to capture robust interference effects as in \cite{Schlagheck18}.

It is worth noting that the derivation of the TWA here relies on the more fundamental semiclassical approximation for MB amplitudes. As such, it is expected that the foundations of the TWA may suffer from ambiguities in systems where either the classical limit and/or the semiclassical regime cannot be precisely defined. Important systems such as spin-one-half chains and Fermi-Hubbard models indeed represent such cases. Progress towards a formal construction of the TWA in these cases has been an active field recently (see \cite{DAVIDSON17} and references therein), with successful applications to SYK models \cite{Schmitt19} and spin chains \cite{Schachenmayer15}.

%%%%%%%%%%%%%%%%%%%%%%%%%%%%%%%%%%%%
\begin{figure*}[th]
\includegraphics[width=16.5 cm]{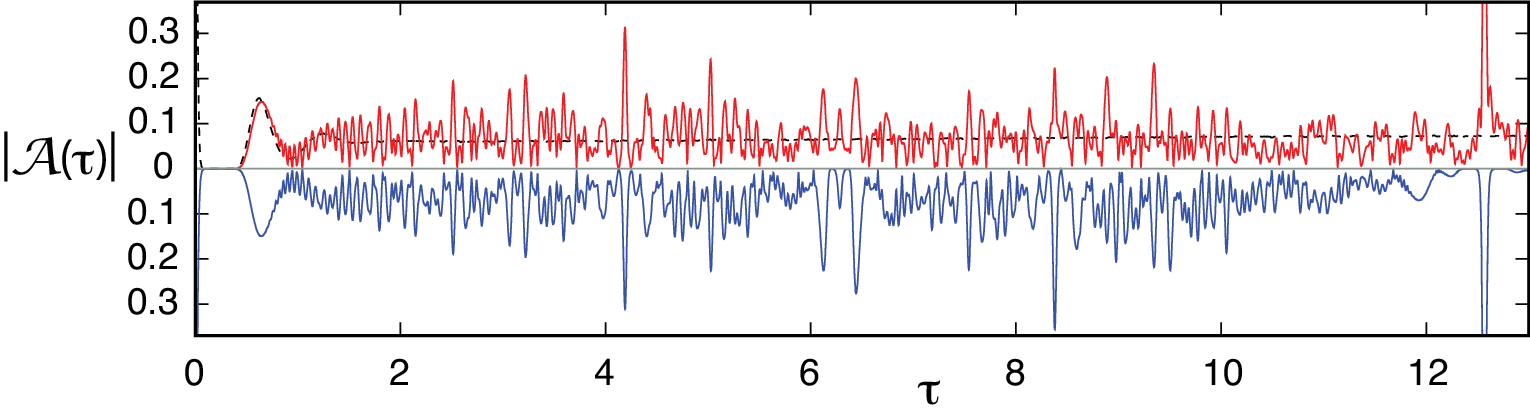}
\caption{Many-body quantum inteferences governing the return probability. Shown is the square root of the quantum time auto-correlation function, Eq.~(\ref{eq:autocor}), compared with the semiclassical result and the truncated Wigner approximation, see text.  The solid (blue), dashed (red), and dotted (black) curves represent the quantum, semiclassical, and TWA results, respectively. The curve showing the quantum mechanical data is reflected at the horizontal axis to better distinguish the curves. 
The numerical calculations shown are for an initial coherent state density wave $|20, 0, 20, 0\rangle$ on a ring with four sites and Bose-Hubbard parameter values $J=0.2$ and $U=0.5$. (Figure taken from Ref.~\cite{Tomsovic2018}, courtesy of the American Institute of Physics) 
\label{fig:autocor} }
\end{figure*}
%%%%%%%%%%%%%%%%%%%%%%%%%%%%%%%%%%%%%%%%%%%%%%%%%%%%%%%%%%%%%%%%%%%%%%%%%%%%%%%%%%%%

As a proof of principle and showcase for the power and accuracy of the semiclassical many-body approach outlined above we consider far out-of-equilibrium MB quantum dynamics, more precisely observables such as the quantum time auto-correlation function and its Fourier transform that yields MB spectra~\cite{Tomsovic2018}. 

Beyond an Ehrenfest time scale~\cite{Ehrenfest27, Berman78}, where quantum and classical expectation values start to diverge, the quantum dynamics is governed by strong interferences.
These play a central role for what is known as scrambling, i.e., the spreading of quantum information across the many degrees of freedom of a non-equilibrium MB system. 
While the truncated Wigner approximation, discussed above, leaves out these interferences, the  semiclassical theory properly incorporates such missing quantum effects.

Ultracold bosonic atomic gases provide a promising setting to explore such phenomena experimentally.  Experiments are nowadays probing dynamical processes far-from-equilibrium induced by a sudden parameter variation in the trapping setting.
Important examples for such non-equilibrium bosonic atoms include dynamical features such as  relaxation~\cite{Trotzky12}, thermalization~\cite{Kaufman16}, and (precursors of) MB-localization~\cite{Choi16}. 

In our context, an experimental regime with mesoscopic populations of lattice sites is relevant, with a few tens of atoms per site~\cite{Esteve08, Gross10}. 
These settings enable one to probe the crossover from the classical mean-field regime, well described by the Gross-Pitaevskii equation~\cite{Pitaevskii03}, to a quantum correlated regime where the mean-field approximation breaks down. The Ehrenfest time marks the limit of validity of the corresponding classical Truncated Wigner approximation (TWA).  Thus, ultracold Bose gases within mesoscopically populated lattices provide a promising setting to explore scrambling effects under well-controlled conditions~\cite{Yao16}.

In the following we use the corresponding theoretical model, a mesoscale Bose-Hubbard model, as the basis for our comparison between semiclassical and quantum MB calculations.
We show below that the semiclassical approach, both, quantitatively accounts for quantum MB interference effects and qualitatively provides insight into the underlying interference mechanisms. 
Furthermore, comparing the semiclassical predictions with those
derived from the TWA clearly signals the onset and presence of post-Ehrenfest MB quantum interferences.

The technical details of the semiclassical approach based on the coherent propagation of a Lagrangian manifold of classical trajectories, here mean-field solutions, can be found in Ref.~\cite{Tomsovic2018}. Here we merely review and interpret the main results.

We consider  Bose-Hubbard systems with tunable tunneling and interaction terms, respectively: 
\begin{equation}
\label{eq:BH}
\hat H = -J \sum_{j=1}^N \left(a^\dagger_j a_{j+1} + h.c.\right) +
\frac{U}{2} \sum_{j=1}^N \hat n_j \left(\hat n_j - 1 \right) 
\end{equation}
with  $N$ sites arranged on a ring. 
$U$ denotes the strength of the two-body interaction and  $J$ controls the tunneling amplitude, which depends on the well depth. 
We focus on the evolution of initial states corresponding to {\em coherent states}, because they begin maximally localized with minimum uncertainty. Hence they correspond to the most {\em classical} states, and therefore are most convenient for investigating the onset of genuinely quantum effects.
More precisely, we consider an initial state~\cite{Tomsovic2018}
\begin{equation}
|{\bf n}\rangle = \prod_{j=1}^N \exp
\left(-\frac{\left|b_j\right|^2}{2} + b_j  \hat a^\dagger_j \right)|{\bf  0}\rangle 
\end{equation}
where each site $j$ is loaded with a coherent state with average number of particles $n_j=\left|b_j\right|^2$. 
Specifically, a coherent state density wave, {\em e.g.}, reads $|n,0,n,0,...,n,0\rangle$.
The central observable, the time autocorrelation function, is denoted, with $\hat U(t)$  the unitary time translation operator,
\begin{equation}
{\cal C}(t) =\left|{\cal A}(t)\right|^2 \quad , \quad {\cal A}(t) = \langle {\bf n }\left| \hat U(t) \right| {\bf n}\rangle \, .
\label{eq:autocor}
\end{equation}

Figure~\ref{fig:autocor} shows the direct comparison of the quantum auto-correlation function along with the time-dependent semiclassical and TWA approximations. Data are shown for a 4-site coherent state density wave on a Bose-Hubbard ring ($J \! =\! 0.2$ and $U\!=\!0.5$) with a mean of  $40$ particles. These parameters correspond to a limiting classical dynamics that is not fully chaotic. 

The TWA is an accurate approximation up to the time scale
$\tau_1 = 2\pi/U n_j = 0.63$,
associated with first return of classical trajectories.
Thereafter, multiple contributing trajectories in the semiclassical approach (not shown) indicate MB quantum interferences. The TWA yields only the classical average of the auto-correlation function. However, the semiclassical approximation provides extremely precise results for long times significantly beyond the time $\tau_1$ that roughly can be associated with an Ehrenfest time. The semiclassical calculation is based on an increasing number (with time) of representative trajectories with interfering amplitudes. 
For instance, to adequately yield the full revival at time 
$\tau_2 = 2\pi/U = 4\pi= 12.57$,
a quantum scale associated with the revival of the initial quantum state,
requires the summation of roughly 600 trajectory contributions, for more details see Ref.~\cite{Tomsovic2018}.

%%%%%%%%%%%%%%%%%%%%%%%%%%%%%%%%%%%%%%%%%%%%%%%%%%%%%%%%%%%%%%%%%%%%%%%%%%%%%%%%%%%%%
\begin{figure}[tbh]
\centering
\includegraphics[width=8.5 cm]{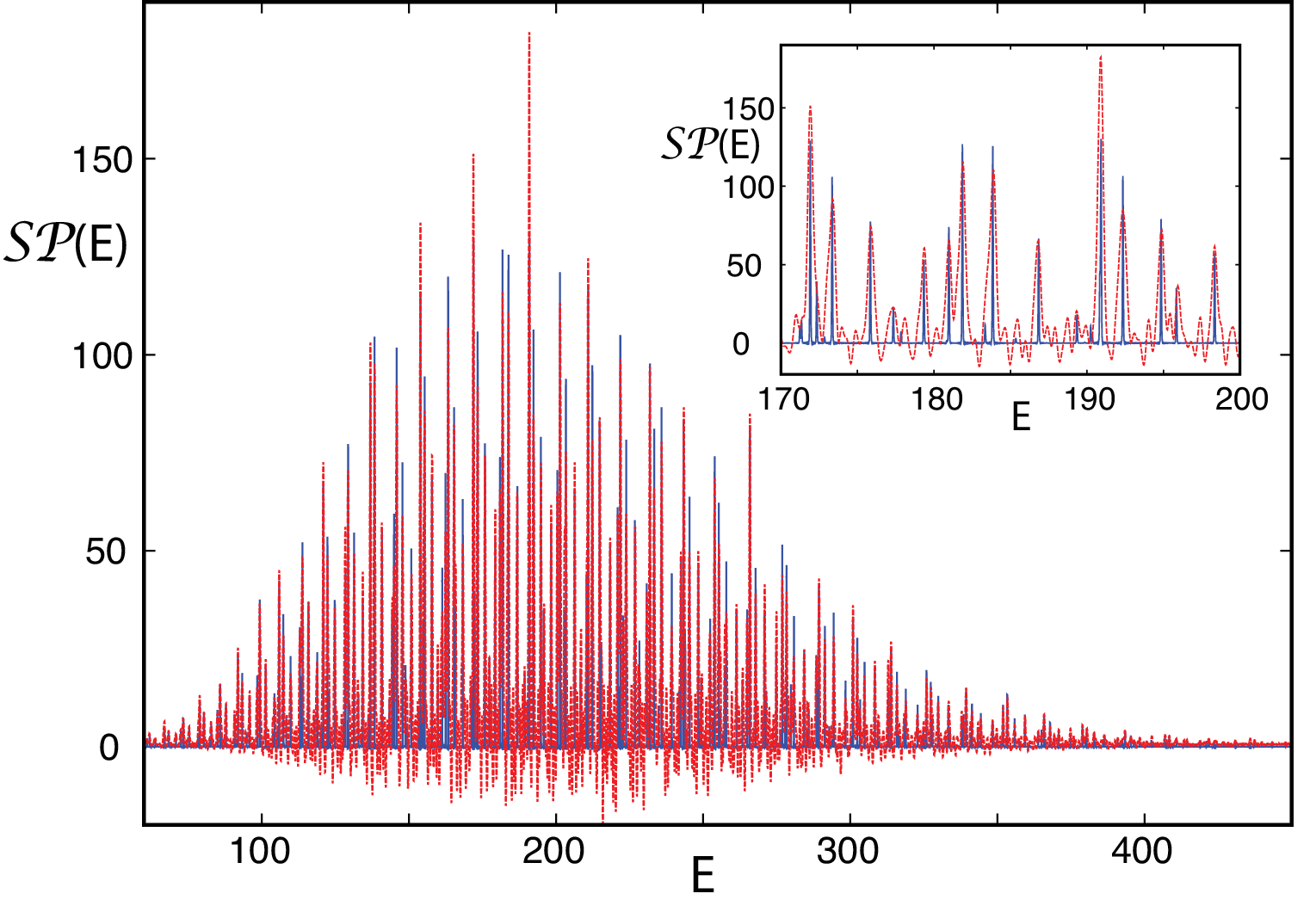}
\caption{Many-body spectra -- comparison of the quantum and semiclassical results (including a constant energy shift of $E_0=0.9$ for the semiclassical data to adjust to the correct energy centroid).
(Figure taken from Ref.~\cite{Tomsovic2018}, courtesy of the American Institute of Physics) 
\label{fig:spectrum} }
\end{figure}
%%%%%%%%%%%%%%%%%%%%%%%%%%%%%%%%%%%%%%%%%%%%%%%%%%%%%%%%%%%%%%%%%%%%%%%%%%%%%%%%%5

It is instructive to consider the Fourier transform of ${\cal A}(\tau)$, including full phase information. It is related via 
\begin{equation}
{\cal SP}(E) = \sum_\nu \left| \langle E_\nu | \vec n \rangle
\right|^2 \delta (E\!-\!E_\nu) \propto \int {\rm d} \tau \ {\rm e}^{iE\tau} {\cal A}(\tau) 
\end{equation}
to the (weighted) spectrum of the quantum system with MB eigenenergies $E_n$.

Figure~\ref{fig:spectrum} shows the spectra generated from the respective quantum and semiclassical autocorrelation functions. The agreement of the quantum mechanical data with the semiclassical theory is excellent. Remarkably, the latter correctly yields individual MB levels which implies a precision well beyond the MB level spacing.

To summarize this section, the semiclassical approximation of the full MB quantum dynamics provides accurate results with fully incorporated MB interferences.  
It is best adapted to systems with mesoscopic populations of particles or more, and most accurate over short to intermediate time scales.  The semiclassical approximation being effectively an expansion in the inverse of the density, the accuracy improves with increasing particle number. Moreover, its accuracy implies rather detailed MB spectroscopic information beyond the  mean-level spacing. 
In stark contrast, the TWA averages over all MB quantum interferences and hence cannot provide spectroscopic information.

The semiclassical method illustrated here is rather general and does not only apply to quantum systems with a fully chaotic limit. The semiclassical calculations have also been performed for $8$-site rings with up to 160 particles (not published), a regime where full quantum calculations get exceedingly challenging or are not yet possible.
Using a semiclassical propagator in fermionic Fock space~\cite{Engl14}), the present approach can be extended to the case of fermionic atoms.

%%%%%%%%%%%%%%%%%%%%%%%%%%%%%%%%%%%%%%%%%%%%%%%%%%%%%

\subsection{Coherent backscattering in Fock space}
\label{sec:cbs}

%%%%%%%%%%%%%%%%%%%%%%%%%%%%%%%%%%%
\begin{figure}
    \centering
  \includegraphics[width=0.7\linewidth]{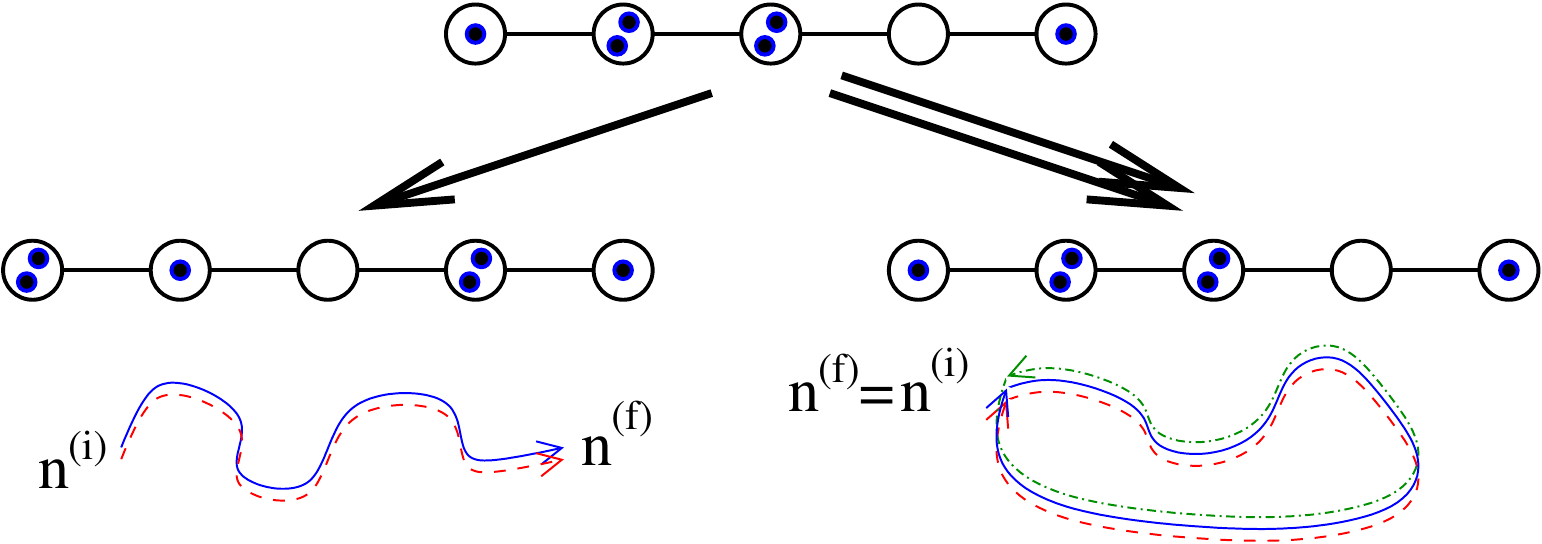}
    \caption{ \label{Fig:CBSII}
   {\bf
   Many-body coherent backscattering in Fock space.
  } The semiclassical propagator used to calculate the transition probability between Fock states $|K({\bf n}^{(i)},{\bf n}^{(f)},t)|^{2}$ naturally leads to the consideration of double sums over solutions of mean-field equations. Under local averaging only robust, smooth contributions survive. Generically, these smooth contributions simply correspond to interference from pairs of identical amplitudes corresponding to the same mean-field solutions. For ${\bf n}^{(f)}={\bf n}^{(i)}$, however, {\it two} different solutions related by time-reversal constructively interfere, a purely quantum effect due to coherent superposition of amplitudes.
  (From Ref.~\cite{Engl14b}.)}
\end{figure}
%%%%%%%%%%%%%%%%%%%%%%%%%%%%%%%%%%%%%%%%%%%%%%%%%%%5

\begin{figure}
    \centering
  \includegraphics[width=0.8\linewidth]{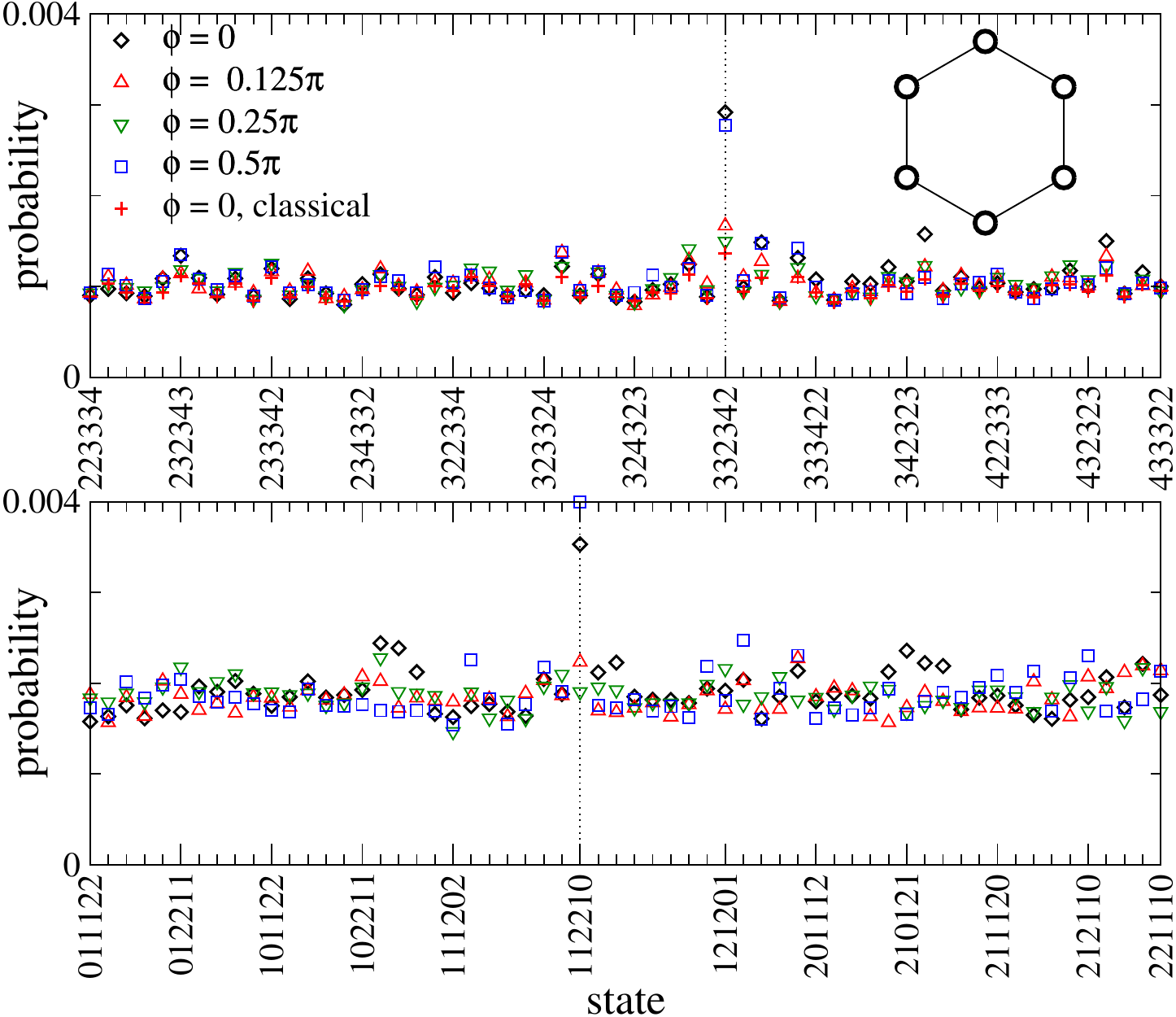}
    \caption{ \label{Fig:CBSI}
   {\bf
   Many-body coherent backscattering in Fock space.
  } Numerical simulation of the transition probability in Fock space for a ring-shaped 6-site Bose-Hubbard ring (upper right inset). The initial state ${\bf n}^{(i)}$ (indicated by the vertical line) is propagated for a time larger than the equilibration time and the probability of finding the system in the final state ${\bf n}^{(f)}$ indicated in the horizontal axis is calculated by exact numerical solution of the quantum dynamics.  The observed enhancement of the transition probability when ${\bf n}^{(f)}={\bf n}^{(i)}$ over the classical uniform background is a purely coherent effect that is suppressed by the application of a gauge field $\phi$ that destroys the time-reversal invariance of the system. (From Ref.~\cite{Engl14b}.)}
\end{figure}
%%%%%%%%%%%%%%%%%%%%%%%%%%%%%%%%%

Following the success of the SP semiclassical approach in describing the leading-order interference term in disordered conductors~\cite{Akkermans07}, the so-called weak localization correction arising from interference between pairs of time-reversed paths in real space, the corresponding MB effect arising from the superposition of amplitudes associated with time-reversed mean-field solutions was proposed in \cite{Engl14b}, where it is shown that such coherent backscattering produces a characteristic enhancement of the return probability in MB Fock space. The schematic picture of this genuinely MB interference effect, characteristic of initial and final states being precisely Fock states is illustrated in Fig.~\ref{Fig:CBSII}.

 The numerical confirmation of this coherent MB effect is shown in Fig.~\ref{Fig:CBSI}. Here, the probability $P({\bf n}^{(i)},{\bf n}^{(f)},t)=|K({\bf n}^{(i)},{\bf n}^{(f)},t)|^{2}$ of finding the system in the final Fock state ${\bf n}^{(f)}$ at time $t$, initially prepared in the state ${\bf n}^{(i)}$, is obtained by numerically solving the quantum dynamics of a 6-site Bose-Hubbard ring in the regime of chaotic mean field dynamics. After a relatively short relaxation time scale, the tendency toward  uniform equilibration is clearly visible where all transition probabilities roughly reach the same classical value (also well described by the TWA). The only exception happens for ${\bf n}^{(f)}={\bf n}^{(i)}$ in which a non-classical enhancement is clearly observed. Furthermore, if time-reversal invariance of the system is broken by means of a gauge field parametrized by $\phi$, this enhancement, a hallmark of coherent backscattering due to quantum interference among classical paths related by time reversal, disappears. The semiclassical explanation of this effect starts with the double-sum over mean field solutions in Eq.~(\ref{eq:doublesum}), and the realization that only for ${\bf n}^{(f)}={\bf n}^{(i)}$, there is an off-diagonal contribution from orbits $\gamma,\gamma'$ related by time reversal as depicted in Fig.~\ref{Fig:CBSII}.

%%%%%%%%%%%%%%%%%%%%%%%%%%%%%%%%%%%%%%%%%%%%%%%%%%%%%%%%%%%%%%%%%%%%%%%%%%%%%%%%%%%%%%

\subsection{Spectral properties I: Many-body trace formulae}
\label{sec:spec-prop}

Beyond its use for calculating dynamical properties of observables,
the MB van Vleck-Gutzwiller propagator (\ref{eq:MB-propagator}) represents the fundamental building block for a semiclassical theory of spectral properties of MB bosonic systems by means of the MB density of states which leads to a MB version of Gutzwiller's trace formula. Following Ref.~\cite{Engl15}, it turns out that
periodic mean-field solutions of the nonlinear wave equations play the role of the
classical SP periodic orbits in Gutzwiller’s original trace formula.
Based on the MB trace formula, RMT-type universal spectral correlations can be addressed through the lens of semiclassical theory as proposed in \cite{Engl15,TF_Remy} where post-Ehrenfest phenomena in the MB context naturally arise again.

The success of Gutzwiller's trace formula (\ref{eq:SP-Gutzwiller}) makes it quite natural to investigate the corresponding MB extension in bosonic systems. The straightforward generalization by means of  increasing the particle number $N$ for the usual semiclassical limit $\hbar \rightarrow 0$, {\em i.e.} the horizontal crossover in Fig~\ref{fig:sc-limits} results, however, in considering periodic orbits in a vast $6N$-dimensional phase space. Furthermore, one has to cope with the (anti-)symmetrization. Hence, in deriving a semiclassical approximation for the MB level density in bosonic systems it is practically preferable to resort to the complementary limit $\heff = 1/N \rightarrow 0$ following \cite{Engl15}.

Under the assumption of chaotic mean field dynamics, a series of further stationary-phase calculations starting from the MB semiclassical propagator (in either quadrature or Fock states representation) leads eventually to the semiclassical approximation for the MB density of states in the form~\cite{Engl15}
\begin{equation}
    \rho(E,N) \simeq \bar{\rho}(E,N) \ + \ \rho^{\rm osc}(E,N) =
    \bar{\rho}(E,N) \ + \ 
    \sum_{\rm pm}A_{\rm pm}{\rm e}^{iNS_{\rm pm}(E)} \, .
    \label{eq:MB-Gutzwiller}
\end{equation}
The first (Weyl) term is given, to leading order in $\heff$, by the phase-space volume of the corresponding mean-field energy shell,
\begin{equation}
    \bar{\rho}(E,N)=\left(\frac{N}{2\pi}\right)^{d}\int \ d{\bf q} \ d{\bf p}\ \delta(E-H_{{\rm cl}} ({\bf q},{\bf p}))\ \delta(N-N({\bf q},{\bf p})) \, ,
\end{equation}
where $H_{{\rm cl}}$ is defined in Eq.~(\ref{eq:Tom2}).
The sum in Eq.~(\ref{eq:MB-Gutzwiller}) extends over all {\it periodic} solutions 
\begin{equation}
    ({\bf q},{\bf p})(t=0)=({\bf q},{\bf p})(t=T) 
\end{equation}
(for some period $T$)
of the classical mean-field  equations 
\begin{equation}
\label{eq:MBequations}
    \hbar \frac{d{\bf q}}{dt}=\frac{\partial H_{\rm cl}({\bf q},{\bf p})}{\partial {\bf p}}, {\rm \ \ \ } \hbar \frac{d{\bf p}}{dt}=-\frac{\partial H_{\rm cl}({\bf q},{\bf p})}{\partial {\bf q}}
\end{equation}
with fixed energy $E$ and particle number 
\begin{equation}
 N({\bf q},{\bf p})=\frac{1}{2}\sum_{i}(q_{i}^{2}+p_{i}^{2}) \, .
\end{equation}
We see that within MB semiclassics, unstable periodic mean-field modes (pm) in Eq.~(\ref{eq:MB-Gutzwiller}) play the role of the classical periodic orbits in the SP context. In close analogy, the actions take the form
\begin{equation}
    S_{\rm pm}(E) = \oint_{\rm pm} {\bf p}({\bf q},E)\cdot d{\bf q}
    \label{eq:MB-action}
\end{equation}
and the amplitudes read, as in Eq.~(\ref{eq:SP-stab}),
\begin{equation}
    A_{\rm pm}(E) = \frac{T_{\rm ppm}(E)}{|{\rm d
    et}({\bf M}_{\rm pm}(E) - {\bf I} |^{1/2}} {\rm e}^{-i\mu_{\rm pm} \frac{\pi}{2}}\ ,
\end{equation}
in terms of the period $T_{\rm ppm}$ of the primitive mean field mode, the monodromy matrix ${\bf M_{\rm pm}}$ depending on the stability of the mode, and its Maslov index $\mu_{\rm pm}$. 

The resulting semiclassical MB trace formula for discrete quantum fields incorporates genuine MB interference including that required to build up the discreteness of the MB spectrum, which arises from the coherent sum over phases associated with periodic mean field solutions.
This is in close analogy to Gutzwiller's expression for the SP case.

Note that the range of validity in energy extends down to lowest energies because $\heff$ and not $\hbar$ controls the semiclassical limit, and thus Eq.~(\ref{eq:MB-Gutzwiller}) holds true even for MB ground states as long as $N \gg 1$. Interestingly, because in the non-interacting case the quantum problem reduces to a harmonic system, the trace formula is still applicable since the corresponding periodic mean-field solutions (of the linear Schrödinger equations) turn out to be isolated in the generic case where the SP energies defining their frequencies are incommensurate. We also note a certain conceptual analogy between the semiclassical MB propagator and the corresponding MB density of states as sums over mean-field solutions on the one hand and configuration interaction methods, constructing MB wave functions as linear combinations of Slater determinants, {\em i.e.} fermionic mean-field solutions on the other hand.

The MB trace formula allows, in principle, for computing an approximate MB density of states\footnote{In Ref.~\cite{Tomsovic18} a spectrum of a Bose-Hubbard system (with $N=$40 atoms) was computed with high accuracy, using corresponding MB semiclassical techniques.} for MB energies and particle numbers that are out of reach of usual numerical MB techniques.  Moreover, the close formal relation between the semiclassical SP, Eq.~(\ref{eq:SP-Gutzwiller}), and MB, Eq.~(\ref{eq:MB-Gutzwiller}), trace formulas implies that many insights and results known for quantum chaotic SP dynamics can swiftly be taken over into the MB context as summarized in the following for spectral fluctuations.

%%%%%%%%%%%%%%%%%%%%%%%%%%%%%%%%%%%%%%%%%%%%%%%%%%%%%%%%%%%%%%%%%%%%%%%%%%%%%%%%%%%%%%

\subsection{Many-body encounters and universal spectral correlations}
\label{sec:spec-stat}

In Sec.~\ref{sec:SPcorrelations} the semiclassical foundations of RMT-type spectral universality are outlined for chaotic SP dynamics, reflected in the BGS conjecture~\cite{Bohigas84}.  
Although it might be evident to consider that this reasoning simply carries over to the quantum chaotic MB case, a formal justification has been missing. 
Also it was not straightforward how the encounter calculus would be generalized to the MB case.

For the usual semiclassical limit $\hbar \rightarrow 0$, the encounter formalism has been shown to be applicable and to lead to RMT results for any phase-space dimensionality~\cite{Turek05}, and hence also to the $6N$ dimensions of an $N$-particle system in 3 spatial dimensions. However, MB generalizations require some care. For instance, for a non-interacting MB system with chaotic single-particle dynamics, {\em e.g.} $N$ non-interacting particles in a billiard, the MB density of states is composed of independent single-particle spectra -- with conserved single-particle energies as associated constants of motion -- and thus do not obey RMT-statistics. The spectral statistics are Poissonian in the infinite dimensional limit; for recent work showing rich spectral features due to finite dimensionalities see \cite{Liao20}.  Correspondingly, in the complementary limit $\heff \rightarrow 0$ the non-interacting case also features non-generic spectral fluctuations that do not correspond to the expected Poissonian spectra of integrable systems. This is a
consequence of the field theoretical context where
the free field corresponds to a peculiar linear system that is non-generic since it is not merely integrable.  There, treating the quasi-integrable case due to the effect of a small interaction within a semiclassical perturbation theory may provide a useful approach.

Consider strongly interacting MB systems with an associated chaotic mean-field dynamics characterized by a largest MB Lyapunov exponent $\lambda$.
For SP dynamics universal spectral correlations arise from interference between periodic orbits with quasi-degenerate actions and periods beyond the Ehrenfest time $\tE^{\rm (sp)} = (1/\lsp) \log (S/\hbar)$. Correspondingly,
for quantum chaotic large-$N$ MB systems in the limit $\heff \rightarrow 0$, correspondingly genuine MB quantum interference is governed by another log-time scale, the Ehrenfest time 
\begin{equation}
    \tE = \frac{1}{\lambda} \ \log N \, , 
    \label{eq:scrambling}
\end{equation}
also referred to as the scrambling time in the MB context~\cite{Swingle16,Xu22}.

This very close formal analogy between the SP $\hbar\rightarrow 0$ and the MB $\heff \rightarrow 0$ regimes -- based on corresponding trace formulas and hierachy of times scales -- allows for the straightforward generalization of the semiclassical calculation of the bosonic MB spectral form factor~\cite{Engl15, TF_Remy} by applying the encounter calculus summarized in Sec.~\ref{sec:SPcorrelations}. This amounts to replacing $\hbar$ by $\heff$, the Lyapunov exponent $\lsp$ by $\lambda$, the Ehrenfest time 
$ \tE^{(sp)}$ by $\tE$, Eq.~(\ref{eq:scrambling}), SP phase space by the $2L$-dimensional phase space of the lattice, and the SP density of states $\rho_{\rm sp}^{\rm osc}(E)$ by
$\rho^{\rm osc}(E;N)$, Eq.~(\ref{eq:MB-Gutzwiller}).
Encounters between different (periodic) mean-field modes take on the role of encounters between classical (periodic) orbits. This implies the interpretation that interfering periodic mean-field solutions of Eqs.~(\ref{eq:MBequations}) with quasi-degenerate actions $S_{\rm pm} (E)$, Eq.~(\ref{eq:MB-action}),
lead to the emergence of universal MB spectral fluctuations, in close correspondence with the reasoning in the SP case. This includes in particular the same RMT-type expressions for the spectral MB two-point correlator $R(\Delta E;N) \sim \langle \rho^{\rm osc}(E;N) \rho^{\rm osc}(E+\Delta E;N)\rangle$ and its associated spectral form factor\footnote{Note that $R(\Delta E;N)$ contains interesting new
parametric correlations with regard to (changes in)
particle number and or interaction strength.}.

These conclusions drawn from the semiclassical MB encounter formalism coincide both with long known results from nuclear shell models and data~\cite{Brody81, Bohigas83}, embedded Gaussian ensembles~\cite{KotaBook}, including those restricted to two-body interactions, as well as with recent results showing random-matrix behavior of the spectral form factor for a periodically kicked Ising spin-1/2 model without a semiclassical limit (to leading orders in $t/t_H$ \cite{Bertini18,Kos18,Bertini19}) and for a Floquet spin model~\cite{Chan18}. Moreover, Wigner-Dyson-type level statistics have recently been numerically shown with appropriate parameters for a discrete Bose-Hubbard system~\cite{MBQC1, TF_Remy} and the SYK-model~\cite{Garcia17} in the large $N$-limit.  They help close an important missing theoretical link in the sense that the k-body restricted interaction ensembles, i.e.~embedded random matrix ensembles, have resisted analytic proofs of their asymptotic behaviors~\cite{Asaga01, Srednicki02} unlike the original classical random matrix ensembles.  Semiclassical results for spectral determinants~\cite{Keating91, Heusler07, Keating07, Mueller09, Waltner09, Waltner20} in terms of pseudo-orbits carry over to the MB case, i.e.~the semiclassical finding \cite{Waltner20} that pseudo-mean-field paths with durations $t > t_H$ necessarily must involve multiple partial traversals that do not contribute to the MB spectrum.

It is worth, highlighting again the relevance of the scrambling time scale $\tE$, Eq.~(\ref{eq:scrambling}), and the role of encounters for entangling mean-field modes: The semiclassical MB propagator and the trace formula, as sums over mean-field paths, contain genuine MB interference thereby giving rise to MB correlations. The encounter calculus is involving ergodic averages distilling out of all these paths, otherwise mostly random interference terms, that prevail for an observable after energy or spatial averages. Each encounter diagram, such as those shown in Fig.~\ref{fig:orbit-pair}, represents all interferences resulting from certain types of coupled mean field trajectory configurations with quasi-degenerate actions. If we think of entanglement as coupling between different product states, as mean field solutions are, then encounters generate entanglement that resists (energy) averaging. After starting to propagate a wave packet along a separable initial (periodic) mean field path, it will split at an encounter, acting like a rail switch and entangling the mean field paths that come close to each other in the encounter region in Fock phase space. The time scale of this entanglement process is given by $\tE$.  Whereas the relevance of encounters for entanglement arises naturally, developing tools to measure the degree of entanglement created through encounter structures remains open. 
This would also distinguish encounter-mediated entanglement growth from some sort of entanglement captured through TWA approaches introduced in Sec.~\ref{sec:twa}.  Quantum unitaries acting as interconnects in quantum networks can be viewed as mimicking certain encounter structures of a quantum chaotic dynamical system. For such random unitary dynamics entanglement growth has been measured~\cite{Nahum17}.

To conclude, MB semiclassical methods developed for systems with underlying chaotic dynamics can provide a direct theoretical derivation of universality in the spectral statistics of large-$N$ MB systems, and the applicability of RMT more generally.  In the regime of exponentially unstable mean field chaos in the classical limit, the local fluctuations of MB spectra comply with RMT predictions.  Though beyond the objective of this contribution, keep in mind that MB semiclassical methods accomplish more than this.  They apply to individual systems and system specific non-statistical quantities as well. In particular, if some system displayed unexpected properties from an RMT perspective, it would still be expected that the semiclassical theory would capture that behavior.  This is especially true of system specific MB dynamics before, up to, and just beyond the Ehrenfest time scale, a time scale that has collapsed to zero in RMT.
%%%%%%%%%%%%%%%%%%%%%%%%%%%%%%%%%%%%%%%

\subsection{Eigenstate properties: the Random Wave Model in Fock space}
\label{sec:RWM}
% \cite{Gut1,*Gut2,Berr1,BB1970,*BB1972,*BB1974,Gutb}.\cite{Berr3,Klaus1,Haak1,*Haak2}.

%%%%%%%%%%%%%%%%%%%%%%%%%%%%%%%%%%%%%%%%%%%%%%
\begin{figure}[ttt]
    %\vspace{-0.97cm}
    \centering
    \includegraphics[width=0.8\linewidth]{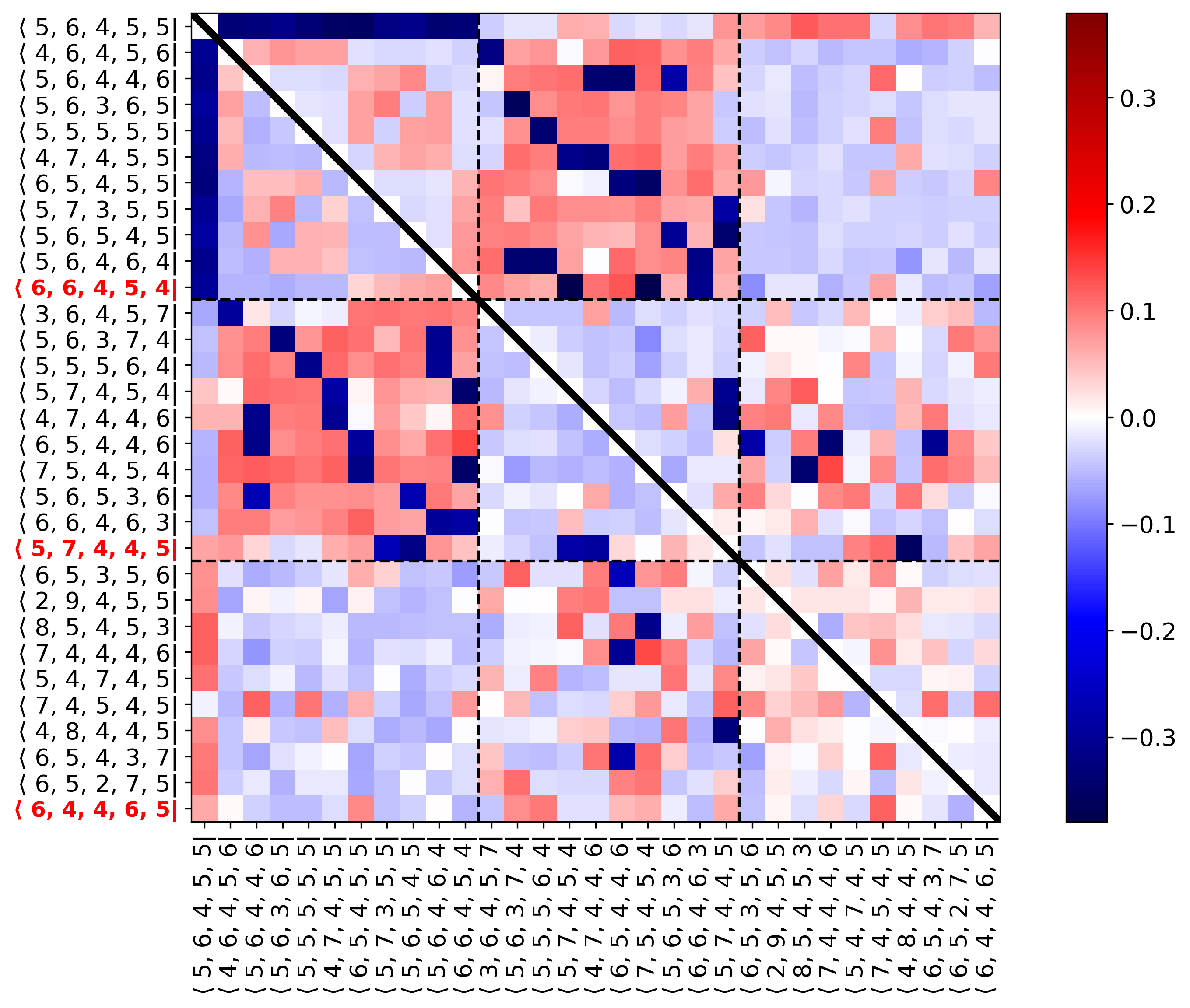}
   \caption{
   {\bf Eigenstate correlations in a Bose-Hubbard chain beyond Random Matrix Theory--}
   Cross correlations $R_{{\bf n},{\bf m}}(E)=\overline{\langle {\bf n} |\psi\rangle \langle \psi |{\bf m} \rangle}_{E}$ between expansion coefficients  (w.r.t  Fock basis states $|\mathbf n\rangle$ and $|\mathbf m\rangle$) of MB eigenstates $|\psi^{(j)}\rangle$ of a chaotic Bose-Hubbard chain, Eq.~(\ref{eq:BH}).
   In the lower triangle results are shown for the normalized correlator $R_{{\bf n},{\bf m}}(E)/\sqrt{R_{{\bf n},{\bf n}}(E)R_{{\bf m},{\bf m}}(E)}$, obtained by averaging numerically calculated eigenstates with energies $E^{(j)}$ over a microcanonical energy window  $E^{(j)} \in [E-\eta/2,E+\eta/2]$,
   Eq.~(\ref{eq:Rnm}). They show clear mesoscopic correlations beyond random matrix theory {\em i.e.} $R^{{\rm RMT}}_{{\bf n},{\bf m}}\sim \delta_{{\bf n},{\bf m}}$, signaling quantum interference, well described by the corresponding semiclassical correlator $R^{{\rm sc}}_{{\bf n},{\bf m}}$, Eq.~(\ref{eq:Rsc}), shown in the upper triangle. For visualization, the representative states $|\mathbf n\rangle$ and $|\mathbf m\rangle$  for $L\!=\!5$ sites and $N\!=\!25$ particles
  are obtained from a reference  "seed" $|5,6,4,5,5\rangle$ by changes in local occupations up to $\pm 3$; see text and \cite{suppmat} for other seed states.
   Adapted from Ref.~\cite{Schoeppl2025}}
	\label{fig:Corr}
\end{figure}

%%%%%%%%%%%%%%%%%%%%%%%%%%%%%%%%%%%%%%%%%%%%%%

It is a well established fact that in SP systems, universality of spectral fluctuations is complemented by universality in the spatial patterns of eigenstates, namely on \emph{eigenstate statistics}. The quantum chaos study of eigenstate morphology leads to a celebrated conjecture, confirmed by the seminal numerical study  \cite{mcdonaldSpectrumEigenfunctionsHamiltonian1979,mcdonaldWaveChaosStadium1988}, and put in firm grounds by  
Berry's semiclassical analysis of chaotic eigenfunctions \cite{Berr4} in $d$-dimensions leading to an amplitude-amplitude correlator between positions ${\bf r}$ and ${\bf r'}$ given by a Bessel function $J_{d/2-1}(x)/x^{d/2-1}$. The characteristic coherent oscillations as a function of the spatial distance $x=k(E,\bar{{\bf r}})|{\bf r}-{\bf r'}|$ are fixed through the local wavenumber $k^{2}(E,\bar{\bf r})=2m(E-V(\bar{\bf r}))/\hbar^{2}$ at the energy $E$ and mean position $\bar{{\bf r}}=({\bf r}+{\bf r'})/2$. Berry's result has the exact same dependence on the external potential $V({\bf r})$ for any chaotic system, an universality that moreover lies beyond  RMT for any finite energy $E < \infty$ . Berry's work goes however well beyond this spatial amplitude correlator by using it as the unique function fixing an amplitude distribution universally given in Gaussian form for chaotic  systems. 

This combination of universal amplitude distribution fixed by a microscopic amplitude correlator constitutes the SP Random Wave Model (RWM) \cite{Berr4,Sied1,JDUre} and turns into a powerful approach to describe the spatial morphology of eigenfunctions in mesoscopic chaotic systems \cite{RWM3,Urbina06,RWM2}. Its validity is supported by a huge amount of numerical \cite{PhysRevA.37.3067,RWMBe3,bogomolny2002percolation} and experimental \cite{HJ1,HJ2,HJ3} results. Moreover, due to its appealing form based on Gaussian random fields, it allows for the explicit analytical computation of even delicate topological features \cite{RWMBe3,RWMBe4,Dennis1,Dennis2} including nodal structures \cite{bogomolny2002percolation,jainNodalPortraitsQuantum2017}.

\begin{figure}
\centering
	\includegraphics[width=0.8\linewidth]{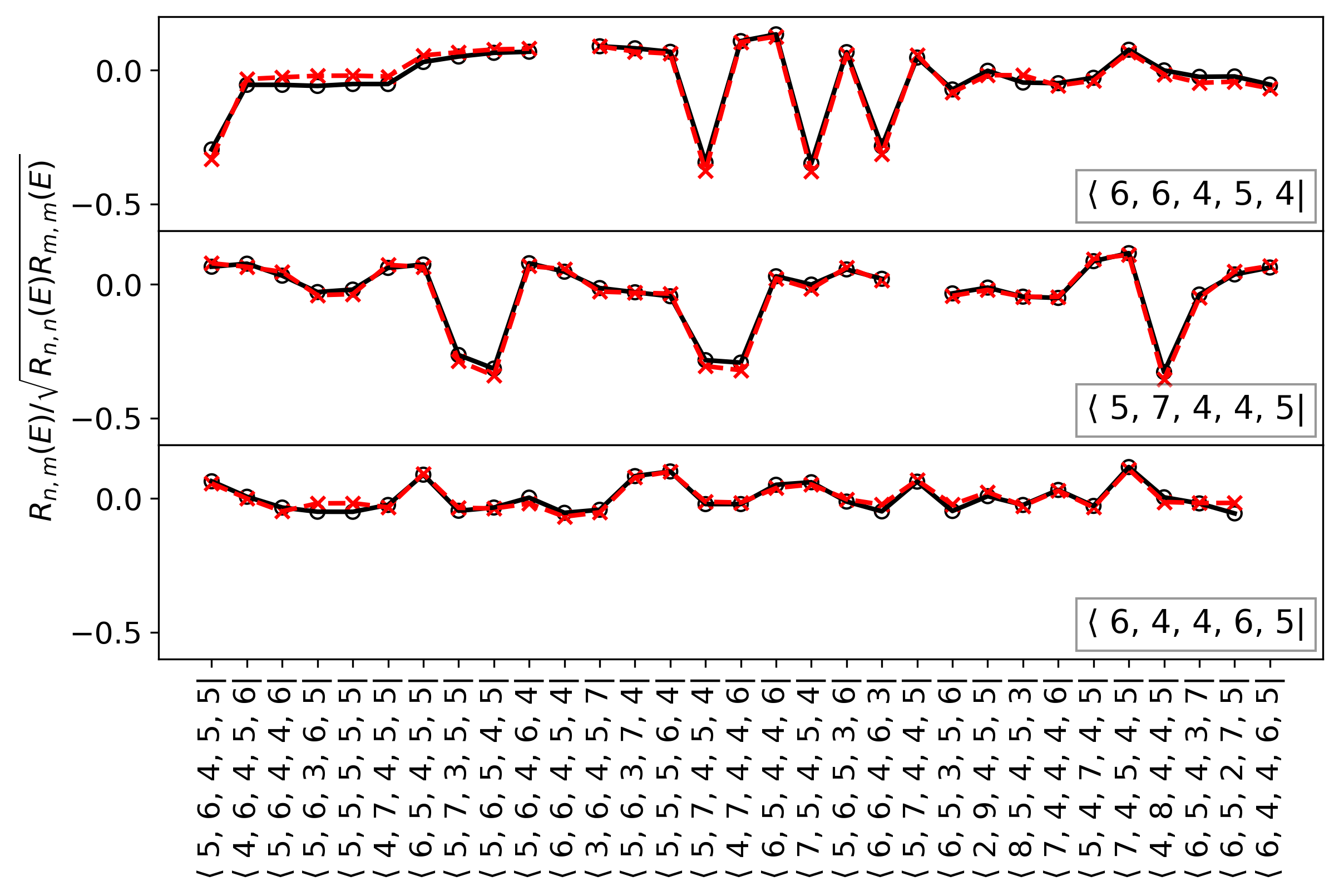}
   \caption{\textbf{Individual normalized correlators --}  The same as in Fig.~\ref{fig:Corr} but for a selection of three horizontal cuts at $|\mathbf n\rangle$ with ${\mathbf n}=6,4,4,6,5$ (bottom), ${\mathbf n}=5,7,4,4,5$ (middle), and ${\mathbf n}=6,6,4,5,4$ (top).
   Along the horizontal axis we mark a representative set of states $|\mathbf m\rangle$ with changes up to $\pm 3$ in their occupation numbers relative to the respective $|\mathbf n\rangle$.
   Black and red symbols mark numerically exact normalized correlators and the prediction (\ref{eq:Rsc}) of the MB semiclassical theory, respectively (lines serve as guides to the eye). The gaps correspond to the diagonal ${\bf n}={\bf m}$, where the normalized correlator is unity by construction.
   Adapted from Ref.~\cite{Schoeppl2025}
   }
\label{fig:cuts}
\end{figure}

Within the resurgence of MB quantum chaos during the last two decades, chaotic MB eigenstates have been mainly described within  pure RMT approaches where they are assumed to behave as random vectors, only weakly correlated through their orthonormality \cite{Rig1}. However, as shown 
in Figs.~\ref{fig:Corr} and~\ref{fig:cuts}, there exist clear eigenstate correlations in the Fock space of bosonic systems with chaotic mean-field limit that are obviously not captured by RMT. Such correlations prevail, even in the ergodic phase, for systems with large but finite particle numbers, making them relevant for quantum simulators operation~\cite{doi:10.1126/science.aal3837,Aidelsburger_2018,RevModPhys.80.885}. The conceptual structure of MB semiclassics should therefore  provide a proper generalization of the RWM and the extension of Berry's ansatz into the mesoscopic regime of chaotic MB systems. 

To introduce the statistical measures to describe eigenstate fluctuations,
consider a  MB system defined by a Hamiltonian $\hat{H}$ with  eigenfunctions $\hat{H}|\psi^{(j)}\rangle=E^{(j)}|\psi^{(j)}\rangle$ and eigenvalues $E^{(0)}\le E^{(1)} \le \ldots$ . Within a spectral region of width $\eta$ around an energy $E$, implemented through a window function $W_{\eta}(x)$ centered at $x=0$, a functional $F(|\psi\rangle)$ associating a complex number to an arbitrary state $|\psi\rangle$ is a fluctuating quantity. We then naturally define
\begin{equation}
\label{eq:Fexact}
\overline{F(|\psi\rangle)}_{E}=\sum_{j=0}^{\infty}\frac{W_{\eta}(E^{(j)}\!-\!E)}{\rho_{\eta}(E)} F(|\psi^{(j)}\rangle)\, 
\end{equation}
% $\overline{F(|\psi\rangle)}_{E}$
as the espectral average of the numbers $F(|\psi^{j}\rangle)$. Demanding $\bar{1}_{E}=1$ gives then
\begin{equation}
 \rho_{\eta}(E)={\rm Tr~}W_{\eta}(E-\hat{H}) \sim {\rm e}^{N}
\end{equation}
 as the mean ({\em i.e.}~coarse-grained) level density, where we assume a large number of eigenstates inside the spectral window. 

A particular important choice for the statistical measure is given by $F(|\psi\rangle)=\langle {\bf n}|\psi\rangle \langle \psi|{\bf m}\rangle$. Substitution of this functional in Eq.~(\ref{eq:Fexact}) defines then the two-point cross-correlation of expansion coefficients $\langle {\bf n}|\psi^{(j)}\rangle$ and $\langle {\bf m}|\psi^{(j)}\rangle$:
\begin{equation}
\label{eq:Rnm}
\overline{\langle {\bf n}|\psi\rangle \langle \psi|{\bf m}\rangle}_{E}=\sum_{j=0}^{\infty}\frac{W_{\eta}(E^{(j)}\!-\!E)}{\rho_{\eta}(E)}\langle {\bf n}|\psi^{(j)}\rangle \langle \psi^{(j)}|{\bf m}\rangle \equiv R_{{\bf n},{\bf m}}(E) \; .
\end{equation}
and consequently the covariance matrix
\begin{equation}
\label{eq:CovR}
    R_{{\bf n},{\bf m}}(E)=\langle {\bf n}| \hat{R}(E)|{\bf m}\rangle,
\end{equation}
nicely expressed in terms of the basis-independent covariance operator
\begin{equation}
  \hat{R}(E)=\overline{|\psi\rangle\langle \psi |}_{E}=\frac{W_{\eta}(E-\hat{H})}{{\rm Tr~}W_{\eta}(E-\hat{H})} \; , 
  \label{eq:R}
\end{equation}
{\it i.e.}~the density operator corresponding to the microcanonical window $W_{\eta}(x)$. 

The implementation of the MB generalization of Berry's construction is performed in two stages: First, we employ for
$\overline{F(|\psi\rangle)}_{E}$
a Gaussian distribution, 
\begin{equation}
\label{eq:Gauss}
\overline{F(|\psi\rangle)}_{E}\!\simeq\!
\overline{F(|\psi\rangle)}_{E}^{\rm G}\!=\!
\frac{1}{\mathcal{N}(\hat{R})}\int_{{\cal H}}\!\!F(|\psi\rangle){\rm e}^{-\langle \psi |\hat{R}(E)^{-1}|\psi\rangle}\ud|\psi\rangle\ ,
\end{equation}
valid in chaotic systems~\cite{Sied1,JDUre,Heller_2007}, where $\mathcal{N}(\hat{R})$ fixes $\overline{1}_{E}^{\rm G}=1$. At a second stage, we use a semiclassical approximation for the correlation matrix $\hat{R}(E)$ instead of its exact and usually untractable microscopic expression \footnote{Knowledge of $\hat{R}(E)$ provides all information required for the calculation of the microcanonical (mc) expectation value of any MB observable $\hat{O}$, as $\langle \hat{O} \rangle^{(\rm mc)}_{E}={\rm Tr}\left[\hat{R}(E)\hat{O}\right]$. Its exact determination is therefore a formidable task in general}, where both universal and system-specific aspects enter.

%%%%%%%%%%%%%%%%%%%%%%%%%%%%%%%%%%%%%%%%%%%%%%

%%%%%%%%%%%%%%%%%%%%%%%%%%%%%%%%%%%%%%%%%%%%%%

To proceed with the MB semiclassical approximation we consider the Wigner transform  of $\hat{R}(E)$, Eq.~(\ref{eq:R}), \footnote{We do not address the issue of defining the Wigner transform on discrete spaces, as it is irrelevant in the semiclassical regime considered here.}
\begin{equation}
\label{eq:Wigner}
    {\cal W}_{E}({\boldsymbol \rho},{\boldsymbol \theta})=\sum_{{\bf m} < {\boldsymbol \rho}} R_{{\boldsymbol \rho}+{\bf m},{\boldsymbol \rho}-{\bf m}}(E) \, {\rm e}^{i{\boldsymbol \theta}\cdot {\bf m}} \,  ,
\end{equation}
where the notation ${\bf m} < {\boldsymbol \rho}$ means $m_{\alpha}<\rho_{\alpha}$ for all $\alpha$. We invoke now the Fock space version of the eigenstate condensation hypothesis of Voros and Berry~\cite{voros1976semi,berry1977semi} in the form ${\cal W}_{E}\simeq {\cal W}_{E}^{{\rm cl}}$, with classical phase space distribution
\begin{equation}
\label{eq:RW}
{\cal W}_{E}^{{\rm cl}}({\boldsymbol \rho},{\boldsymbol \theta})=\frac{1}{\rho_{\eta}^{{\rm cl}}(E)}W_{\eta}\left(E\!-\!H_{\rm cl}({\boldsymbol \rho},{\boldsymbol \theta})\right) \, .
\end{equation}

The level density $
\rho_{\eta}^{{\rm cl}}(E)=\int W_{\eta}\left(E\!-\!H_{\rm cl}({\boldsymbol \rho},{\boldsymbol \theta})\right) d{\boldsymbol \rho}d{\boldsymbol \theta}$ is
 defined by the volume of the classical energy shell of width $\eta$. Inverting relation~(\ref{eq:Wigner}) to obtain the corresponding semiclassical form of the covariance matrix $R^{\rm sc}$ is in general a formidable task, but thanks to the local character of the Bose-Hubbard type of Hamiltonian $\hat{H}$, Eq.~(\ref{eq:BH}, we get as our main result (see~\cite{suppmat} for details)
\begin{equation}
\label{eq:Rsc}
\hspace{-0.3cm}R^{{\rm sc}}_{{\bf n},{\bf m}}(E)=\frac{1}{\rho_{\eta}^{{\rm cl}}(E)}\sum_{Q\in Z}  \int_{-\infty}^\infty\frac{\ud \tau}{2\pi} \tilde{W}(\tau)\exp{\{{\rm i} \tau [E\!-\! U \sum_{\alpha} I_\alpha(I_\alpha\!-\!1) \!-\! \sum_{\alpha} \varepsilon_\alpha I_\alpha]}\} \prod_{\alpha=1}^L
  e^{{\rm i} \frac{\pi}{2}(\delta_{\alpha}+ Q)} J_{\delta_{\alpha}+ Q}(2t\tau\sqrt{I_\alpha I_{\alpha+1}}) \, .
\end{equation}
Here $\tilde{W}(\tau)$ is the Fourier transform of $W_{\eta}(E)$,
$J_{l}(x)$ the $l$th Bessel function, ${\bf I}= ({\bf n}+{\bf m})/2$ with $I_{L+1}=I_{1}$, and $\delta_\alpha = \sum_{\beta=1}^\alpha (n_\beta - m_\beta)$.
    
Together with Eq.~(\ref{eq:Gauss}) the  expression~(\ref{eq:Rsc}) allows us to compute the average of \textit{any} functional $F$ in Eq.~(\ref{eq:Fexact}) within our MB RWM. It is therefore important to compare this central analytical prediction with corresponding numerical results for the Bose Hubbard model, Eq.~(\ref{eq:BH}). For definiteness we choose $L=5,N=25$ and $UN=2t$, where the classical dynamics is predominantly chaotic. In Figs.~\ref{fig:Corr} and~\ref{fig:cuts} we compare the normalized correlators $R_{{\bf n},{\bf m}}(E)/\sqrt{R_{{\bf n},{\bf n}}(E)R_{{\bf m},{\bf m}}(E)}$ obtained by exact diagonalization and averaged over a spectral window comprising $\sim 10^{3}$ MB states, with the MB semiclassical approximation (\ref{eq:Rsc}). In Fig.~(\ref{fig:cuts}) we show the correlations between reference Fock states  $|\mathbf n\rangle =|6,6,4,5,4\rangle, |5,7,4,4,5\rangle,|6,4,4,6,5\rangle$ and representative states $|\mathbf m\rangle$ indicated along the horizontal axis. In all three cases we find remarkable agreement between the semiclassical prediction (red symbols) and the numerical results (black symbols). Instead of fixing a single reference basis state $|\mathbf n\rangle$,  the color matrix in Fig. \ref{fig:Corr} represents cross correlations among all Fock basis states around $|5,6,4,5,5\rangle$. The upper right and lower left triangle shows the semiclassical and the numerical results, respectively. The apparent symmetry, quantitatively confirmed by the cuts shown in Fig.~(\ref{fig:cuts}), again reflects the excellent agreement, see also \cite{suppmat} for further comparisons. 

Very much as with the SP case, the universality of this result comes from its robustness with respect to changes of the form of the interaction term (kin to the role of external potential $V({\mathbf r})$ in the SP context. In fact, any function of the occupations, being diagonal in the Fock basis, will let $R^{{\rm sc}}_{{\bf n},{\bf m}}(E)$ in Eq.~\ref{eq:Rsc}) unchanged. Also, again, the full RMT result of uncorrelated random vectors emerges only in the limit $E\to \infty$. The universal, coherent mesoscopic fluctuations predicted by the Fock sace version of the RWM are beyond any RMT approach.

%%%%%%%%%%%%%%%%%%%%%%%%%%%%%%%%%%%%%%%%%%%%%%%%%%%%%%%%%%%%%%%%%%%%%%%%%%%%%%%%%%%%

\subsection{Scrambling in bosonic MB systems}
\label{sec:OTOC}

%%%%%%%%%%%%%%%%%%%%%%%%%%%%%%%%%%%%%%%%%%%%%%%%%%%%%%%%%%%%%%%%%%%%%%%%%%%%%%%%%%%%%%

\subsubsection{Concept and heuristic quantum-classical correspondence}
\label{sec:concept}

The introduction of quantum chaos concepts in the MB context demanded a semiclassical understanding of mechanisms genuinely beyond the SP picture with MB interference beyond the Ehrenfest time being a prime example. In this section we go one step further and show how genuinely MB interference affects the scrambling of correlations, a fundamental MB mechanism that has emerged as a signature of complex dynamics in several fields. 

MB scrambling is quantified through so-called out-of-time-order correlators (OTOCs)~\cite{Larkin69,Shenker14,Maldacena16} 
\begin{equation}
  C(t) = 
\langle \Psi | 
  \left[ \hat{W}(t), \hat{V}(0)  \right]^\dagger
  \left[ \hat{W}(t),\hat{V}(0)  \right] | \Psi \rangle \, 
  \label{eq:OTOComm_definition}
\end{equation}
comprising two forward and two backward propagations by means of the Heisenberg operator $\hat{W}(t) = \exp{(-i\hat{H}t/\hbar)} \hat{W}(0)  \exp{(i\hat{H}t/\hbar)}$. Initial local (quantum) information is spread and scrambled across the huge Hilbert space of the interacting MB system with its vast number of degrees of freedom~\cite{Sekino08}, making OTOCs the measures of choice for quantifying growing complexity and instability of MB systems, thereby with relevance for quantum computing \cite{Mi21} and several measurement protocols available~\cite{Zhu16,Swingle16,Li16,Garttner16,Dominguez21}. See \cite{Xu22}) for a recent tutorial.

For the SP case invoking a heuristic classical-to-quantum correspondence for small $\hbar$ by replacing the commutator (\ref{eq:OTOComm_definition})
for pre-Ehrenfest times by Poisson brackets $\{\cdot,\cdot\}$ generates a leading-order Moyal expansion, e.g.~$\hat{W}=\hat{p}_i$ and $\hat{V}=\hat{q}_j$
\cite{Larkin69,Maldacena16},
\begin{equation}
 C(t) \longrightarrow
  |{\rm i} \hbar|^2 \left\{p_i,q_j(t) \right\}^{\!2}
  \! \simeq \! \hbar^2 \! 
  \left(\frac{\partial q_j(t)}{\partial q_i}\!
  \right)^{\!2} \propto
   \hbar^{2} {\rm e}^{2\lambda t} \, .
  \label{eq:OTOC_Moyal}
\end{equation}
The leading off-diagonal monodromy matrix element $ \partial q_j(t) / \partial q_i$
is replaced by an exponential growth determined by the classical single-particle Lyapunov exponent $\lambda_{SP}$.  This close quantum-to-classical correspondence for quantum chaotic single-particle dynamics is intriguing, and thus there is also the quest for establishing a corresponding MB quantum-to-classical correspondence, {\em i.e.}, a MB version of such a quantum butterfly effect. This, in particular, includes an interpretation of the quantum mechanical growth rate of OTOCs for MB systems with a classical limit.  Most notably their growth is bounded in time and OTOCs saturate due to a MB interference mechanism setting in at the Ehrenfest time, i.e.~scrambling time. It gives rise to an interference term that is of the same order as the corresponding diagonal contribution \cite{Rammensee2018}.  Such distinct features at $\tE$ render OTOC evolution a hallmark of Ehrenfest phenomena.
%

%%%%%%%%%%%%%%%%%%%%%%%%%%%%%%%%%%%%%%%%%%%%%%%%%%%%%%%%%%%%%%%%%%%%%%%%%%%%%%%%%%%%%%

\subsubsection{Pre- and post-Ehrenfest times: exponential growth and saturation}
\label{sec:prepost}

Considering  again Bose-Hubbard-like systems describing $N$ interacting bosons the OTOC, Eq.~(\ref{eq:OTOComm_definition}), defined by position and momentum quadratures reads
\begin{equation}
  C(t)=\langle \Psi \left|
  \left[ \hat{p}_i,\hat{U}^\dagger(t)\hat{q}_j\hat{U}(t)\right] 
  \left[ \hat{U}^\dagger(t)\hat{q}_j\hat{U}(t),\hat{p}_i\right] 
  \right|\Psi \rangle
  \label{eq:pq_OTOC_w_Ut}
\end{equation}
in terms of the MB time evolution operator
$\hat{U}(t)=\exp(-{\rm i} \hat{H} t / \hbar)$.
In Eq.~(\ref{eq:pq_OTOC_w_Ut}) and in the following, we consider an initial coherent state $\Psi$ localized in both quadratures. This is an important point as another possible choice, namely an initial thermal state fixing the temperature instead of the average energy, leads to substantially different manifestation of scrambling. 

The semiclassical derivation is based on approximating $\hat{U}(t)$ by its asymptotic form for
small $\heff$, the MB version~\cite{Engl14b,Engl15}, Eq.~(\ref {eq:MB-propagator}), of the van Vleck-Gutzwiller propagator.  The corresponding sum runs over all mean-field solutions $\gamma$ of the
equations of motion
${\rm i} \hbar \partial \Phi/\partial t = \partial H_{\mathrm{cl}}/\partial \Phi^\ast$
of the classical Hamilton function~(\ref{eq:Tom2})
that denotes the mean-field limit of $\hat{H}$ 
for $\heff=1/N\rightarrow 0$:
 \begin{equation}
 \label{eq:BHham}
        H_{\mathrm{cl}}
        \left(\vec{q},\vec{p}\right)
        = \sum_{i, j}h_{ij}  \Phi_i^*\Phi_j
        +\sum_{i, j, i', j'}V_{i j i' j'}
        \Phi_i^*\Phi_{i'} \Phi_j^*\Phi_{j'} \, .
       \end{equation}
In the coherent sum over mean-field paths in Eq.~(\ref{eq:semquad}) the phases are given by classical actions
\begin{equation}
    R_\gamma(\vecfin{q},\vecinit{q};t) \! = \!
  \int_0^t {\rm d} t
    [
       \vec{p}_\gamma(t)\cdot\vec{q}_\gamma(t)
      -H^{\mathrm{cl}}
      \left(\vec{q}_\gamma(t),\vec{p}_\gamma(t)\right)/\hbar
    ]
\end{equation}
along $\gamma$, and the weights $A_\gamma$ reflect their classical (in)stability. Finally, in order to make a connection to RMT-type universality, assume that the mean-field limit exhibits uniformly hyperbolic, chaotic dynamics with the same Lyapunov exponent $\lambda$ at any phase space point.
Evaluating Eq.~(\ref{eq:pq_OTOC_w_Ut}) in position quadrature and using Eq.~(\ref{eq:semquad}) for the propagator $K$ gives the starting point of the semiclassical MB representation of the OTOC. Following standard methods of asymptotic analysis, to leading order in $\hbar_{\rm eff}$, the derivatives $\hat{p}_i = -{\rm i} \heff \partial /  \partial q_i$ only act on the phases $R_\gamma$ in $K$. The classical identities 
$
 \vecinit{p}_\gamma = -
  \partial R_{\gamma}/\partial \vecinit{q}
$
finally yield for the OTOC~\cite{Rammensee2018}, Eq.~(\ref{eq:pq_OTOC_w_Ut}), 
\begin{eqnarray}
    C(t) \simeq 
    \!\! \int\! {\rm d}^n q_1'\! \int \! {\rm d}^n q_2\!
    \int \! {\rm d}^n q_3'\! \int \! {\rm d}^n q_4\!
    \int {\rm d}^n q_5'
    \Psi^{*}\!\left(\vec{q}_1'\right)\Psi\!\left(\vec{q}_5'\right)
     &\times&\sum_{\alpha': \vec{q}_1' {\rightarrow}\vec{q}_2}
    \sum_{\alpha : \vec{q}_3' {\rightarrow}\vec{q}_2}
    A_{\alpha'}^*  A_{\alpha}
    {\rm e}^{({\rm i}/\heff)\left(\!-R_{\alpha'}+R_{\alpha}\right)}
    \left(\init{p}_{\alpha',i}\!-\!\init{p}_{\alpha,i} \right)\fin{q}_{2,j} \nonumber \\
     &\times&\sum_{\beta': \vec{q}_3' {\rightarrow}\vec{q}_4}
   \sum_{\beta : \vec{q}_5' {\rightarrow}\vec{q}_4}
    A_{\beta'}^*  A_{\beta}
    {\rm e}^{({\rm i}/\heff) \left(\!-R_{\beta'}+R_{\beta}\right)} \!
    \left(\init{p}_{\beta,i}\!-\!\init{p}_{\beta',i} \right)\fin{q}_{4,j} \, .
  \label{eq:OTOC_sc_integral_representation}
\end{eqnarray}

\begin{figure}
  \begin{center} 
  \begin{tabular}{ccc}
  \includegraphics[width=0.5\linewidth]{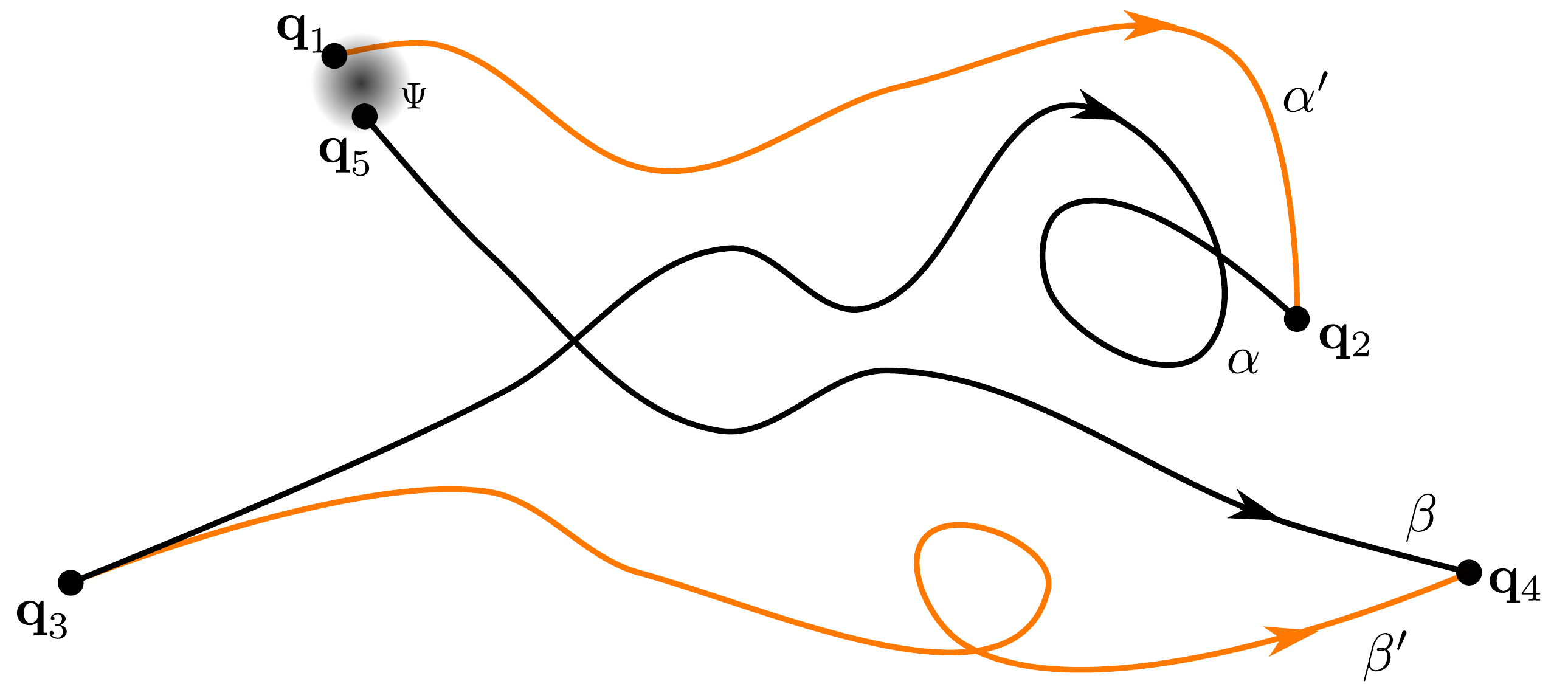}
      & &
  \includegraphics[width=0.4\linewidth]{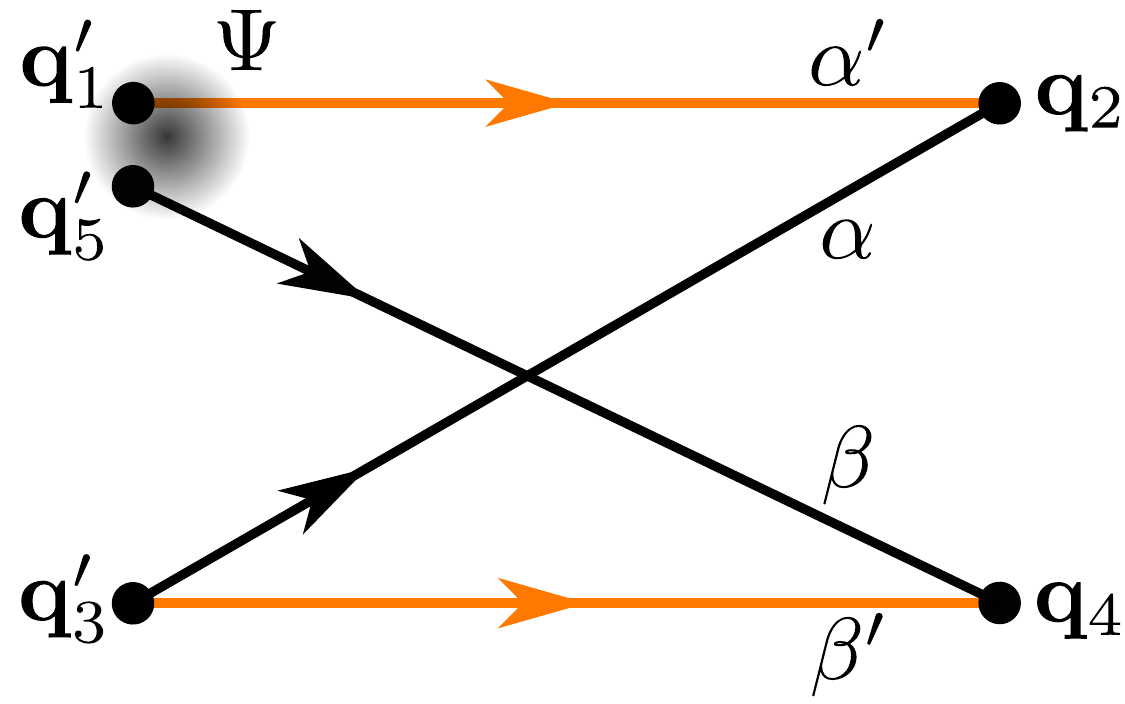} 
     \\ (a)&&(b) \\
 \includegraphics[width=0.4\linewidth]{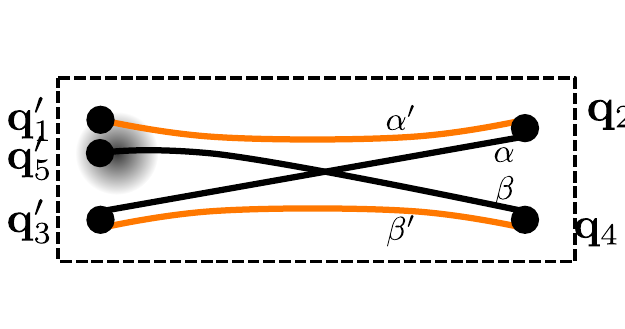}
  & & 
\includegraphics[width=0.4\linewidth]{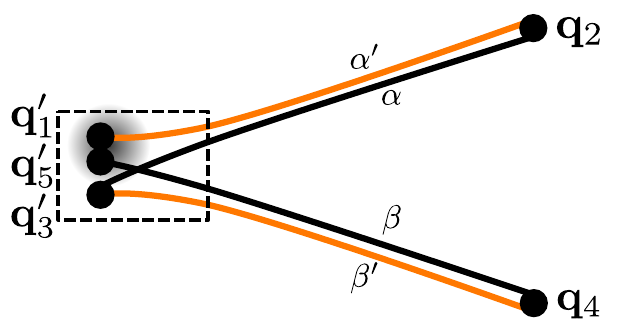}
\\   (c)&&(d) \\
\includegraphics[width=0.4\linewidth]{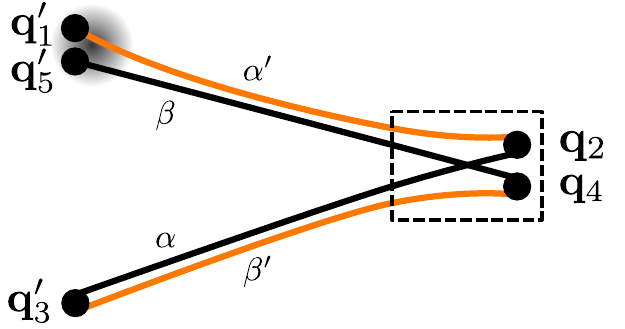}
   &  &
 \includegraphics[width=0.4\linewidth]{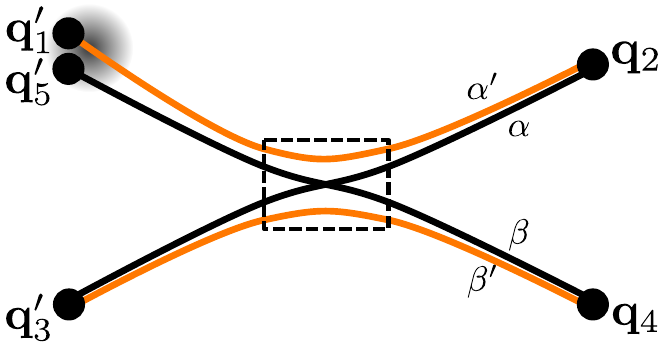}
      \\    (e)&&(f)
     \end{tabular}
  \end{center}
      \caption{
      {\bf Configurations of interfering mean-field paths that contribute to the OTOC $C(t)$,} Eq.~(\ref{eq:OTOC_sc_integral_representation}).
      (a) arbitrary trajectory quadruple and (b) corresponding general diagram denoting
       forward and backward propagation along black and orange mean-field paths. (c)-(f): Relevant configurations contributing predominantly to $C(t)$:
     The trajectory quadruples reside (c) inside an encounter (marked by dashed box), form a "two-leg"-diagram
     with an encounter (d) at the beginning or (e) at the end, or (f) build a "four-leg"-diagram with the encounter
     in between. (From Ref.~\cite{Rammensee2018}.)}
     \label{fig:OTOC_diagrams}
\end{figure}

The geometric connections among the trajectories quadruples involved are sketched in Fig.~\ref{fig:OTOC_diagrams}.
Panel (a) shows an arbitrary orbit quadruple and (b) the corresponding diagram. Black and orange arrows refer to contributions to $K$ and $K^\ast$, respectively,  i.e.~forward and backward propagation in time. The grey shaded spots mimic the initial state $|\Psi\rangle$.

In the semiclassical limit $R_\gamma(\vecfin{q},\vecinit{q};t) \! \gg \! \heff$. Hence, the corresponding phase factors in Eq.~(\ref{eq:OTOC_sc_integral_representation})
are highly oscillatory with respect to initial and final positions.  Thus, contributions from arbitrary trajectory quadruples are usually suppressed whereas
correlated trajectory quadruples with action differences such that
$R_\alpha\!-\!R_{\alpha'}\!+\!R_{\beta}\!-\!R_{\beta'} \simeq O(\heff)$ are not averaged out and contribute dominantly to $C(t)$.  For post-Ehrenfest times these are quadruples where for most of the propagation the four paths are pairwise nearly identical except for 
encounter regions where trajectory pairs approach each other, evolve in a 
correlated manner, and exchange their partners. The encounter calculus applies in the high-dimensional phase space associated with the MB Fock space.

To leading order in $\heff$, the relevant quadruples for OTOCs involve a single encounter.  These can be subdivided
into four classes depicted in Fig.~\ref{fig:OTOC_diagrams}(c)-(f).
Diagram (c) represents a bundle of four trajectories staying in close vicinity to each other throughout time $t$, i.e.~forming an encounter marked by the dashed box. This diagram turns out to be dominant for times $t<\tE$, Eq.~(\ref{eq:scrambling}), the time scale for traversing an encounter region, if the associated action differences are of order $\heff$.  Due to exponential divergences in chaotic phase space the dynamics merges beyond the encounter boundary into uncorrelated time evolution of the individual trajectories.  However, the symplectic Hamiltonian structure implies that the exponential separation along unstable phase space manifolds is complemented by motion near stable manifolds.  This enables the formation of pairs of exponentially close trajectories~\cite{Sieber01}, e.g.~paths $\alpha'$ and $\alpha$ or $\beta$ and $\beta'$ in Figs.~\ref{fig:OTOC_diagrams}(d,f), a mechanism that becomes quantum mechanically relevant for times beyond the MB Ehrenfest time $t_{\rm E}$, and here it is crucial for understanding post-Ehrenfest OTOC saturation.
Panels (c, d) display diagrams with an encounter at the beginning or end of two such trajectory pairs.
The diagrams in (f) are characterized by uncorrelated motion of four trajectory pairs
before and after the encounter.

The evaluation of Eq.~(\ref{eq:OTOC_sc_integral_representation}) requires a thorough consideration of the
dynamics in and outside the encounter regions.
Inside an encounter, Fig.~\ref{fig:OTOC_diagrams}(c),
the hyperbolic dynamics essentially follows a common mean-field solution: linearization in the vicinity of one reference path allows for expressing contributions from the remaining three trajectories.
The detailed evaluation of the diagrams (d-f) in Fig.~\ref{fig:OTOC_diagrams} is given in Ref.~\cite{Rammensee2018}.
It involves the calculation of corresponding encounter integrals based on phase space averages invoking ergodicity.
Diagrams similar to class (f) 
have been earlier considered in the context of shot noise ~\cite{Lassl03,Schanz03,Braun06}
and observables based on quantum chaotic single-particle~\cite{Kuipers11,Kuipers13} and MB \cite{Urbina16} scattering.
However, the evaluation of such encounter integrals for OTOCs requires a generalization to high-dimensional MB phase spaces. The occurrence of operators (positions and momenta) in the case of OTOCs demand a generalization of the encounter calculus and special treatment,
 depending on whether the initial or final position quadratures are inside an encounter.

Using the amplitudes $A_\gamma$ in
 Eq.~(\ref{eq:OTOC_sc_integral_representation})
to convert integrals over final positions into initial momenta, the OTOC contribution from each diagram is conveniently represented as an ergodic phase-space average
\begin{equation}
%  C^{\textrm{generic}}(t) =
%  C^{\textrm{(a-d)}}(t) =
  C(t) \simeq
  \int {\rm d}^n q' \int {\rm d}^n p' W(\vec{q}',\vec{p'})
%  h(\vec{q}',\vec{p}',t)
%  h^{\textrm{(a-d)}}(\vec{q}',\vec{p}',t)
 I(\vec{q}',\vec{p}';t) \, .
  \label{eq:PS_average}
\end{equation}
Here,
\begin{equation}
W(\vec{q}',\vec{p'}) \!=\! \int {\rm d}^n y /  (2\pi)^n
  \Psi^*\!\left(\vec{q}'\!+\! \vec{y}/2 \right)
  \Psi\left(\vec{q}'\!-\!\vec{y}/2 \right)
  \exp[({\rm i})\vec{y}\vec{p}']
\end{equation} 
is the Wigner function~\cite{OzorioBook} and $I(\vec{q}',\vec{p}',t)$ comprises all encounter integrals.  The detailed evaluation of the encounter integrals $I$ represented by the different diagrams in Fig.~\ref{fig:OTOC_diagrams}, and
thereby $C(t)$, yields the following results for pre- and post-Ehrenfest time evolution~\cite{Rammensee2018}:
From diagram (c) it follows for $\lambda^{-1} < t < \tE$
upon ergodic averaging in the semiclassical limit
\begin{equation}
    I(\vec{q}',\vec{p'};t) \simeq F_<(t)  \quad ; \quad 
    F_<(t) \approx e^{2\lambda(t - \tE)} \Theta(\tE-t)
    = \heff^2 e^{2\lambda t} \Theta(\tE-t) \, .
\label{eq:F<}
\end{equation}
Diagram (d) turns out to be negligible, diagrams (e, f) together yield for $t > \tE$
\begin{equation}
    I (\vec{q}',\vec{p'};t) \simeq  F_>(t) \langle
      \left(p_i'-p_i\right)^2 \rangle ( \langle q_j^2\rangle - \langle q_j^2 \rangle )  \quad ; \quad F_>(t) = \Theta(t-\tE) \, .
\label{eq:F>}
\end{equation}
Here,
\begin{equation}
    \langle f \rangle =  \frac{1}{\Sigma(E)} \int{\rm d}^n q \int {\rm d}^n p f(\vec{q},\vec{p}) \delta\!\left(E \!-\!
      \mathcal{H}^{\rm cl}\left(\vec{p},\vec{q}\right)
    \right)
\end{equation}
is the ergodic average with $\Sigma(E)$ the phase space volume of the energy shell at energy $E$.

%%%%%%%%%%%%%%%%%%%%%%%%%%%%%%%%%%%%%%%%%%%%%%%%%%%%%%%%%%%%%%%%%%

\begin{figure}
\centering{
  \includegraphics[width=0.8\linewidth]{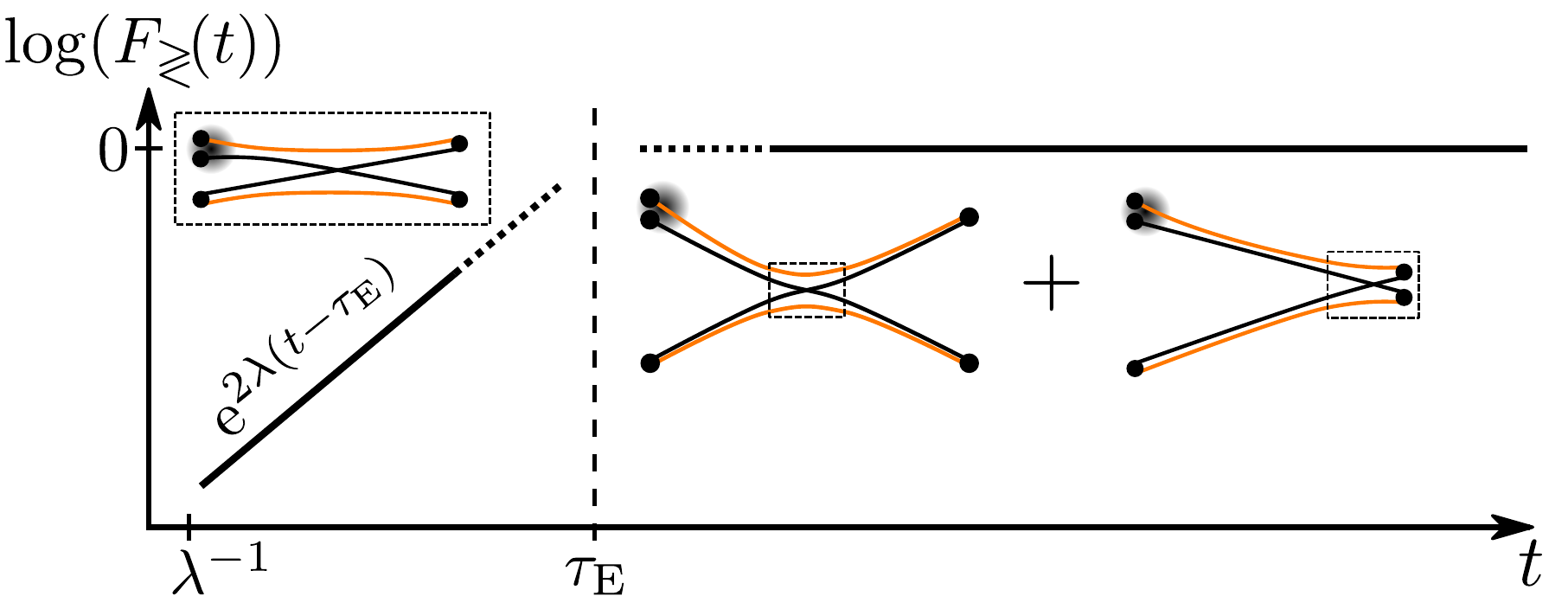}
  }
    \caption{
    \label{fig:OTOCscheme}
 {\bf  Universal contribution to the time evolution of out-of-time-order commutators} | 
Exponential increase according to $F_<(t)$,
Eq.~(\ref{eq:F<}), before and according to $F_>(t)$, Eq.~(\ref{eq:F>}), after the 
 Ehrenfest time $\tE = (1/\lambda) \log N$ marked by the vertical dashed line.
Insets depict diagrams (c), (f) and (e)
from Fig.~\ref{fig:OTOC_diagrams} representing interfering mean-field trajectories.
  (From Ref.~\cite{Rammensee2018}.)
  }
\end{figure}
%%%%%%%%%%%%%%%%%%%%%%%%%%%%%%%%%%%%%%%%%%%%%%%%%%%%%%%%%%%%%%%%%%%

The time-dependences of the universal functions $F_<$ and $F_>$ are sketched in Fig.~\ref{fig:OTOCscheme}.
For $t<\tE$ the  semiclassical evaluation for MB systems confirms the heuristic result, Eq.~(\ref{eq:OTOC_Moyal}). The careful treatment of the encounter dynamics, diagram (c), provides a natural  cut-off (exponential suppression) at $\tE$, absent in Eq.~(\ref{eq:OTOC_Moyal}). It results from the mechanism that the initial phase space area enabling four trajectories to stay close to each other is exponentially shrinking for $t > \tE$. 
The fact that for $t<\tE$ all four mean-field solutions essentially follow in the linearizable vicinity of a common one, see diagramm (c),  indicates that the initial exponential increase of an OTOC of a chaotic MB system can be considered as a property of unstable mean-field dynamics that would also be captured by a truncated Wigner approach. 

%%%%%%%%%%%%%%%%%%%%%%%%%%%%%%%%%%%%%%%%%%%%%%%%%%%%%%%%%%%%%%%%%%

\begin{figure}
\centering{
  \includegraphics[width=0.8\linewidth]{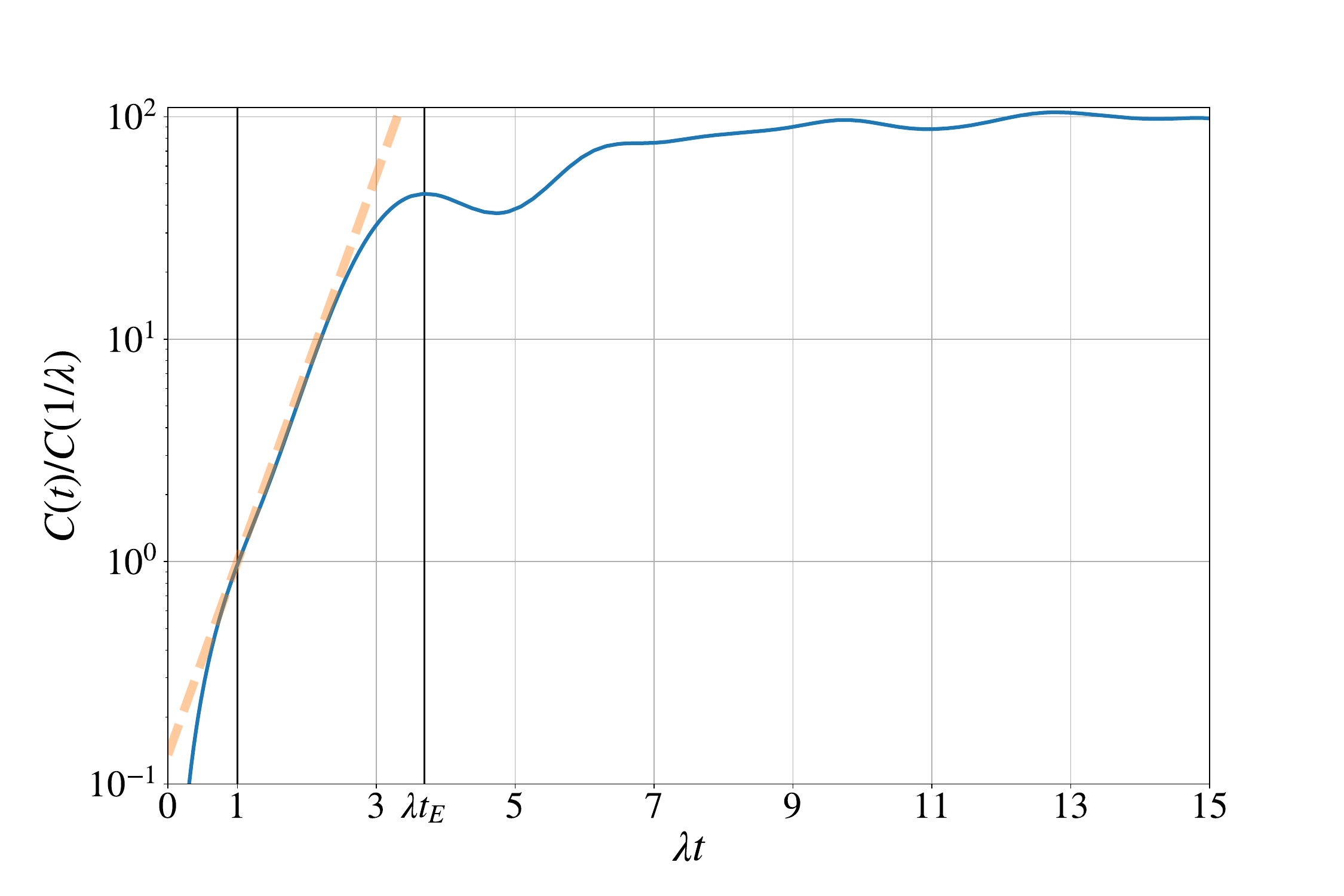}
  }
    \caption{
    \label{fig:OTOCBH}
 {\bf Out-of-time-order commutator of a Bose-Hubbard (BH) system} | Numerically exact calculation of Eq.~(\ref{eq:OTOComm_definition}) for a BH system with four sites and $N=40$ particles. The system is initially described by a coherent state localized near a hyperbolic fixed point of the classical mean-field dynamics. For the choice of parameters $J/NU\simeq \pi/2$ in Eq.~(\ref{eq:BHham}) with hopping $h_{i,j}=J(\delta_{i,j+1}+\delta_{i+1,j})$ and local interactions $V_{i,j,i',j'}=U\delta_{i,j}\delta_{i',j'}\delta_{i,j}$, the corresponding stability exponent is given by $(J/\hbar)\lambda$. For times $1<\lambda t < \log(N)$ (time in units of the typical hopping time between sites $\sim \hbar/J$) a clear exponential growth due to this local hyperbolicity can be observed. 
 (courtesy of Mathias~Steinhuber). }
\end{figure}
%%%%%%%%%%%%%%%%%%%%%%%%%%%%%%%%%%%%%%%%%%%%%%%%%%%%%%%%%%%%%%%%%%%

On the contrary, the term $F_>(t)$ in Eq,~(\ref{eq:F>}) is suppressed for $t\!<\!\tE$, but is indeed responsible for OTOC saturation.  After the scrambling time $t>\tE$ genuine MB interference sets in captured by encounter diagrams such as in panel (f). This diagram represents successive forward and backward dynamics swapping back and  forth along different encounter-coupled mean-field trajectories. This involves correlated
quantum MB dynamics and the temporal build up of entanglement between mean-field modes.
This mechanism is evidently in a regime where mean-field approaches fail~\cite{Han16}. 
Thus, genuine MB interference
is the quantum mechanism behind the commonly observed saturation of OTOCs at the scrambling or Ehrenfest time.

Note that the expression, Eq.~(\ref{eq:F>}), for the OTOC 
 contains variances of classical quantities, 
 e.g.~the variance of the $j$-th final position quadrature that determine the OTOC saturation level.
Here different types of classical MB dynamics at post-scrambling times, e.g.~diffusive versus chaotic evolution, may lead to a different time-evolution of these classical variances.
As shown in \cite{Rammensee2018}, diffusive dynamics implies a linear increase with time, whereas a calculation assuming ergodic dynamics yields $C(t) \approx 2/L^2$ for $t \gg \tE$ with $L$ the number of sites of a Bose-Hubbard system (for $|\Psi\rangle$ being either a localized wave packet or extended chaotic MB state) corresponding to the flat plateau in Fig.~\ref{fig:OTOCscheme}.

Figure~\ref{fig:OTOCBH} shows the OTOC Eq.~(\ref{eq:OTOComm_definition}) with $\hat{V}= \hat{n}_0$ and $\hat{W}=\hat{n}_1$ denoting occupation operators for adjacent sites
obtained quantum mechanically for a 4-site BH system with $N=40$ particles. These numerics confirm the semiclassical predictions.  Up to $\tE$ the OTOC increases exponentially with slope $2\lambda$ where $\lambda$ agrees with the Lyapunov exponent of the (here locally) unstable MB mean-field dynamics of the specfifc BH system. At $t \simeq \tE$ saturation sets in.

The present semiclassical analysis of MB OTOCs in the large-$N$ limit, the vertical limit in Fig.~\ref{fig:sc-limits}, can be readily generalized to systems of $N$ particles in $d$ spatial dimensions in the complementary limit of small $\hbar$, in particular to
the quantum chaotic single-particle case.
Invoking the corresponding Gutzwiller propagator, Eq.~(\ref{eq:vVG}), in $n=d\times N$ dimensions the exponential increase of the OTOC $C_N(t)$ in Eq.~(\ref{eq:F<}) is then governed by the leading Lyapunov exponent $\lambda_N$ of the corresponding classical $N$-particle system.  Saturation sets in at the corresponding Ehrenfest time $(1/\lambda_N) \log S/\hbar$ with $S$ a typical classical action.  For a chaotic phase space average the usual semiclassical limit the saturation value  $C(t) \approx \hbar^2 N/d$ results~\cite{Rammensee2018}.  For $N=1$ this short-time growth with $\lambda_1$ has also been independently semiclassically derived in \cite{Kurchan18,Jalabert18}.
The exponential increase and saturation of such single-particle OTOCs was considered in detail in numerical case studies of the kicked rotor~\cite{Rozenbaum17} and quantum maps~\cite{Garcia-Mata18}.
%%%%%%%%%%%%%%%%%%%%%%%%%
\section{Outlook}
\label{sec:persp}

During the past two decades considerable progress has been made and various breakthroughs have been achieved in laying the foundations of a semiclassical theory that successfully provides the understanding of quantum chaotic  (universal spectral) features of MB quantum systems exhibiting chaotic classical limits. 
The now existing theoretical framework, reviewed here, opens various interesting perspectives and challenges. 
These include, among many others, the semiclassical treatment of fermionic MB systems, the idea to control chaotic MB quantum systems using chaos as a resource and the question whether one can use these techniques to consider models of quantum gravity that are know to be "chaotic" through the lense of quantum chaos. 
%%%%%%%%%%%%%%%%%%%%%%%%%%%%%%%%%

\subsection{Acknowledgements}

A larger part of the results reviewed here arose in collaborations with many colleagues. Hence we particularly thank our (former) PhD students in Regensburg, Th. Engl, J. Rammensee, F. Sch\"oppl and M. Steinhuber.
We further thank our colleagues R. Dubertrand, P. Schlagheck, D. Ullmo and S. Tomsovic as coworkers and knowlegable discussion partners on topics of many-body semiclassical methods.

We acknowledge financial support from the Deutsche Forschungsgemeinschaft (German Research Foundation) through Projects Ri681/14-1, through Ri681/15-1 within the Reinhart-Koselleck Programme, as well as funding through Vielberth Foundation in Regensburg.

%%%%%%%%%%%%%%%%%%%%%%%%%%%%%%

\bibliography{quantumchaos,general_ref,manybody,molecular,refs,refsRW}

\providecommand{\href}[2]{#2}\begingroup\raggedright\begin{thebibliography}{100}

\bibitem{Bohr36}
N.~Bohr, \emph{Neutron capture and nuclear constitution}, {\emph{Nature} {\bfseries 137} (1936) 344}.

\bibitem{Wigner55}
E.P.~Wigner, \emph{Characteristic vectors of bordered matrices with infinite dimensions}, {\emph{Ann.~of Math.} {\bfseries 62} (1955) 548}.

\bibitem{Wigner58}
E.P.~Wigner, \emph{On the distribution of the roots of certain symmetric matrices}, {\emph{Ann.~of Math.} {\bfseries 67} (1958) 325}.

\bibitem{Haake06}
F.~Haake, \emph{Quantum Signatures of Chaos}, Springer-Verlag, Berlin, Heidelberg (2006).

\bibitem{Bohigas88}
O.~Bohigas and H.A.~Weidenmueller, \emph{Aspects of chaos in nuclear physics}, {\emph{Ann.~Rev.~Nucl.~Part.~Sci.} {\bfseries 38} (1988) 421}.

\bibitem{Guhr98}
T.~Guhr, A.~M{\"u}ller-Groeling and H.A.~Weidenm{\"u}ller, \emph{Random-matrix theories in quantum physics: common concepts}, {\emph{Phys.~Rep.} {\bfseries 299} (1998) 189}.

\bibitem{RevModPhys.69.731}
C.W.J.~Beenakker, \emph{Random-matrix theory of quantum transport}, {\emph{Review of Modern Physics} {\bfseries 69} (1997) 731}.

\bibitem{StockmannBook}
H.-J.~St\"ockmann, \emph{Quantum Chaos: An Introduction}, Cambridge University Press, Cambridge (1999).

\bibitem{Mehta04}
M.L.~Mehta, \emph{Random Matrices (Third Edition)}, Elsevier, Amsterdam (2004).

\bibitem{Gutzwiller70}
M.C.~Gutzwiller, \emph{Energy spectrum according to classical mechanics}, {\emph{Journal of Mathematical Physics} {\bfseries 11} (1970) 1791}.

\bibitem{Gutzwiller71}
M.C.~Gutzwiller, \emph{Periodic orbits and classical quantization conditions}, {\emph{J.~Math.~Phys.} {\bfseries 12} (1971) 343}.

\bibitem{Gutzwiller71a}
M.C.~Gutzwiller, \emph{Periodic orbits and classical quantization conditions}, {\emph{Journal of Mathematical Physics} {\bfseries 12} (1971) 343}.

\bibitem{Gutzwiller90}
M.C.~Gutzwiller, \emph{Chaos in Classical and Quantum Mechanics}, Springer-Verlag, New York (1990).

\bibitem{BraBha03Book}
M.~Brack and R.~Bhaduri, \emph{Semiclassical Physics}, Frontiers in physics, Westview (2003).

\bibitem{Ezra91}
G.S.~Ezra, K.~Richter, G.~Tanner and D.~Wintgen, \emph{Semiclassical cycle expansion for the helium atom}, \href{https://doi.org/10.1088/0953-4075/24/17/001}{\emph{Journal of Physics B: Atomic, Molecular and Optical Physics} {\bfseries 24} (1991) L413–L420}.

\bibitem{Wintgen87}
D.~Wintgen, \emph{Connection between long-range correlations in quantum spectra and classical periodic orbits}, \href{https://doi.org/10.1103/PhysRevLett.58.1589}{\emph{Phys. Rev. Lett.} {\bfseries 58} (1987) 1589}.

\bibitem{Klaus1}
M.~Sieber and K.~Richter, \emph{Correlations between periodic orbits and their r{\^o}le in spectral statistics}, {\emph{Physica Scripta} {\bfseries T90} (2001) 128}.

\bibitem{Berry76}
M.V.~Berry and M.~Tabor, \emph{Closed orbits and the regular bound spectrum}, {\emph{Proc.~R.~Soc.~Lond.~A} {\bfseries 349} (1976) 101}.

\bibitem{Reichl21}
L.~Reichl, \emph{The Transition to Chaos: Conservative Classical and Quantum Systems}, Springer (2021).

\bibitem{Bohr13a}
N.~Bohr, \emph{I.~on the constitution of atoms and molecules}, {\emph{Phil.~Mag.} {\bfseries 26} (1913) 1}.

\bibitem{KlausH1}
G.~Tanner, K.~Richter and J.-M.~Rost, \emph{The theory of two-electron atoms: between ground state and complete fragmentation}, \href{https://doi.org/10.1103/RevModPhys.72.497}{\emph{Review of Modern Physics} {\bfseries 72} (2000) 497}.

\bibitem{Kragh12}
H.~Kragh, \emph{Niels Bohr and the Quantum Atom}, Oxford University Press (2012).

\bibitem{Primack98}
H.~Primack and U.~Smilansky, \emph{On the accuracy of the semiclassical trace formula}, \href{https://doi.org/10.1088/0305-4470/31/29/016}{\emph{Journal of Physics A: Mathematical and General} {\bfseries 31} (1998) 6253}.

\bibitem{Richter14}
M.~Richter, S.~Lange, A.~B\"acker and R.~Ketzmerick, \emph{Visualization and comparison of classical structures and quantum states of four-dimensional maps}, {\emph{Phys.~Rev.~E} {\bfseries 89} (2014) 022902}.

\bibitem{Akila17}
M.~Akila, D.~Waltner, B.~Gutkin, P.~Braun and T.~Guhr, \emph{Semiclassical identification of periodic orbits in a quantum many-body system}, \href{https://doi.org/10.1103/PhysRevLett.118.164101}{\emph{Phys. Rev. Lett.} {\bfseries 118} (2017) 164101}.

\bibitem{Waltner17}
D.~Waltner, P.~Braun, M.~Akila and T.~Guhr, \emph{Trace formula for interacting spins}, \href{https://doi.org/10.1088/1751-8121/aa5533}{\emph{Journal of Physics A: Mathematical and Theoretical} {\bfseries 50} (2017) 085304}.

\bibitem{Akila16}
M.~Akila, D.~Waltner, B.~Gutkin and T.~Guhr, \emph{Particle-time duality in the kicked ising spin chain}, \href{https://doi.org/10.1088/1751-8113/49/37/375101}{\emph{Journal of Physics A: Mathematical and Theoretical} {\bfseries 49} (2016) 375101}.

\bibitem{Weidenmueller1993}
H.A.~Weidenm\"uller, \emph{Semiclassical periodic-orbit theory for identical particles}, \href{https://doi.org/10.1103/PhysRevA.48.1819}{\emph{Phys. Rev. A} {\bfseries 48} (1993) 1819}.

\bibitem{Ullmo_2008}
D.~Ullmo, \emph{Many-body physics and quantum chaos}, \href{https://doi.org/10.1088/0034-4885/71/2/026001}{\emph{Reports on Progress in Physics} {\bfseries 71} (2008) 026001}.

\bibitem{Ullmo98}
D.~Ullmo, H.U.~Baranger, K.~Richter, F.~von Oppen and R.A.~Jalabert, \emph{Chaos and interacting electrons in ballistic quantum dots}, \href{https://doi.org/10.1103/PhysRevLett.80.895}{\emph{Phys. Rev. Lett.} {\bfseries 80} (1998) 895}.

\bibitem{Urbina16}
J.-D.~Urbina, J.~Kuipers, S.~Matsumoto, Q.~Hummel and K.~Richter, \emph{Multiparticle correlations in mesoscopic scattering: Boson sampling, birthday paradox, and hong-ou-mandel profiles}, \href{https://doi.org/10.1103/PhysRevLett.116.100401}{\emph{Phys. Rev. Lett.} {\bfseries 116} (2016) 100401}.

\bibitem{Richter02}
K.~Richter and M.~Sieber, \emph{Semiclassical theory of chaotic quantum transport}, \href{https://doi.org/10.1103/PhysRevLett.89.206801}{\emph{Phys.~Rev.~Lett.} {\bfseries 89} (2002) 206801}.

\bibitem{Mueller09}
S.~Müller, S.~Heusler, A.~Altland, P.~Braun and F.~Haake, \emph{Periodic-orbit theory of universal level correlations in quantum chaos}, \href{https://doi.org/10.1088/1367-2630/11/10/103025}{\emph{New Journal of Physics} {\bfseries 11} (2009) 103025}.

\bibitem{Berkolaiko12}
G.~Berkolaiko and J.~Kuipers, \emph{Universality in chaotic quantum transport: The concordance between random-matrix and semiclassical theories}, \href{https://doi.org/10.1103/PhysRevE.85.045201}{\emph{Phys. Rev. E} {\bfseries 85} (2012) 045201}.

\bibitem{Aaronson10}
S.~Aaronson and A.~Arkhipov, \emph{The computational complexity of linear optics},  2010.

\bibitem{Hong87}
C.K.~Hong, Z.~Ou and L.~Mandel, \emph{Measurement of subpicosecond time intervals between two photons by interference}, {\emph{Phys.~Rev.~Lett.} {\bfseries 59} (1987) 2044}.

\bibitem{Gutkin16}
B.~Gutkin and V.~Osipov, \emph{Classical foundations of many-particle quantum chaos}, \href{https://doi.org/10.1088/0951-7715/29/2/325}{\emph{Nonlinearity} {\bfseries 29} (2016) 325}.

\bibitem{Gutkin21}
B.~Gutkin, P.~Cvitanovi\'c, R.~Jafari, A.K.~Saremi and L.~Han, \emph{{Linear encoding of the spatiotemporal cat}}, \href{https://doi.org/10.1088/1361-6544/abd7c8}{\emph{Nonlinearity} {\bfseries 34} (2021) 2800} [\href{https://arxiv.org/abs/1912.02940}{{\ttfamily 1912.02940}}].

\bibitem{Laksh11}
A.~Lakshminarayan and S.~Tomsovic, \emph{Kolmogorov-sinai entropy of many-body hamiltonian systems}, \href{https://doi.org/10.1103/PhysRevE.84.016218}{\emph{Phys. Rev. E} {\bfseries 84} (2011) 016218}.

\bibitem{Gutb}
M.C.~Gutzwiller, \emph{Chaos in classical and quantum mechanics}, vol.~1, Springer Science \& Business Media (2013).

\bibitem{Vanvleck28}
J.H.V.~Vleck, \emph{The correspondence principle in the statistical interpretation of quantum mechanics}, {\emph{Proc.~Natl.~Acad.~Sci.~U.S.A.} {\bfseries 14} (1928) 178}.

\bibitem{Engl14}
T.~Engl, P.~Pl\"o$\beta$l, J.D.~Urbina and K.~Richter, \emph{The semiclassical propagator in fermionic fock space}, {\emph{Theoretical Chemistry Accounts} {\bfseries 133} (2014) 1563}.

\bibitem{Fulde95}
P.~Fulde, \emph{Electron Correlations in Molecules and Solids}, Springer (1995).

\bibitem{Engl15}
T.~Engl, J.D.~Urbina and K.~Richter, \emph{Periodic mean-field solutions and the spectra of discrete bosonic fields: Trace formula for bose-hubbard models}, {\emph{Phys.~Rev.~E} {\bfseries 92} (2015) 062907}.

\bibitem{TF_Remy}
R.~Dubertrand and S.~Müller, \emph{Spectral statistics of chaotic many-body systems}, \href{https://doi.org/10.1088/1367-2630/18/3/033009}{\emph{New Journal of Physics} {\bfseries 18} (2016) 033009}.

\bibitem{Rammensee2018}
J.~Rammensee, J.-D.~Urbina and K.~Richter, \emph{{Many-Body Quantum Interference and the Saturation of Out-of-Time-Order Correlators}},  May, 2018.

\bibitem{Schoeppl2025}
F.~Schoeppl, R.~Dubertrand, J.-D.~Urbina and K.~Richter, \emph{Universal correlations in chaotic many-body quantum states: Fock-space formulation of berry's random wave model}, \href{https://doi.org/10.1103/PhysRevLett.134.010404}{\emph{Phys. Rev. Lett.} {\bfseries 134} (2025) 010404}.

\bibitem{Haake11}
F.~Haake and K.~Richter, \emph{Pfade, phasen, fluktuationen}, {\emph{Physik Journal} {\bfseries 10} (2011) 35}.

\bibitem{Ehrenfest27}
P.~Ehrenfest, \emph{Bemerkung \"uber die angen\"aherte g\"ultigkeit der klassischen mechanik innerhalb der quantenmechanik}, {\emph{Zeit.~der Phys.} {\bfseries 45} (1927) 455}.

\bibitem{Haake91}
F.~Haake, \emph{Quantum signatures of chaos}, Springer, Berlin (1991).

\bibitem{Keller58}
J.B.~Keller, \emph{Corrected bohr-sommerfeld quantum conditions for nonseparable systems}, {\emph{Ann.~Phys. (N.Y.)} {\bfseries 4} (1958) 180}.

\bibitem{Balian1970}
R.~Balian and C.~Bloch, \emph{{Distribution of Eigenfrequencies for the Wave Equation in a Finite Domain. I. Three-dimensional problem with smooth boundary surface}}, {\emph{Ann. Phys. (NY)} {\bfseries 60} (1970) 401}.

\bibitem{Cvitanovic89}
P.~Cvitanovi\'{c} and B.~Eckhardt, \emph{Periodic-orbit quantization of chaotic systems}, {\emph{Phys.~Rev.~Lett.} {\bfseries 63} (1989) 823}.

\bibitem{Mueller2009}
S.~M{\"u}ller, S.~Heusler, A.~Altland, P.~Braun and F.~Haake, \emph{{Periodic-orbit theory of universal level correlations in quantum chaos}}, {\emph{New J. Phys.} {\bfseries 11} (2009) 103025}.

\bibitem{Hannay84}
J.H.~Hannay and A.M.O.~de~Almeida, \emph{Periodic orbits and a correlation function for the semiclassical density of states}, {\emph{J.~Phys.~A} {\bfseries 17} (1984) 3429}.

\bibitem{Berry85}
M.V.~Berry, \emph{Semiclassical theory of spectral rigidity}, {\emph{Proc.~R.~Soc.~A} {\bfseries 400} (1985) 229}.

\bibitem{Bohigas91}
O.~Bohigas, \emph{Random matrix theories and chaotics dynamics},  in \emph{Chaos and Quantum Physics}, M.-J.~Giannoni, A.~Voros and J.~Zinn-Justin, eds., (Amsterdam), pp.~87--199, North-Holland (1991).

\bibitem{Argaman93}
N.~Argaman, F.-M.~Dittes, E.~Doron, J.P.~Keating, A.Y.~Kitaev, M.~Sieber et~al., \emph{Correlations in the actions of periodic orbits derived from quantum chaos}, {\emph{Phys.~Rev.~Lett.} {\bfseries 71} (1993) 4326}.

\bibitem{Sieber01}
M.~Sieber and K.~Richter, \emph{Correlations between periodic orbits and their role in spectral statistics}, {\emph{Physica Scripta} {\bfseries T90} (2001) 128}.

\bibitem{Sieber02}
K.~Richter and M.~Sieber, \emph{Semiclassical theory of chaotic quantum transport}, {\emph{Phys. Rev. Lett.} {\bfseries 89} (2002) 206801}.

\bibitem{Mueller04}
S.~M\"uller, S.~Heusler, P.~Braun, F.~Haake and A.~Altland, \emph{Semiclassical foundation of universality in quantum chaos}, \href{https://doi.org/10.1103/PhysRevLett.93.014103}{\emph{Phys. Rev. Lett.} {\bfseries 93} (2004) 014103}.

\bibitem{Heusler06}
S.~Heusler, S.~M\"uller, P.~Braun and F.~Haake, \emph{Semiclassical theory of chaotic conductors}, {\emph{Phys. Rev. Lett.} {\bfseries 96} (2006) 066804}.

\bibitem{Mueller07}
S.~Müller, S.~Heusler, P.~Braun and F.~Haake, \emph{Semiclassical approach to chaotic quantum transport}, \href{https://doi.org/10.1088/1367-2630/9/1/012}{\emph{New Journal of Physics} {\bfseries 9} (2007) 12}.

\bibitem{Haake10}
F.~Haake, \emph{Quantum signatures of chaos, third edition}, Springer, Heidelberg (2010).

\bibitem{Berr4}
M.V.~Berry, \emph{Regular and irregular semiclassical wavefunctions}, \href{https://doi.org/10.1088/0305-4470/10/12/016}{\emph{Journal of Physics A: Mathematical and General} {\bfseries 10} (1977) 2083}.

\bibitem{Polkovnikov10}
A.~Polkovnikov, \emph{Phase space representation of quantum dynamics}, {\emph{Ann.~Phys. (N.Y.)} {\bfseries 35} (2010) 1790}.

\bibitem{Sakurai94}
J.J.~Sakurai, \emph{Modern quantum mechanics}, Addison-Wesley (1994).

\bibitem{Dyson67a}
F.J.~Dyson and A.~Lenard, \emph{Stability of matter {I}}, {\emph{J.~Math.~Phys.} {\bfseries 8} (1967) 423}.

\bibitem{Dyson67b}
F.J.~Dyson and A.~Lenard, \emph{Stability of matter {II}}, {\emph{J.~Math.~Phys.} {\bfseries 9} (1967) 698}.

\bibitem{Lieb76}
E.H.~Lieb, \emph{The stability of matter}, {\emph{Rev.~Mod.~Phys.} {\bfseries 48} (1976) 553}.

\bibitem{Bose24}
S.N.~Bose, \emph{Plancks gesetz und lichtquantenhypothese}, {\emph{Zeit.~der Phys.} {\bfseries 26} (1924) 178}.

\bibitem{Einstein24}
A.~Einstein, \emph{Quantentheorie des einatomigen idealen gases}, {\emph{K\"onigliche Preu{\ss}ische Akademie der Wissenschaften} {\bfseries Sitzungsberichte:} (1924) 261}.

\bibitem{richterSemiclassicalTheoryMesoscopic2000}
K.~Richter, \emph{Semiclassical Theory of Mesoscopic Quantum Systems}, no.~161 in Springer Tracts in Modern Physics, Springer.

\bibitem{Rodolfo}
R.A.~Jalabert, G.~Casati, I.~Guarneri and U.~Smilansky, \emph{The semiclassical tool in mesoscopic physics},  in \emph{Proceedings-International School of Physics Enrico Fermi}, vol.~143, pp.~145--222, IOS Press; Ohmsha; 1999, 2000.

\bibitem{Negele98}
J.~Negele and H.~Orland, \emph{Quantum Many-particle Systems}, Advanced Books Classics, Westview Press (1998).

\bibitem{Klauder78}
J.R.~Klauder, \emph{Continuous representations and path integrals, revisited},  in \emph{Proceedings of the NATO Advanced Study Institute on Path Integrals and their Applications in Quantum, Statistical, and Solid State Physics}, G.J.~Papadopoulos and J.T.~Devresse, eds., (New York), pp.~5--38, Plenum (1978).

\bibitem{Baranger01}
M.~Baranger, M.A.M.~de~Aguiar, F.~Keck, H.J.~Korsch and B.~Schellhaass, \emph{Semiclassical approximations in phase space with coherent states}, {\emph{J.~Phys.~A:~Math.~Gen.} {\bfseries 34} (2001) 7227}.

\bibitem{Scully97}
M.O.~Scully and M.S.~Zubairy, \emph{Quantum Optics}, Cambridge University Press, Cambridge, UK (1997).

\bibitem{Engl2016}
T.~Engl, J.D.~Urbina and K.~Richter, \emph{{The semiclassical propagator in Fock space: dynamical echo and many-body interference}}, {\emph{Philos. Trans. Royal Soc. A} {\bfseries 374} (2016) 20150159}.

\bibitem{Bartlett07}
S.D.~Bartlett, T.~Rudolph and R.W.~Spekkens, \emph{Reference frames, superselection rules, and quantum information}, \href{https://doi.org/10.1103/RevModPhys.79.555}{\emph{Rev. Mod. Phys.} {\bfseries 79} (2007) 555}.

\bibitem{Schulman81}
L.S.~Schulman, \emph{Techniques and Applications of Path Integration}, John Wiley and Sons, Inc., New York (1981).

\bibitem{Gradshteyn00}
I.S.~Gradshteyn and I.M.~Ryzhik, \emph{Table of Integrals, Series, and Products}, Academic Press, San Diego (2000).

\bibitem{Gutzwiller07}
M.C.~Gutzwiller, \emph{Quantum chaos}, {\emph{Scholarpedia} {\bfseries 2} (2007) 3146}.

\bibitem{Engl14b}
T.~Engl, J.~Dujardin, A.~Arg\"uelles, P.~Schlagheck, K.~Richter and J.D.~Urbina, \emph{Coherent backscattering in fock space: a signature of quantum many-body interference in interacting bosonic systems}, {\emph{Phys.~Rev.~Lett.} {\bfseries 112} (2014) 140403}.

\bibitem{Engl2015}
T.~Engl, \emph{{A Semiclassical Approach to Many-Body Interference in Fock-Space}}, phd thesis, Universit{\"{a}}t Regensburg, 2015.

\bibitem{OzorioBook}
A.M.~Ozorio~de Almeida, \emph{Hamiltonian systems: Chaos and quantization}, Cambridge University Press, Cambridge (1988).

\bibitem{Tomsovic18}
S.~Tomsovic, \emph{Complex saddle trajectories for multidimensional quantum wave packet and coherent state propagation: Application to a many-body system}, \href{https://doi.org/10.1103/PhysRevE.98.023301}{\emph{Phys. Rev. E} {\bfseries 98} (2018) 023301}.

\bibitem{Schlagheck18}
P.~Schlagheck, D.~Ullmo, J.D.~Urbina, K.~Richter and S.~Tomsovic, \emph{Enhancement of many-body quantum interference in chaotic bosonic systems: The role of symmetry and dynamics}, \href{https://doi.org/10.1103/PhysRevLett.123.215302}{\emph{Phys. Rev. Lett.} {\bfseries 123} (2019) 215302}.

\bibitem{DAVIDSON17}
S.~Davidson, D.~Sels and A.~Polkovnikov, \emph{Semiclassical approach to dynamics of interacting fermions}, \href{https://doi.org/https://doi.org/10.1016/j.aop.2017.07.003}{\emph{Annals of Physics} {\bfseries 384} (2017) 128}.

\bibitem{Schmitt19}
M.~Schmitt, D.~Sels, S.~Kehrein and A.~Polkovnikov, \emph{Semiclassical echo dynamics in the sachdev-ye-kitaev model}, \href{https://doi.org/10.1103/PhysRevB.99.134301}{\emph{Phys. Rev. B} {\bfseries 99} (2019) 134301}.

\bibitem{Schachenmayer15}
J.~Schachenmayer, A.~Pikovski and A.M.~Rey, \emph{Many-body quantum spin dynamics with monte carlo trajectories on a discrete phase space}, \href{https://doi.org/10.1103/PhysRevX.5.011022}{\emph{Phys. Rev. X} {\bfseries 5} (2015) 011022}.

\bibitem{Tomsovic2018}
S.~Tomsovic, P.~Schlagheck, D.~Ullmo, J.-D.~Urbina and K.~Richter, \emph{{Post-Ehrenfest many-body quantum interferences in ultracold atoms far out of equilibrium}}, \href{https://doi.org/10.1103/PhysRevA.97.061606}{\emph{Phys. Rev. A} {\bfseries 97} (2018) 061606}.

\bibitem{Berman78}
G.P.~Berman and G.M.~Zaslavsky, \emph{Condition of stochasticity of quantum nonlinear systems}, {\emph{Physica A} {\bfseries 91} (1978) 450}.

\bibitem{Trotzky12}
S.~Trotzky, Y.-A.~Chen, A.~Flesch, I.P.~McCulloch, U.~Schollw\"ock, J.~Eisert et~al., \emph{Probing the relaxation towards equilibrium in an isolated strongly correlated 1d bose gas}, {\emph{Nature Physics} {\bfseries 8} (2012) 325}.

\bibitem{Kaufman16}
A.M.~Kaufman, M.E.~Tai, A.~Lukin, M.~Rispoli, R.~Schittko, P.M.~Preiss et~al., \emph{Quantum thermalization through entanglement in an isolated many-body system}, {\emph{Science} {\bfseries 353} (2002) 794}.

\bibitem{Choi16}
J.~Choi, S.~Hild, J.~Zeiher, P.~Schau$\beta$, A.~Rubio-Abadal, T.~Yefsah et~al., \emph{Exploring the many-body localization transition in two dimensions}, {\emph{Science} {\bfseries 352} (2016) 1547}.

\bibitem{Esteve08}
J.~Estève, C.~Gross, A.~Weller, S.~Giovanazzi and M.K.~Oberthaler, \emph{Squeezing and entanglement in a bose-einstein condensate}, {\emph{Nature} {\bfseries 455} (2008) 1216}.

\bibitem{Gross10}
C.~Gross, T.~Zibold, E.~Nicklas, J.~Est\'eve and M.K.~Oberthaler, \emph{Nonlinear atom interferometer surpasses classical precision limit}, {\emph{Nature} {\bfseries 464} (2002) 1165}.

\bibitem{Pitaevskii03}
L.P.~Pitaevskii and S.~Stringari, \emph{Bose-Einstein Condensation}, Oxford University Press, Oxford (2003).

\bibitem{Yao16}
N.Y.~Yao, F.~Grusdt, B.~Swingle, M.D.~Lukin and E.A.D.~Dan M.~Stamper-Kurn, Joel E.~Moore, \emph{Dmrg and periodic boundary conditions: a quantum information perspective}, {\emph{arXiv:1607.01801v1 [quant-ph]} (2016) }.

\bibitem{Akkermans07}
E.~Akkermans and G.~Montambaux, \emph{Mesoscopic Physics of Electrons and Photons}, Cambridge University Press, Cambridge (2007).

\bibitem{Bohigas84}
O.~Bohigas, M.-J.~Giannoni and C.~Schmit, \emph{Characterization of chaotic quantum spectra and universality of level fluctuation laws}, {\emph{Phys.~Rev.~Lett.} {\bfseries 52} (1984) 1}.

\bibitem{Turek05}
M.~Turek, D.~Spehner, S.~M\"{u}ller and K.~Richter, \emph{Semiclassical form factor for spectral and matrix element fluctuations of multidimensional chaotic systems}, {\emph{Phys.~Rev.~E} {\bfseries 71} (2005) 016210}.

\bibitem{Liao20}
Y.~Liao, A.~Vikram and V.~Galitski, \emph{Many-body level statistics of single-particle quantum chaos}, \href{https://doi.org/10.1103/PhysRevLett.125.250601}{\emph{Phys. Rev. Lett.} {\bfseries 125} (2020) 250601}.

\bibitem{Swingle16}
B.~Swingle, G.~Bentsen, M.~Schleier-Smith and P.~Hayden, \emph{Measuring the scrambling of quantum information}, {\emph{Phys.~Rev.~A} {\bfseries 94} (2016) 040302(R)}.

\bibitem{Xu22}
S.~Xu and B.~Swingle, \emph{Scrambling dynamics and out-of-time ordered correlators in quantum many-body systems: a tutorial},  2022.

\bibitem{Brody81}
T.A.~Brody, J.~Flores, J.B.~French, P.A.~Mello, A.~Pandey and S.S.M.~Wong, \emph{Random-matrix physics: spectrum and strength fluctuations}, {\emph{Rev.~Mod.~Phys.} {\bfseries 53} (1981) 385}.

\bibitem{Bohigas83}
O.~Bohigas, R.~ul~Haq and A.~Pandey, \emph{Fluctuation properties of nuclear energy levels and widths: comparison of theory with experiment},  in \emph{Nuclear data for science and technology}, K.H.~B\"ockhoff, ed., (Dordrecht), pp.~809--814, Reidel (1983).

\bibitem{KotaBook}
V.K.B.~Kota and R.~ul~Haq, eds., \emph{Spectral distributions in nuclei and statistical spectroscopy}, World Scientific, Singapore (2010).

\bibitem{Bertini18}
B.~Bertini, P.~Kos and T.~Prosen, \emph{Exact spectral form factor in a minimal model of many-body quantum chaos}, \href{https://doi.org/10.1103/PhysRevLett.121.264101}{\emph{Phys. Rev. Lett.} {\bfseries 121} (2018) 264101}.

\bibitem{Kos18}
P.~Kos, M.~Ljubotina and T.~Prosen, \emph{Many-body quantum chaos: Analytic connection to random matrix theory}, \href{https://doi.org/10.1103/PhysRevX.8.021062}{\emph{Phys. Rev. X} {\bfseries 8} (2018) 021062}.

\bibitem{Bertini19}
B.~Bertini, P.~Kos and T.~Prosen, \emph{Exact correlation functions for dual-unitary lattice models in 1+ 1 dimensions}, {\emph{Phys.~Rev.~Lett.} {\bfseries 123} (2019) 210601}.

\bibitem{Chan18}
A.~Chan, A.D.~Luca and J.T.~Chalker, \emph{Solution of a minimal model for many-body quantum chaos}, .

\bibitem{MBQC1}
A.R.~Kolovsky and A.~Buchleitner, \emph{Quantum chaos in the bose-hubbard model}, \href{https://doi.org/10.1209/epl/i2004-10265-7}{\emph{Europhysics Letters} {\bfseries 68} (2004) 632}.

\bibitem{Garcia17}
A.M.~Garc\'{\i}a-Garc\'{\i}a and J.J.M.~Verbaarschot, \emph{Analytical spectral density of the {Sachdev}-{Ye}-{Kitaev} model at finite {$N$}}, \href{https://doi.org/10.1103/PhysRevD.96.066012}{\emph{Phys. Rev. D} {\bfseries 96} (2017) 066012}.

\bibitem{Asaga01}
T.~Asaga, L.~Benet, T.~Rupp and H.A.~Weidenm\"uller, \emph{Non-ergodic behavior of the k-body embedded {Gaussian} random ensembles for bosons}, {\emph{Europhys.~Lett.} {\bfseries 56} (2001) 340}.

\bibitem{Srednicki02}
M.~Srednicki, \emph{Spectral statistics of the k-body random-interaction model}, {\emph{Phys.~Rev.~E} {\bfseries 66} (2002) 046138}.

\bibitem{Keating91}
J.P.~Keller, \emph{The cat maps: quantum mechanics and classical motion}, {\emph{Nonlinearity} {\bfseries 4} (1991) 309}.

\bibitem{Heusler07}
S.~Heusler, S.~M\"uller, A.~Altland, P.~Braun and F.~Haake, \emph{Periodic-orbit theory of level correlations}, {\emph{Phys.~Rev.~Lett.} {\bfseries 98} (2007) 044103}.

\bibitem{Keating07}
J.P.~Keating and S.~Müller, \emph{Resummation and the semiclassical theory of spectral statistics}, \href{https://doi.org/10.1098/rspa.2007.0178}{\emph{Proceedings of the Royal Society A: Mathematical, Physical and Engineering Sciences} {\bfseries 463} (2007) 3241} [\href{https://arxiv.org/abs/https://royalsocietypublishing.org/doi/pdf/10.1098/rspa.2007.0178}{{\ttfamily https://royalsocietypublishing.org/doi/pdf/10.1098/rspa.2007.0178}}].

\bibitem{Waltner09}
D.~Waltner, S.~Heusler, J.D.~Urbina and K.~Richter, \emph{The semiclassical origin of curvature effects in universal spectral statistics}, \href{https://doi.org/10.1088/1751-8113/42/29/292001}{\emph{Journal of Physics A: Mathematical and Theoretical} {\bfseries 42} (2009) 292001}.

\bibitem{Waltner20}
D.~Waltner and K.~Richter, \emph{Towards a semiclassical understanding of chaotic single- and many-particle quantum dynamics at post-heisenberg time scales}, \href{https://doi.org/10.1103/PhysRevE.100.042212}{\emph{Phys. Rev. E} {\bfseries 100} (2019) 042212}.

\bibitem{Nahum17}
A.~Nahum, J.~Ruhman, S.~Vijay and J.~Haah, \emph{Quantum entanglement growth under random unitary dynamics}, \href{https://doi.org/10.1103/PhysRevX.7.031016}{\emph{Phys. Rev. X} {\bfseries 7} (2017) 031016}.

\bibitem{suppmat}
F.~Schoeppl, R.~Dubertrand, J.D.~Urbina and K.~Richter{\emph{Supplemental material} (2024) }.

\bibitem{mcdonaldSpectrumEigenfunctionsHamiltonian1979}
S.W.~McDonald and A.N.~Kaufman, \emph{Spectrum and {{Eigenfunctions}} for a {{Hamiltonian}} with {{Stochastic Trajectories}}}, \href{https://doi.org/10.1103/physrevlett.42.1189}{\emph{Physical Review Letters} {\bfseries 42} (1979) 1189}.

\bibitem{mcdonaldWaveChaosStadium1988}
S.W.~McDonald and A.N.~Kaufman, \emph{Wave chaos in the stadium: {{Statistical}} properties of short-wave solutions of the {{Helmholtz}} equation}, \href{https://doi.org/10.1103/physreva.37.3067}{\emph{Physical Review A} {\bfseries 37} (1988) 3067}.

\bibitem{Sied1}
S.~Hortikar and M.~Srednicki, \emph{Correlations in chaotic eigenfunctions at large separation}, \href{https://doi.org/10.1103/PhysRevLett.80.1646}{\emph{Physical Review Letters} {\bfseries 80} (1998) 1646}.

\bibitem{JDUre}
J.D.~Urbina and K.~Richter, \emph{Random quantum states: recent developments and applications}, \href{https://doi.org/10.1080/00018732.2013.860277}{\emph{Advances in Physics} {\bfseries 62} (2013) 363}.

\bibitem{RWM3}
A.M.~Chang, H.U.~Baranger, L.N.~Pfeiffer, K.W.~West and T.Y.~Chang, \emph{Non-gaussian distribution of coulomb blockade peak heights in quantum dots}, \href{https://doi.org/10.1103/PhysRevLett.76.1695}{\emph{Physical Review Letters} {\bfseries 76} (1996) 1695}.

\bibitem{Urbina06}
J.D.~Urbina and K.~Richter, \emph{Statistical description of eigenfunctions in chaotic and weakly disordered systems beyond universality}, \href{https://doi.org/10.1103/PhysRevLett.97.214101}{\emph{Physical Review Letters} {\bfseries 97} (2006) 214101}.

\bibitem{RWM2}
M.R.~Dennis, \emph{Nodal densities of planar gaussian random waves}, {\emph{The European Physical Journal Special Topics} {\bfseries 145} (2007) 191}.

\bibitem{PhysRevA.37.3067}
S.W.~McDonald and A.N.~Kaufman, \emph{Wave chaos in the stadium: Statistical properties of short-wave solutions of the helmholtz equation}, \href{https://doi.org/10.1103/PhysRevA.37.3067}{\emph{Physical Review A} {\bfseries 37} (1988) 3067}.

\bibitem{RWMBe3}
M.V.~Berry and M.R.~Dennis, \emph{Phase singularities in isotropic random waves}, \href{https://doi.org/10.1098/rspa.2000.0602}{\emph{Proceedings of the Royal Society of London. Series A: Mathematical, Physical and Engineering Sciences} {\bfseries 456} (2000) 2059}.

\bibitem{bogomolny2002percolation}
E.~Bogomolny and C.~Schmit, \emph{Percolation model for nodal domains of chaotic wave functions}, {\emph{Physical Review Letters} {\bfseries 88} (2002) 114102}.

\bibitem{HJ1}
M.~Barth and H.-J.~St\"ockmann, \emph{Current and vortex statistics in microwave billiards}, \href{https://doi.org/10.1103/PhysRevE.65.066208}{\emph{Physical Review E} {\bfseries 65} (2002) 066208}.

\bibitem{HJ2}
A.~B\"arnthaler, S.~Rotter, F.~Libisch, J.~Burgd\"orfer, S.~Gehler, U.~Kuhl et~al., \emph{Probing decoherence through fano resonances}, \href{https://doi.org/10.1103/PhysRevLett.105.056801}{\emph{Physical Review Letters} {\bfseries 105} (2010) 056801}.

\bibitem{HJ3}
H.-J.~St{\"o}ckmann, \emph{Chaos in Microwave resonators}, Springer (2013).

\bibitem{RWMBe4}
M.~Berry and M.~Dennis, \emph{Polarization singularities in isotropic random vector waves}, \href{https://doi.org/10.1098/rspa.2000.0660}{\emph{Proceedings of the Royal Society of London. Series A: Mathematical, Physical and Engineering Sciences} {\bfseries 457} (2001) 141}.

\bibitem{Dennis1}
R.~H{\"o}hmann, U.~Kuhl, H.-J.~St{\"o}ckmann, J.~Urbina and M.~Dennis, \emph{Density and correlation functions of vortex and saddle points in open billiard systems}, {\emph{Physical Review E} {\bfseries 79} (2009) 016203}.

\bibitem{Dennis2}
A.J.~Taylor and M.R.~Dennis, \emph{Vortex knots in tangled quantum eigenfunctions}, {\emph{Nature communications} {\bfseries 7} (2016) 1}.

\bibitem{jainNodalPortraitsQuantum2017}
S.R.~Jain and R.~Samajdar, \emph{Nodal portraits of quantum billiards: {{Domains}}, lines, and statistics}, \href{https://doi.org/10.1103/revmodphys.89.045005}{\emph{Reviews of Modern Physics} {\bfseries 89} (2017) }.

\bibitem{Rig1}
M.~Rigol, V.~Dunjko and M.~Olshanii, \emph{Thermalization and its mechanism for generic isolated quantum systems}, {\emph{Nature} {\bfseries 452} (2008) 854}.

\bibitem{doi:10.1126/science.aal3837}
C.~Gross and I.~Bloch, \emph{Quantum simulations with ultracold atoms in optical lattices}, {\emph{Science} {\bfseries 357} (2017) 995}.

\bibitem{Aidelsburger_2018}
M.~Aidelsburger, \emph{Artificial gauge fields and topology with ultracold atoms in optical lattices}, {\emph{Journal of Physics B: Atomic, Molecular and Optical Physics} {\bfseries 51} (2018) 193001}.

\bibitem{RevModPhys.80.885}
I.~Bloch, J.~Dalibard and W.~Zwerger, \emph{Many-body physics with ultracold gases}, {\emph{Review of Modern Physics} {\bfseries 80} (2008) 885}.

\bibitem{Heller_2007}
E.J.~Heller and B.R.~Landry, \emph{Statistical properties of many particle eigenfunctions}, \href{https://doi.org/10.1088/1751-8113/40/31/006}{\emph{Journal of Physics A: Mathematical and Theoretical} {\bfseries 40} (2007) 9259}.

\bibitem{voros1976semi}
A.~Voros, \emph{Semi-classical approximations.[wigner representation, wkb-maslov method, bohr-like quantization]}, {\emph{Annales de l'Institut Henri Poincar\'e} {\bfseries 24} (1976) }.

\bibitem{berry1977semi}
M.V.~Berry, \emph{Semi-classical mechanics in phase space: a study of wigner’s function}, {\emph{Philosophical Transactions of the Royal Society of London. Series A, Mathematical and Physical Sciences} {\bfseries 287} (1977) 237}.

\bibitem{Larkin69}
A.I.~Larkin and Y.N.~Ovchinnikov, \emph{Quasiclassical method in the theory of superconductivity}, {\emph{Soviet Physics JETP} {\bfseries 28} (1969) 1200}.

\bibitem{Shenker14}
S.H.~Shenker and D.~Stanford, \emph{Black holes and the butterfly effect}, {\emph{JHEP} {\bfseries 3} (2014) 67}.

\bibitem{Maldacena16}
J.~Maldacena, S.H.~Shenker and D.~Stanford, \emph{{A bound on chaos}}, \href{https://doi.org/10.1007/JHEP08(2016)106}{\emph{J. High Energy Phys.} {\bfseries 2016} (2016) 106}.

\bibitem{Sekino08}
Y.~Sekino and L.~Susskind, \emph{Fast scramblers}, {\emph{JHEP} {\bfseries 10} (2008) 065}.

\bibitem{Mi21}
X.M.~et~al., \emph{Information scrambling in quantum circuits}, \href{https://doi.org/10.1126/science.abg5029}{\emph{Science} {\bfseries 374} (2021) 1479} [\href{https://arxiv.org/abs/https://www.science.org/doi/pdf/10.1126/science.abg5029}{{\ttfamily https://www.science.org/doi/pdf/10.1126/science.abg5029}}].

\bibitem{Zhu16}
G.~Zhu, M.~Hafezi and T.~Grover, \emph{{Measurement of many-body chaos using a quantum clock}}, \href{https://doi.org/10.1103/PhysRevA.94.062329}{\emph{Physical Review A} {\bfseries 94} (2016) 062329} [\href{https://arxiv.org/abs/1607.00079}{{\ttfamily 1607.00079}}].

\bibitem{Li16}
J.~Li, R.~Fan, H.~Wang, B.~Ye, B.~Zeng, H.~Zhai et~al., \emph{{Measuring out-of-time-order correlators on a nuclear magnetic resonance quantum simulator}}, \href{https://doi.org/10.1103/PhysRevX.7.031011}{\emph{Physical Review X} {\bfseries 7} (2017) 1} [\href{https://arxiv.org/abs/1609.01246}{{\ttfamily 1609.01246}}].

\bibitem{Garttner16}
M.~Garttner, J.G.~Bohnet, A.~Safavi-Naini, M.L.~Wall, J.J.~Bollinger and A.M.~Rey, \emph{{Measuring out-of-time-order correlations and multiple quantum spectra in a trapped-ion quantum magnet}}, \href{https://doi.org/10.1038/NPHYS4119}{\emph{Nature Physics} {\bfseries 13} (2017) 781} [\href{https://arxiv.org/abs/1608.08938}{{\ttfamily 1608.08938}}].

\bibitem{Dominguez21}
F.D.~Dom\'{\i}nguez and G.A.~\'Alvarez, \emph{Dynamics of quantum information scrambling under decoherence effects measured via active spin clusters}, \href{https://doi.org/10.1103/PhysRevA.104.062406}{\emph{Phys. Rev. A} {\bfseries 104} (2021) 062406}.

\bibitem{Lassl03}
A.~Lassl, \emph{Semiklassik jenseits der diagonalnäherung: Anwendung auf ballistische mesoskopische systeme}, {\emph{diploma thesis} (2003) }.

\bibitem{Schanz03}
H.~Schanz, M.~Puhlmann and T.~Geisel, \emph{Shot noise in chaotic cavities from action correlations}, \href{https://doi.org/10.1103/PhysRevLett.91.134101}{\emph{Phys. Rev. Lett.} {\bfseries 91} (2003) 134101}.

\bibitem{Braun06}
P.~Braun, S.~Heusler, S.~Müller and F.~Haake, \emph{Semiclassical prediction for shot noise in chaotic cavities}, \href{https://doi.org/10.1088/0305-4470/39/11/l01}{\emph{Journal of Physics A: Mathematical and General} {\bfseries 39} (2006) L159}.

\bibitem{Kuipers11}
J.~Kuipers, T.~Engl, G.~Berkolaiko, C.~Petitjean, D.~Waltner and K.~Richter, \emph{Density of states of chaotic andreev billiards}, \href{https://doi.org/10.1103/PhysRevB.83.195316}{\emph{Phys. Rev. B} {\bfseries 83} (2011) 195316}.

\bibitem{Kuipers13}
J.~Kuipers and K.~Richter, \emph{Transport moments and andreev billiards with tunnel barriers}, \href{https://doi.org/10.1088/1751-8113/46/5/055101}{\emph{Journal of Physics A: Mathematical and Theoretical} {\bfseries 46} (2013) 055101}.

\bibitem{Han16}
X.~Han and B.~Wu, \emph{Ehrenfest breakdown of the mean-field dynamics of bose gases}, \href{https://doi.org/10.1103/PhysRevA.93.023621}{\emph{Phys. Rev. A} {\bfseries 93} (2016) 023621}.

\bibitem{Kurchan18}
J.~Kurchan, \emph{Quantum bound to chaos and the semiclassical limit}, {\emph{J. Stat. Phys.} {\bfseries 171} (12018) 965–979}.

\bibitem{Jalabert18}
R.A.~Jalabert, I.~Garc\'{\i}a-Mata and D.A.~Wisniacki, \emph{Semiclassical theory of out-of-time-order correlators for low-dimensional classically chaotic systems}, \href{https://doi.org/10.1103/PhysRevE.98.062218}{\emph{Phys. Rev. E} {\bfseries 98} (2018) 062218}.

\bibitem{Rozenbaum17}
E.B.~Rozenbaum, S.~Ganeshan and V.~Galitski, \emph{Lyapunov exponent and out-of-time-ordered correlator's growth rate in a chaotic system}, {\emph{Phys.~Rev.~Lett.} {\bfseries 118} (2017) 086801}.

\bibitem{Garcia-Mata18}
I.~Garc\'{\i}a-Mata, M.~Saraceno, R.A.~Jalabert, A.J.~Roncaglia and D.A.~Wisniacki, \emph{Chaos signatures in the short and long time behavior of the out-of-time ordered correlator}, \href{https://doi.org/10.1103/PhysRevLett.121.210601}{\emph{Phys. Rev. Lett.} {\bfseries 121} (2018) 210601}.

\end{thebibliography}\endgroup

\bibliographystyle{JHEP}%
%\bibliography{references}
%\bibliography{refs,refsRW}

\end{document}